\documentclass[aps,eqsecnum,nofootinbib,showpacs,preprintnumbers]{revtex4-2}

\usepackage[utf8]{inputenc}
\usepackage[T1]{fontenc}
\usepackage{amsmath,amssymb,amsfonts}
\usepackage{mathrsfs}
\usepackage{physics}
\usepackage{bm}
\usepackage{enumitem}
\usepackage{graphicx}
\usepackage{comment}
\usepackage{braket}
\usepackage{url}
\usepackage[colorlinks=true, citecolor=blue, urlcolor=blue]{hyperref}

\usepackage{graphicx}  
\usepackage{bbold}
\usepackage{slashed}
\usepackage{moresize}
\usepackage[normalem]{ulem}
\newcommand{\bL}{\begin{Large}}
\newcommand{\eL}{\end{Large}}

\newcommand{\bea}{\begin{eqnarray}}
\newcommand{\eea}{\end{eqnarray}}
\newcommand{\be}{\begin{equation}}
\newcommand{\ee}{\end{equation}}

\newcommand{\cma}{\color{black}}
\newcommand{\cmaa}{\color{black}}

\newcommand{\cbl}{\color{black}}
\usepackage[colorinlistoftodos]{todonotes}
\usepackage{appendix}

\def\eq{\eqref}

\begin{document}
\preprint{\leftline{KCL-PH-TH/2025-{\bf 26}}}
\title{Squeezed gravitons from superradiant axion fields around rotating black holes}

\author{Panagiotis Dorlis$^a$}

\author{Nick E. Mavromatos$^{b,c,d}$}

\author{Sarben Sarkar$^c$}

\author{Sotirios-Neilos Vlachos$^a$}
\medskip
\affiliation{$^a$ Physics Division, School of Applied Mathematical and Physical Sciences, National Technical University of Athens, Zografou Campus, Athens 157 80, Greece}

\affiliation{$^b$ Department of Theoretical Physics and IFIC, University of Valencia and CSIC, E-46100, Valencia, Spain}

\affiliation{$^c$Theoretical Particle Physics and Cosmology Group, Department of Physics, King's College London, London, WC2R 2LS, UK}

\affiliation{$^d$ Currently on leave from: Physics Division, School of Applied Mathematical and Physical Sciences, National Technical University of Athens, Zografou Campus, Athens 157 80, Greece}


\begin{abstract}
We propose, in (3+1)-dimensional spacetimes, a novel astrophysical source of squeezed graviton states,  due to  superradiant axionic clouds surrounding rotating (Kerr-type) black holes (BH). The microscopic origin of these axions is diverse, ranging from the Kalb-Ramond (model-independent) axions and compactification axions in string theory, to   contorted geometries exemplified by a totally antisymmetric component of torsion in Einstein-Cartan theory. 
The axion fields couple to chiral gauge and gravitational  Chern-Simons (CS) anomaly terms in the effective gravitational actions. In the presence of a Kerr BH background, such axions lead, upon acquiring a mass, to superradiance and the production of pairs of entangled gravitons in a squeezed state. The specific microscopic origin of the axions  is not important, provided they are massive. 
\color{black} We explain, by means of simplified but representative examples, how this multimode squeezed-graviton state can be studied via an Autonne-Takagi decomposition,  used in  quantum optics\color{black}.  In the effective action it is shown  that  squeezing effects associated with conventional general relativity (GR) dominate, by many orders of magnitude, the corresponding effects due to the CS gravitational anomaly terms. For a sufficiently long lifetime of the axionic cloud of the BH, we find that significant squeezing (quantified through the average number of gravitons with respect to the appropriate vacuum) can be produced from the GR effects. \color{black} Our approach also allows for an imposition of phenomenological upper bounds on axion-cloud life times, due to the non-observation of squeezed graviton states in current interferometers\color{black}. In addition, it is demonstrated explicitly that the structure of the entangled states (when the latter are expressed in a left-right polarization basis) depends highly on whether GR or the anomalous CS effects produce the entanglement.

\end{abstract}

\maketitle

\section{Introduction}

Detection of signatures of quantum gravity (QG) remains an exciting but elusive aim in physics. This search for new physics has been vastly enhanced by development of the technology of detectors~\cite{Arun2022-oe}. The discovery of gravitational waves~\cite{LIGOScientific:2016aoc,RevModPhys.90.040503}, which  only change distances by $10^{-21}$ m,  is a triumph of experiment, specially in a noisy environment. We  are emboldened  by the rapid advances in detection technology  to look for quantum aspects of gravity~\cite{Addazi:2021xuf}. However, gravity as a force is not just another force but is intertwined with the notion of space and time. This,  leads to a debate whether we should be quantising gravity in any conventional sense \cite{Bianconi:2024aju}. In the approach of \cite{Verlinde,Verlinde2}, for instance, gravity is not a fundamental, but an emerging, entropic in nature, force, associated with information carried by the positions of material degrees of freedom. In such an approach, there are no quantum gravitons. We, however, take the view that QG entails  quantising the metric tensor, regarded as a field on spacetime. In particular, we adopt the (rather conservative) approach of semiclassical gravity~\cite{Birrell2012-xt}, which as we shall argue will suffice for our purposes of identifying potential astrophysical sources of graviton squeezing. In this respect, our approach will pertain to the weak-graviton, low-energy limit of a fundamental theory of gravity, which could be string theory~\cite{str1,str2,pol1,pol2}, loop-quantum gravity~\cite{loop,loop2}, {\it etc}. Certainly our low-energy approach will be compatible with the asymptotic safety continuum approach to QG~\cite{as,as2}.

  Since the gravitational coupling constant is forty orders of magnitude smaller than the electromagnetic one, detecting individual gravitons is close to impossible~\cite{Dyson:2013hbl,Carney:2023nzz,singlegrav}. We consider {\it{non-classical}} collective states of gravitons, similar to collective states of photons in quantum optics~\cite{WallsMilburn2008,Boyd2008,Boyd_optics}. Our effective model of QG is inspired by: (i) the gravitational multiplet of string theory~\cite{str1,str2,pol1,pol2}  and (ii) gravity augmented with dynamical torsion. There is significant overlap between the two models~\cite{Duncan:1992vz}. 
 
 There is a parallel between spacetime metrics in GR and dielectric media in \color{black} electromagnetism~\cite{Scully:1982fn,Barcelo:2005fc}\color{black}. In classical electrodynamics light propagates through a dielectric medium with modified behaviour due to refractive index  and polarisation tensors. So the dielectric modifies the effective geometry experienced by electromagnetic waves. In GR, matter and light follow geodesics determined by the metric $g_{\mu \nu}$. The metric modifies the causal structure since light cones bend, distances warp and time dilates. Transformation optics \cite{Ginis2015-mk}, for example,  uses spatially varying media to guide light, mimicking horizons in a curved spacetime.

The analogy between spacetime metrics as gravitational dielectrics and optical media has implications also for QG and quantum optics~\cite{WallsMilburn2008,Boyd2008,Boyd_optics}. Quantum optics gives a toolbox for considering field quantisation and interactions in media through theoretical concepts and methodologies for the generation and characterisation of nonclassical states. In perturbative QG~\cite{Hooft2003-xp} we expand around a fixed background spacetime $g^{(0)} _{\mu \nu}$:
\begin{align} \label{pert_metr}   
g_{\mu \nu }=g^{(0)} _{\mu \nu }+\kappa h_{\mu \nu }\,,
\end{align}
with $\kappa =\sqrt{8\pi G}$, where $G = M_{\rm Pl}^{-2}$ is the (3+1)-dimensional Newton's constant, and 
$M_{\rm Pl}= 2.435 \times 10^{18}$~GeV the reduced Planck mass.\footnote{Throughout this article we work in natural units, $\hbar=c=1$.} The graviton field $h_{\mu \nu }$ mediates the gravitational interaction.\footnote{ This approach has received some support from the study of decoherence around black holes \cite{Danielson:2024yru}. } Just as electromagnetic fields respond to polarizable media, the quantum vacuum in gravity can be treated as a polarisable gravitational medium. The metric perturbation $h_{\mu \nu }$ is analogous to a fluctuation in a refractive index. Metric perturbations scatter gravitons, just as dielectric fluctuations scatter photons.  Classical gravitational waves can be modeled by coherent quantum states of gravitons. A squeezed vacuum of the electromagnetic field~\cite{WallsMilburn2008,Boyd2008} is mathematically analogous to primordial gravitational wave fluctuations (e.g. from inflation) \cite{hjlj-sv2t}.\footnote{The analogy between gravity and optics, as far as non-trivial optical properties of the vacuum are concerned, goes even further to cover the exotic case of quantum-gravity-induced spacetime foam, which might exhibit a non-trivial refractive index, responsible for modified dispersion relations of radiation and matter probes, propagating in it~\cite{aemn1,aemns,efmmn}. This conjecture can be falsified in principle using, for instance,  multimessenger observations~\cite{Addazi:2021xuf}.} 

In our approach we shall be interested in squeezed graviton states produced around a rotating (Kerr type~\cite{Kerr:1963ud}) astrophysical black hole (BH). 
It is known that such BHs are  accompanied by  clouds of massive pseudoscalar fields (axions or axion-like particles (ALP)) outside the exterior BH horizon, which can become superradiant under certain conditions~\cite{Detweiller,BritoCardoso,BritoScales1,BritoScales2}. In such a case, as we argue briefly in a recent work~\cite{Dorlis:2025zzz},  entangled polarization states of two quantum gravitons are produced, to leading orders in a weak-graviton expansion;  the production is caused   by either 
the fusion of two axions within the context of GR, or the decay of an axion in the case of Chern-Simons (CS) gravity~\cite{Jackiw:2003pm,Alexander:2009tp}, which is a modification of GR that includes an anomalous gravitational CS (gCS) interaction coupled linearly  to an axion-like field. As argued in \cite{Dorlis:2025zzz}, the squeezing effects (quantified as the average number of squeezed graviton states with respect to an appropriate vacuum) due to the GR terms in the gravitational effective action are much larger than the corresponding effects due to the gCS anomaly terms. We shall also see that the presence of gCS terms affects the form of the entangled graviton states. 

There are many analogies between squeezing in  QG and quantum optics~\cite{WallsMilburn2008,Boyd2008,Boyd_optics}. 
 An optical beam splitter is analogous to graviton mixing or mode coupling. A Kerr Black-Hole (BH) 
 medium in optics can correspond to graviton self-interaction. This fruitful interplay happens, not just at the theoretical level, but also at the experimental level, spawning laboratory analogues of QG via metamaterials \cite{PhysRevA.88.033843}, Bose Einstein Condensates (BECs)~\cite{PhysRevLett.126.041105} and nonlinear optics. BH Horizons are, for example, simulated by moving media or modulated dielectrics. 

The microscopic origin of axion fields in our framework can be diverse.  The fields may   come from  the cancellation of  quantum anomalies in the context of effective gravitational field theories arising from string theory (string-model-independent axions)~\cite{str1,str2,pol1,pol2,Duncan:1992vz,Svrcek:2006yi}, or be  moduli fields characterising extra dimensions in string theory compactification (compactification axions)~\cite{Svrcek:2006yi} , or even be associated with the totally antisymmetric components of torsion in Einstein-Cartan theory of fermions in contorted geometries~\cite{Cartan:1938ph,hehl}. \emph{All }such axions can lead to BH superradiance, provided there exist mechanisms which render them {\it massive}. We assume (without further discussion) that the standard (non-perturbative) mechanisms~\cite{Kim:2008hd,Borsanyi:2016ksw}, are responsible for  axion mass generation. The possible origin of axions in our framework is further discussed in section \ref{sec:originaxion}; their massive nature is of primary importance for the formation of a condensate-like cloud around a Kerr BH.

Even though we do not know the UV complete theory involving the Standard Model (SM) and gravity, we can adopt an effective field theory approach, which encodes all low-energy observables in one functional. So the generating functional $Z[J]$ has the schematic form:
\[Z\left[ J \right] \sim \int D\varphi e^{iS_{eff}\left[ \varphi \right] +i\int J\varphi}\]
for generic fields $\varphi$; correlation functions and S-matrix elements are obtained by taking functional derivatives with respect to the source $J$. The unknown UV theory is organised through a series
\[S_{eff}=\int d^{4}x\sum_{n\ge 0} \frac{c_{n}}{\Lambda^{n-1}} O_{n}\left( x \right)\]
 where $\Lambda$ is the cut-off energy scale for the effective theory. The Wilson coefficients ($c_n$) are dimensionless and so could be numerical constants or functions of dimensionless ratios of physical constants of the theory. The low energy effective theories, derived from string theory for gravitational degrees of freedom, are of this type with $\Lambda = M_s \lesssim M_{\rm Pl}$, where $M_s$ is the string scale~\cite{str1,str2,pol1,pol2}. Higher dimensional terms, at energy scale $E$, are suppressed by powers of $E/\Lambda$. When gravity is included in the effective theory through the Riemann tensor, say, the only pertinent length scale near a non-rotating horizon is the gravitational radius, $r_g= G \mathcal{M}$, where $\mathcal M$ is the black-hole mass. For a Kerr BH ({\it cf.} \eqref{Kerr} in Appendix \ref{app:BHSuperradiance}), of interest to us here, there is an additional length, the Boyer-Lindquist spin parameter $\alpha=\mathcal{J}_H/{\mathcal M}$, where $\mathcal{J}_H$ is the angular momentum of the BH. Hence, we can construct  a dimensionless spin $\alpha/r_g$, and so, when we compare different operators in our effective action, it is not possible to rely just on dimensional analysis to compare the effect on physical quantities, such as the expectation of the number of gravitons in a quantum state.

 Before discussing theoretical issues, we give some empirical arguments for the validity of semiclassical analysis near the horizon of a Kerr BH. The key idea is that, if curvature scales remain small compared to the Planck scale and quantum fields are in a regular state (or vacuum, to be elaborated later), quantum field theory on a classical background is a good approximation.

 The semiclassical Einstein equation is 
 \begin{align}\label{Eeq}
 G_{\mu \nu}\left[ g \right] =8\pi G\, \left< T_{\mu \nu} \right> \ ,
 \end{align}
 where 
 \begin{align}\label{ET}
 G_{\mu \nu} =  R_{\mu\nu}- \frac{1}{2}g_{\mu\nu} R\,
 \end{align}
 is the Einstein tensor, $g_{\mu \nu}$ is the classical Kerr metric, $\left< T_{\mu \nu} \right>$ is the expectation value of the quantum stress tensor (e.g. from a scalar field or linearised gravitational  field) with respect to an appropriate vacuum state in the presence of gravity~\cite{Birrell2012-xt}. 
 Semiclassical arguments will be valid provided $\frac{\left| \left< T_{\mu \nu} \right> \right|}{M_{Pl}^{2}} \ll \left| G_{\mu \nu} \right|$. The Kerr spacetime ({\it cf.} \eqref{Kerr}) has curvature scale set by the (ADM) mass $\mathcal{M}$; 
 so typical components of the Riemann tensor are: $\left| R_{\alpha \beta \gamma \delta} \right| \sim \frac{\mathcal{M}}{r^{3}}$. Near the event horizon $r \sim r_{+}$,  $\left| R \right| \sim \frac{1}{\mathcal{M}^{2}}$ (in Planck units) and, provided  we have $\mathcal{M} \gg M_{\rm Pl} $, then $\left| R \right| \ll\frac{1}{M_{\rm Pl}^{2}}$. This is valid for astrophysical BH. Semiclassical gravity treats fields as having typical wavelengths $\lambda \gg l_{\rm Pl}=M_{\rm Pl}^{-1}$, where $l_{\rm Pl}$ is the (reduced) Planck length.

 The above argument relies on $\left< T_{\mu \nu} \right>_{vacuum} <\infty$ (as $r \to r_{+}$). In general, a vacuum state $|0\rangle$ in quantum field theory is a state with no particles, as defined by a choice of mode decomposition of the field (labeled by $k$) and an associated set of annihilation operators $\alpha_k$ such that $\alpha_{k}|0\rangle =0$. In curved spacetime, specially in the presence of a BH, there is no unique time coordinate, and hence no unique frequency decomposition. This leads to inequivalent definitions of vacuum, depending on the observer and the region of spacetime. These considerations apply to a scalar field as well as to gravitons, i.e. quantised perturbations of the metric \eqref{pert_metr}. The vacuum for gravity around a BH background refers to a choice of quantum state of the graviton field $h_{\mu \nu}$ and is defined by specifying which graviton modes are unexcited in a particular region (e.g. near the horizon or at infinity). There are  common choices of gravitational vacua in black  hole spacetimes \cite{Birrell2012-xt}. For astrophysical BH, the Unruh vacuum~\cite{PhysRevD.10.3194}, which mimics a BH formed by collapse, seems the appropriate choice. For a scalar field, the Unruh vacuum leads to the following regions for boundary conditions: near $\mathscr{I}^{-}$, past null infinity, ingoing radiation comes from infinity; near $\mathscr{H}^{-}$, the past event horizon, radiation may emerge from the past horizon; near $\mathscr{I}^{+}$ (future null infinity) radiation escapes to infinity; near $\mathscr{H}^{+}$, the future event horizon, radiation may fall into the BH.  The Unruh vacuum is crucial for processes that involve definite particle content near a black hole horizon, understanding spontaneous emission from the vacuum, evolving field modes across the horizon smoothly and calculating  $\left< T_{\mu \nu} \right>$. If we were to focus on gravitons spontaneously emitted by the BH into empty space, the Unruh vacuum is needed, since  it is regular on $\mathcal{H}^+$ and asymptotically empty at $\mathcal{I}^{-}$. In our case we do not study vacuum decay, interaction region is outside the event horizon and the axion cloud is localised well outside the exterior horizon $r \gg r_{+}$~\cite{Detweiller,BritoCardoso}; we can therefore assume local flatness.

In the current work, we consider quantum graviton perturbations of gravitational wave (GW) type, around Kerr BH backgrounds, in axionic quantum field theory (QFT) cases, with a string inspired KR axion field~\cite{Duncan:1992vz}, or compactification axion~\cite{Svrcek:2006yi}, or an axion associated with the totally antisymmetric components of torsion in Einstein Cartan theories~\cite{Cartan:1938ph,hehl}. If these axions are massive, they can trigger a superradiant instability~\cite{BritoCardoso,Detweiller} of the neighbouring rotating (Kerr) BH, and form axionic clouds/condensates in the surrounding exterior area of the BH. In the presence of axions, an important feature of a rotating BH background  is a non-vanishing gravitational Chern-Simons (gCS) anomaly term~\cite{Duncan:1992vz,Jackiw:2003pm,Alexander:2009tp} in the effective Lagrangian. Upon quantization of weak GW in these theories, we  show that squeezed states of quantum gravitons can be produced by the coherent field of the axionic condensate surrounding the BH, either by fusion of two axions into two gravitons, induced by the GR part of the effective action, or by the decay of a single axion into two gravitons coming from the gCS anomaly term. \color{black} As we restrict our attention to regions of spacetime far away from the exterior horizon of the BH, we consider the quantisation procedure of weak graviton perturbations around the Minkowski metric, {\it i.e.} we set $g_{\mu\nu}^{(0)} = \eta_{\mu\nu}$ in \eqref{pert_metr}. This implies that 
in our approach it suffices to use the Minkowski vacuum rather than the aforementioned Unruh vacuum.\color{black}

 In contrast to the linear-graviton coupling to the axions stress tensor $(h_{\mu \nu} \ T^{(b)\,\mu\nu})$, which corresponds to coherent GW, the aforementioned interactions involving two gravitons, lead to squeezed graviton states, that lack classical analogues. As we prove after detailed calculations, the squeezing effect coming from GR terms dominates, by many orders of magnitude, that induced by the gCS anomaly terms, as  expected   from an effective theory point of view, due to the higher-derivative nature of the gCS anomaly, \color{black} but also from the large number of axions in the cloud.\color{black}

We remark that in future  GW interferometric detectors, quantum fluctuations of the gravitational field might be detectable as an appropriate kind of {\it{quantum}} noise, depending on the state of gravitons~\cite{Amelino-Camelia:1998mjq,Amelino-Camelia:1999vks,Parikh:2020nrd,Parikh:2020kfh,Parikh:2020fhy}. Squeezed states have no classical analogue, and so, their detection 
will be a signature of QG.  The detection of squeezed states, with sufficiently large squeezing parameters, in interferometers is much easier than  detecting single graviton states.  For example, in the case of GW, non-classicality can in principle be explored efficiently in a Hanbury-Brown-Twiss (HBT) interferometer \cite{Kanno:2018cuk, Kanno:2019gqw},  which  can reveal sub-Poissonian statistics for the number of gravitons in squeezed coherent states.  However, HBT analysis is purely theoretical, with no experimental implementation so far, since it relies on delicate intensity-correlation techniques associated with indirect graviton counting.  Another approach uses the electromagnetic field cavity as an optomechanical GW detector \cite{Guerreiro:2019vbq,Coradeschi:2021szx}, where quantum properties of the graviton states can be revealed by statistical measurements. \color{black} A \emph{single-mode} squeezed vacuum state could have an observational signature, provided it has a squeezing parameter~\cite{Scully_Zubairy_1997} 
which is sufficiently larger than one. For instance, in \cite{Coradeschi:2021szx} it has been conjectured that a GW squeezing parameter 
of order $r\sim78$, corresponding to an excited vacuum containing $\langle N\rangle =\sinh^2 r\sim 10^{67}$ gravitons, could be detectable in future GW detectors. However, the non-observation of squeezed GW states by the LIGO/Virgo GW interferometer~\cite{LIGOScientific:2016aoc,McCuller:2021mbn} imposes a stringent bound on the GW-squeezing parameter $r < 41$~\cite{Hertzberg:2021rbl}.\footnote{Since a GW involves a continuum of modes, the authors of \cite{Hertzberg:2021rbl}, in deriving their bound on the squeezing parameter of a GW, have considered (the more realistic case of) a Gaussian profile of squeezed modes, with the peak corresponding to the characteristic frequency of the classical GW as measured by the detector in a given event.} 
The allowed region still implies sufficiently large squeezing parameters, and, as we shall see later in section~\ref{sec:number_of_gravitons}, is compatible with the two-mode-squeezed-graviton state considered in this work. \color{black} A direct operational link between the theoretical supermodes corresponding to the entangled squeezed graviton states, produced from axions in the cloud,  and the practical spectral modes accessible in an interferometric detector is provided by means of Autonne-Takagi-decomposition analysis~\cite{autonne},\cite{Takagi1933} (for a recent review on applications to quantum optics see also \cite{Houde:2024mkj}), discussed in section \ref{sec:Takagi} and in Appendix \ref{appC:takagi}.
\color{black} 

Squeezed graviton states arise naturally during inflation. At cosmological scales, the expansion of the universe produces pairs of particles, forming a two-mode squeezed state with opposite momenta \cite{Mukhanov:2007zz,Kanno:2018cuk,Kanno:2019gqw}. 
The inflationary vacuum appears as a squeezed vacuum from the viewpoint of the radiation-era, via appropriate Bogoliubov transformations. From a non-cosmological perspective, \emph{astrophysical sources} of non-classical GWs are crucial, though rarely discussed in the literature~\cite{Guerreiro_astro};  such sources  will  increase the abundance of observational sources (beyond those predicted by inflation).  Exploring astrophysical  frameworks  for producing  sufficient  ``squeezing'' is the aim of the letter \cite{Dorlis:2025zzz} and the current article. \color{black} Producing squeezed graviton states from macroscopic phenomena, such as superradiance of the axionic BH clouds,  enhances the potential signal significantly, as compared to the single-graviton case~\cite{Dyson:2013hbl}, due to the large number of ALPs  involved\color{black}.

The structure of the article is the following: in section  \ref{sec:originaxion}, we discuss briefly the diverse microscopic origins of the axion-like fields in two 
 frameworks, where axions stem from quantum anomalies: string theory and contorted spacetime geometries. \color{black} We give further technical details on each of these frameworks in Appendix \ref{app:axion}.\color{black}~In section \ref{sec:CST}, we discuss  a resulting   effective Chern-Simons gravity model, for which we derive the axion-graviton interactions up to, and including, second order graviton perturbations. In section \ref{sec:QGW}, we discuss the canonical quantization of GW in a flat background spacetime, which is a valid approximation when considering non-relativistic axionic clouds, forming far away from the BH exterior horizon. In section \ref{sec:BHSuperradiance}, we review in detail the emergence of the superradiant instability of a rotating (Kerr) BH in the presence of the axion field (``gravitational atom''), which is manifested by the appearance of BH modes with a positive imaginary part in frequency. In our analysis we restrict ourselves to the non-relativistic approximation, following the analysis of~\cite{Detweiller}. \color{black}
  For the general reader we review details of the BH superradiant phenomenon in Appendix \ref{app:BHSuperradiance}. \color{black}
  In section \ref{sec:Multimode}, we demonstrate the existence of multimode squeezed states of quantum entangled gravitons in the above scenario, which stem either from the GR part of the action or the gCS anomaly. \color{black} In subsections \ref{sec:VIA}, \ref{sec:cs}, we demonstrate the dependence of the  polarisation structure of the entangled graviton states on the nature of their production, while in subsection \ref{sec:VIC} we draw analogies between this feature and entanglement phenomena in particle physics\color{black}.
 We also  point out  analogues with a quantum optics situation, in which the GR-induced  squeezing corresponds to the Spontaneous Four-Wave Mixing (SFWM)~\cite{SFWM1,SFWM2,SFWM_3}, while the gCS-anomaly-induced squeezing corresponds to Spontaneous Parametric Down Conversion (SPDC)~\cite{PhysRevA.31.2409}. In section \ref{sec:number_of_gravitons}, we show that, for sufficiently long-lived axionic clouds, the GR-induced squeezing effect  dominates the gCS one, and leads to a significant production of squeezed gravitons. This could have potential observational prospects in future interferometers. \color{black} In section \ref{sec:Takagi}, we discuss the role of the Takagi mode decomposition on the interferometric detection of squeezed gravitons, by restricting our attention to some simplified but quite representative examples, that include all important features of our full quantum graviton case. A more complete treatment, including the complete picture, together with observational prospects in current and future experiments, will be presented in a future publication.
  Finally, section \ref{sec:Conclusions} contains our conclusions and outlook. In several Appendices we review, for the general reader, details of relevant concepts and formalism used in our approach. In Appendix \ref{app:axion}, we discuss in detail the emergence of axion-like fields in the frameworks of Einstein-Cartan torsion and strings.
In Appendix \ref{app:BHSuperradiance}, we review the basic characteristics of the theory of superradiance of axionic clouds  surrounding rotating BHs. In Appendix  \ref{appC:takagi}, we give a discussion on analogies of our entangled graviton correlations with Takagi modes~\cite{Takagi1933}. Specifically, 
in subsection \ref{subsec:ATdecomp}, we describe the Takagi decomposition of multimode squeezed photon states in quantum optics, and draw an analogy with our multimode squeezed graviton states. In 
subsection \ref{subsec:covmatrix}, we discuss the concept of covariance matrix of Gaussian states, while in the next subsection \ref{subsec:reduct} we discuss the reduction of Gaussian system, by 
dividing it into two parts, and then integrating out the degrees of freedom in one subsystem.
In subsection 
\ref{sec:appD}, we discuss the r\^ole of symplectic transformations in the calculation of the von Neuman entropy of  subsystems of multimode Gaussian squeezed states, in terms of the relevant density matrices, which serves as a measure of their entanglement. We demonstrate our main points in a  simplified but illustrative example of a two-mode squeezing, which is quite relevant to our purposes here of studying GW polarization entanglement. Finally, in subsection \ref{subsec:2mode}, we discuss some analogues from quantum optics of our squeezed-graviton-state kernels, \eqref{G_GR_General}, \eqref{CSkern},
involving two-mode graviton squeezing, specifically, the so-called parametric down conversion and four-wave mixing, which, as our analysis in the current work 
demonstrates, share qualitatively similar behaviours with our graviton case.

  \color{black}

\color{black} 
\section{On the microscopic origins of Axion-like fields in our scenario}\label{sec:originaxion}

Axions (and axion-like particles, ALPs) can arise due to a variety of  different physical mechanisms and are generic in string compactifications \cite{Cicoli:2012sz}.   One much studied example is the QCD axion introduced by Peccei and Quinn~\cite{Peccei:1977hh,Peccei:1977ur} (where axion mass is generated~\cite{Kim:2008hd} through non-perturbative (instanton) effects of the SU(3)$_C$ colour gauge group of SM~\cite{RevModPhys.71.S96}). 
ALPs also characterise  string theories~\cite{pol1,pol2,str1,str2}, arising from appropriate compactification~\cite{Svrcek:2006yi}.

Of more immediate interest to us are axions/ALPs associated with quantum anomalies in two contexts:
(i) generalisation of Einstein gravity theory to spacetimes which have dynamical torsion~\cite{Cartan:1938ph,hehl,Duncan:1992vz}, and 
(ii) higher form gauge fields which are required in UV completions of gravitational field theories, stemming from strings theories~\cite{Duncan:1992vz}; such axions are known as string-model-independent axions~\cite{Svrcek:2006yi}, and may co-exist with  axions arising from compactification.
As discussed in Appendix 
\ref{app:axion}, in  (3+1)-dimensional-spacetime frameworks, these axion-like fields are associated with a totally antisymmetric three form. In the case of contorted geometries the axion is the totally antisymmetric covariant component of torsion $T_{\mu\nu\rho}$. In the string case (due to the  anomaly cancellation mechanism proposed by Green and Schwarz~\cite{GS}) the axion is related to the generalised  field strength of the spin-one antisymmetric tensor, the Kalb-Ramond (KR) gauge field $B_{\mu\nu}=-B_{\nu\mu}$ of the bosonic massless superstring ground state~\cite{pol1,pol2,str1,str2}.

If we use the generic notation $\mathcal F_{\mu\nu\rho}$ for 
the appropriate three form in each of the above cases, then the axion-like-field $b(x)$, in (3+1)-dimensional spacetimes, is:
\begin{align}\label{gen3form}
\partial_\mu b(x) \propto \varepsilon_{\mu\nu\rho\sigma} \, \mathcal F^{\nu\rho\sigma}\,.    
\end{align}
Here $\varepsilon_{\mu\nu\rho\sigma}$
is the gravitationally covariant Levi-Civita tensor~\cite{eguchi}, totally antisymmetric in its indices ({\it cf.} next section), and the proportionality symbol $\propto$ denotes the appropriate renormalization constant for each case.
In  both frameworks the axion field $b(x)$ is introduced 
as an initially non-propagating pseudoscalar degree of freedom, implementing an appropriate constraint. As explained in Appendix \ref{app:axion}, the constraint either (i) expresses the conservation of torsion charge, order by order in a quantum theory, in which the torsion is integrated out in the appropriate path integral, or (ii) is a consequence of a Bianchi identity for the Green-Schwarz-modified  KR antisymmetric-tensor field strength in the case of strings. The resulting effective gravitational theory (on integrating out the torsion of the KR field strength in the path integral)
contains a dynamically propagating $B(x)$ coupled to Chern-Simons-type anomalies, of both gauge and gravitational type~\cite{Alvarez-Gaume:1983ihn}.
In both frameworks, the initially generated dynamical axion-like field $b(x)$ is massless. However, in common with axion fields more generally, it can acquire its mass through appropriate non-perturbative mechanisms~\cite{Kim:2008hd}, which we assume. 
In what follows we ignore the explicit coupling of axions to gauge anomalies, assuming that the role of the latter is to provide masses $\mu_b$ for the $b(x)$ axion fields, through appropriate  instantons~\cite{Kim:2008hd} for a non-Abelian gauge-group.
This suffices for our purposes. 
As shown in Appendix \ref{app:axion}, the resulting effective gravitational theory is a Chern-Simons gravity~\cite{Jackiw:2003pm,Alexander:2009tp} with a massive axion coupled to gCS terms; in the background of a rotating (Kerr-type~\cite{Kerr:1963ud}) black hole, the theory is non-trivial.\footnote{\color{black}  Any back reaction effects of the axionic cloud surrounding the black hole on the Kerr geometry itself~\cite{Chatzifotis:2022mob,Chatzifotis:2022ene} are ignored as subleading.\color{black}} In the next section we proceed to discuss the quantization of tensor perturbations $h_{\mu\nu}$ about a flat spacetime background \eqref{pert_metr} in such a theory to quadratic order in the weak-graviton approximation, taking into account effects of superradiance. 

\color{black}

\section{Gravitational Chern-Simons (${\bf {\rm g}}$CS) Theory}\label{sec:CST}

The Chern - Simons (CS) gravitational theory \cite{Jackiw:2003pm,Duncan:1992vz,Alexander:2009tp} introduces a linear coupling of the pseudo-scalar field $b$, with 
the gravitational Chern - Simons term (gCS), $R_{CS}$:\footnote{Our conventions and definitions used throughout this work are: signature of metric $(-, +,+,+ )$, Riemann Curvature tensor:
$R^\lambda_{\,\,\,\,\mu \nu \sigma} = \partial_\nu \, \Gamma^\lambda_{\,\,\mu\sigma} + \Gamma^\rho_{\,\, \mu\sigma} \, \Gamma^\lambda_{\,\, \rho\nu} - (\nu \leftrightarrow \sigma)$, Ricci tensor $R_{\mu\nu} = R^\lambda_{\,\,\,\,\mu \lambda \nu}$, and Ricci scalar $R_{\mu\nu}g^{\mu\nu}$. We also work in units $\hbar=c=1$.}
\begin{equation}
	S=\int d^4x \,\sqrt{-g}\,  \left[\frac{R}{2\kappa^2}-\frac{1}{2}(\partial_\mu b)(\partial^\mu b) -\frac{1}{2}\mu^{2}_b \  b^2 - A\, b\,R_{CS} \right] \ ,
\label{eq:Action}
\end{equation}
where  
 $A$ is a coupling constant, of mass dimension $\left[A\right]=-1$, $\kappa=\sqrt{8\pi G}=M_{\rm Pl}^{-1}$ is the inverse of the reduced Planck mass $M_{\rm Pl} = 2.435 \times 10^{18}~\rm GeV$ and $\mu_b$ denotes the mass of the axion. The gCS term $R_{CS}$ in \eqref{eq:Action} is given by:
\begin{equation}
	\label{RCS}
	R_{CS}= \frac{1}{2}R^{\mu}_{\,\,\,\nu\rho\sigma}\widetilde{R}^{\nu\,\,\,\,\rho\sigma}_{\,\,\,\mu},
	\end{equation}
with the symbol $\widetilde{(\dots)}$ denoting the dual of the Riemann tensor, defined as
\begin{equation}
\label{dualriem2}
\widetilde{R}_{\alpha\beta\gamma\delta}=\frac{1}{2}R_{\alpha\beta}^{\,\,\,\,\,\,\,\,\rho\sigma}\varepsilon_{\rho\sigma\gamma\delta}\,,
\end{equation}
to be contrasted with the Hodge-dual 
\begin{equation}
\label{hodgedualriem}
^\star{R}_{\alpha\beta\gamma\delta}=\frac{1}{2}R_{\alpha\beta}^{\,\,\,\,\,\,\,\,\rho\sigma}\hat \epsilon_{\rho\sigma\gamma\delta}\,,
\end{equation}
 where $\varepsilon_{\mu\nu\alpha\beta} = \sqrt{-g(x)} \, \hat \epsilon_{\mu\nu\alpha\beta} $ is the covariant Levi-Civita tensor, and $\hat \epsilon_{\mu\nu\rho\sigma}$ denotes the Minkoswki space-time Levi-Civita totally antisymmetric symbol, with the convention 
$\hat \epsilon_{0123}=1$, {\it etc}. We can therefore make use of both duals, exploring the identity 
\begin{align}\label{identity}
\sqrt{-g} \, R_{\mu\nu\rho\sigma} \, \widetilde R^{\mu\nu\rho\sigma} = R_{\mu\nu\rho\sigma} \, ^\star R^{\mu\nu\rho\sigma} \,.
\end{align}
The gCS term is a total derivative, and can be written as $R_{CS}=\nabla_\mu\mathcal{J}^{\mu}$, where $\mathcal{J}^{\mu}$ denotes the topological current. Furthermore, the gCS term $b\,R_{CS}$ is characterized by a non trivial variation with respect to the metric tensor, yielding the Cotton-like tensor $C_{\mu\nu}$~\cite{Jackiw:2003pm}: 
\begin{equation}
\label{cottdef}
	C_{\mu\nu}=-\frac{1}{2}\nabla^{\alpha}\left[(\nabla^{\beta} b) \widetilde{R}_{\alpha\mu\beta\nu}+(\nabla^{\beta} b) \widetilde{R}_{\alpha\nu\beta\mu}\right]~.
\end{equation}
Variation of the action with respect to the metric  and the axion field yields the following equations of motion:
\begin{align}
	\label{grav}
	&G_{\mu\nu}=\kappa^2 T^{(b)}_{\mu\nu}+4 \kappa^2 A C_{\mu\nu}~,\\
	\label{Axion}
	&\left(\square - \mu^{2}_b\right)b=A \, R_{CS}~,
\end{align}
where $T^{(b)}_{\mu\nu}$ is the  stress energy-momentum tensor associated with the kinetic term of a matter field,
\begin{equation}\label{stressb}
	T^{(b)}_{\mu\nu}=\nabla_\mu b\nabla_\nu b-\frac{1}{2}g_{\mu\nu}(\nabla b)^2 - \frac{1}{2}g_{\mu\nu}\mu^{2}_b \ b^2~.
\end{equation}
The property of the Cotton tensor $C_{\mu\nu}^{\quad ;\mu} = -\frac{1}{4} \, (\partial_\nu b)\, R_{CS}$, where $;$ denotes gravitational covariant derivative (with respect to a torsion-free connection),  implies that the naive covariant conservation of the axion-matter stress tensor fails, 
\begin{align}\label{nonconsstress}
T^{(b)\,\,;\mu}_{\mu\nu} = A (\partial_\nu b) \, R_{CS}\,.
\end{align}
This implies  the non conservation of the naive axion stress tensor, which is physically interpreted as indicating a non trivial exchange of energy between the axion matter and the gravitational field. Such an exchange of energy preserves diffeomorphism invariance of the theory, since it is the modified stress-energy momentum tensor that is conserved, 
\begin{equation}
    \left(T^{(b)}_{\mu\nu}+4AC_{\mu\nu}\right)^{;\mu}=0\ .
\end{equation}

The $R_{CS}$ term is CP violating,\footnote{The reader should note that, if the field $b$ is a scalar, then the coupled $b$-mixed-anomaly interaction term in the action \eqref{eq:Action} violates CP. Par contrast, the latter symmetry is preserved in the case of a pseudo-scalar $b$ field, which will be the focus of our interest here.} and, as such, it vanishes for spherically symmetric or isotropic and homogeneous spacetime backgrounds. This assures the persistence of the vacuum Schwarzschild and Friedman-Lemaitre-Robertson-Walker (FLRW) spacetime background. However, this is not the case for rotating (stationary and axi-symmetric) spacetimes, for example, described by the Kerr metric. Thus, $R_{CS}\neq0$ for a Kerr background; so the effects of back-reaction of the axion field on the underlying geometry needs to be taken into account. This leads to BH with axionic hair~\cite{Alexander:2009tp,Chatzifotis:2022mob}. We mention here that the Cotton tensor  (a contribution to the stress-energy tensor from non physical matter (gravitational) degrees of freedom)     sources the backreaction to the gravitational equations of motion (and also violates the energy conditions,  common in quantum field theory ).\footnote{Another situation where the parity-violating CS term is non vanishing is in the presence of chiral GW, with left-handed and right-handed polarisations propagating differently. This leads to birefringence through modified dispersion  relations, with interesting cosmological implications~\cite{Lue:1998mq}. The quadrupole formula for GW-emission may also be modified. Recently, Gauss-Bonnet quadratic curvature terms (which are non-trivial in (3+1)-dimensions, if non-constant dilatons are present) have been added and their effect on gravitational waves considered \cite{PhysRevD.109.124012}.}

We proceed with the perturbative expansion of the effective theory up to, and including, second order terms in GW perturbations $\kappa h_{\mu\nu}$. On perturbing the background metric as:
\begin{align}
\label{weak_field_expansion}
    g_{\mu\nu}&\rightarrow g_{\mu\nu}+\kappa h_{\mu\nu} \ , \\ 
    g^{\mu\nu}&\rightarrow g^{\mu\nu}-\kappa h^{\mu\nu} +\kappa^2 h^{\mu \alpha}h_{\alpha}^{\nu} \ ,
\end{align}
we obtain the following Einstein-Hilbert Lagrangian, {\it i.e.} in the absence of gCS anomalies, 
\begin{align} \label{GravitonLgrangian2}
    \delta\mathcal{L}^{(2)}_{EH}=& \frac{1}{4}\sqrt{-g}G_{\mu\nu}h_\alpha^\mu h^{\alpha\nu}-\frac{1}{4}\sqrt{-g}\,h\,h^{\alpha\beta}G_{\alpha\beta} \nonumber \\
    &+\frac{1}{2}\sqrt{-g}\Big[ -\frac{1}{8}h^2\, R-\frac{1}{2}R^\alpha_{\,\lambda\beta\gamma}h^\gamma_\alpha  h^{\lambda\beta}+\frac{1}{2}\nabla^\gamma h_{\alpha\gamma}\nabla_\beta h^{\alpha\beta}+\frac{1}{4}\nabla_\beta h \nabla^\beta h \nonumber \\ &-\frac{1}{2}\nabla_\beta h\nabla_\alpha h^{\alpha\beta}-\frac{1}{4}\nabla_\gamma h_{\alpha\beta}\nabla^\gamma h^{\alpha\beta}\Big] \ ,    
    \end{align}
while the matter Lagrangian up to second order to the metric perturbations reads,
\begin{align}\label{dl2}
\delta\mathcal{L}_{matter}^{(2)} &= \sqrt{-g}\Big[\frac{
\kappa}{2}T_{\mu\nu}h^{\mu\nu}+\frac{\kappa^2}{4}h\,h_{\alpha\beta}\nabla^\alpha b\nabla^\beta b+\frac{\kappa^2}{4}\left(h_{\alpha}^\mu h^{\alpha\nu}-\frac{1}{2}h h^{\mu\nu}\right)g_{\mu\nu}\ \mathcal{L}_{matter} \nonumber \\ &-\frac{\kappa^2}{2}h_{\alpha\beta}h_{\mu}^{\beta} \nabla^\alpha b\nabla^\mu b\Big] \ ,
\end{align}
where the energy-momentum tensor of the axion is given by:
\begin{equation}\label{stresstens}
 T_{\mu\nu}= \partial_\mu b\,\partial_\nu b -\frac{1}{2}g_{\mu\nu}\left(\eta^{\rho\sigma}\partial_\rho b\,\partial_\sigma b -\mu^2 b^2\right)\, .
\end{equation}
The Lagrangian given by \eqref{GravitonLgrangian2} and \eqref{dl2} is invariant under the following gauge  (general coordinate) transformations of the axion and graviton fields ~\cite{tHooft:1974toh},
\begin{align}
b&\to b+\xi_\alpha\partial^\alpha b\\
\kappa h_{\mu\nu} &\to \kappa h_{\mu\nu} + ( g_{\alpha \nu} + \kappa h_{\alpha\nu}) \, \nabla_\mu \xi^\alpha + (g_{\mu\alpha} +\kappa  h_{\mu\alpha} )\, \nabla_\nu \xi^\alpha + \kappa\,  \xi^\alpha \, \nabla_\alpha \, h_{\mu\nu}\,,
\end{align}
where $\nabla_\mu$ is the gravitational covariant derivative with respect to the background metric $g_{\mu\nu}$, and $\xi^\mu$ is the general diffeomorphism parameter. Note that in order to keep the Lagrangian gauge invariant to all orders of perturbations, higher order gauge transformations have to be included, resembling the diffeomorphism invariance of the full theory.

For a  background metric, obeying the vacuum equations of motion, $G_{\mu\nu}=0$ and in  the traceless-transverse gauge (TT-gauge), $h^{0\mu}=0\ , h=0 \ \ \text{and}\ \ \nabla_\mu h^{\mu\nu}=0$,  \eqref{GravitonLgrangian2} reduces to, 
\begin{equation}
\begin{aligned}
    \delta\mathcal{L}^{(2)}_{EH}=    -\frac{1}{4\kappa^2}\sqrt{-g}\left[\frac{1}{2}\nabla_\gamma h_{\alpha\beta}\nabla^\gamma h^{\alpha\beta}+R^\alpha_{\,\lambda\beta\gamma}h^\gamma_\alpha  h^{\lambda\beta}\right] \ .
    \end{aligned}
    \label{GravitonLgrangian2_TT}
\end{equation}
The curvature contributes to the Lagrangian as source of a quadratic interaction for the graviton. Moreover, considering the axion field as a source term, i.e. it is on-shell and satisfies the Klein-Gordon equation with curved spacetime background, we obtain the following interacting Lagrangian in the TT-gauge for the graviton, 
\begin{equation}\label{dl2_TT}
\delta\mathcal{L}_{matter}^{(2)}= \sqrt{-g}\left[\frac{\kappa}{2}T_{\mu\nu}h^{\mu\nu}-\frac{\kappa^2}{2}h_{\alpha\beta}h_{\mu}^{\beta} \nabla^\alpha b\nabla^\mu b\right].
\end{equation}
since the on-shell condition implies $\mathcal{L}_{matter}=0$. \par 

Including the gCS term, up to second order in the graviton perturbations\footnote{One can show that the gCS term is exactly gauge invariant with respect to the linearized gauge transformation,  $h_{\mu\nu}\rightarrow h_{\mu\nu}+\partial_\mu \xi_{\nu}+\partial_\mu\xi_\nu$, as seen from a straightforward calculation, 
\begin{equation}
    \begin{aligned}
        \frac{1}{2}\ \epsilon_{\beta\mu\rho\sigma}\left(  \partial_\alpha\partial^\sigma h^\rho_\nu-\partial_\nu\partial^\sigma h^\rho_\alpha     \right)\partial^\mu\partial^\nu h^{\alpha\beta} \rightarrow& \frac{1}{2} \epsilon_{\beta\mu\rho\sigma}\left( \partial_\alpha\partial^\sigma\partial^\rho\xi_\nu+\partial_\alpha\partial^\sigma\partial_\nu\xi^\rho-\partial_\nu\partial^\sigma\partial^\rho\xi_\alpha -\partial_\nu\partial^\sigma\partial_\alpha\xi^\rho    \right)\partial^\mu\partial^\nu h^{\alpha\beta}\\
        +&\frac{1}{2}\epsilon_{\beta\mu\rho\sigma}\left(  \partial_\alpha\partial^\sigma h^\rho_\nu-\partial_\nu\partial^\sigma h^\rho_\alpha     \right) \left( \partial^\mu\partial^\nu\partial^\alpha\xi^\beta+\partial^\mu\partial^\nu\partial^\beta\xi^\alpha    \right)=0\ . 
    \end{aligned}
\end{equation}
This is a manifestation of the anomaly being exact at all orders to the curvature.},
\begin{equation}
    R_{CS}=\frac{\kappa^2}{2}\ \epsilon_{\beta\mu\rho\sigma}\left(  \partial_\alpha\partial^\sigma h^\rho_\nu-\partial_\nu\partial^\sigma h^\rho_\alpha     \right)\partial^\mu\partial^\nu h^{\alpha\beta} + \text{higher orders} \ ,
\end{equation}
and the relevant interaction part in the TT-gauge is given by:
\begin{equation}
\mathcal{L}_I=\mathcal{L}^{(1)}_{I}+\mathcal{L}^{(2)}_{I}+\mathcal{L}^{(2)}_{I,CS}
    \label{interaction_lagrnagian}
\end{equation}
where
\begin{align}
   \label{Interaction_Coherent_GR} 
   &\mathcal{L}_{I}^{(1)}=\frac{\kappa}{2}T^{ij}h_{ij} \ , \\
   \label{Interaction_second_order_GR} 
   &\mathcal{L}_{I}^{(2)}=  - \frac{\kappa^2}{2}\ h_{im}h^{m}_{\ j} \ \partial^i b\,\partial^j b \ , \\ 
   \label{Interaction_CS} 
   &\mathcal{L}^{(2)}_{I,CS}=A\kappa^2 \ b \ \epsilon_{ijk} \ \Bigg( \partial_l\partial^k  h^j_m  \partial^m\dot{h}^{li} +\ddot{h}^{li}\partial^k\dot{h}^j_l 
    -\partial_m\partial^kh^j_l\partial^m\dot{h}^{li}  \Bigg)  \ .
\end{align}
These are the remaining axion-graviton interactions in the TT-gauge. The first one, linear in the graviton field $h_{\mu\nu}$ is responsible for the production of coherent GW, while the second and the third ones, being quadratic to $h_{\mu\nu}$ produce the \emph{squeezed graviton states}, of interest to us in this work.

\section{Quantization of Gravitational Waves in ${\rm g}$CS Theory}\label{sec:QGW}

In order to express the perturbation $h_{\mu\nu}$, we  define a tetrad $\left\{e^\mu_{(0)},e^\mu_{(1)},e^\mu_{(2)},e^\mu_{(3)} \right\}$, obeying the handedness condition, 
\begin{equation}
    e^\mu_{(0)}e^\nu_{(1)}e^\rho_{(2)}e^\sigma_{(3)}\epsilon_{\mu\nu\rho\sigma}=1 \   ,
    \label{handness}
\end{equation}
the orthogonality,
\begin{equation}
    \eta_{\mu\nu}e^\mu_{(a)}e^\nu_{(b)}=\eta_{(a)(b)} \ ,
    \label{orthogonality}
\end{equation}
and the completeness relation,
\begin{equation}
    \sum_{a}\eta^{(a)(a)}e^\mu_{(a)}e^\nu_{(a)}=\eta^{\mu\nu}\ .
    \label{completeness}
\end{equation}
In this tetrad $e^\mu_{(0)}=u^\mu$, with $u^\mu$ the four velocity of the observer. We choose the lab-frame by taking the four-velocity to be $u^\mu=(1,\vec{0})$ and then, by the orthogonality condition, we can easily obtain for the other vectors of the tetrad,
\begin{equation}
    e^0_{(i)}=0\ ,\  \forall i=1,2,3\ \ \ \ ,
\end{equation}
i.e. the temporal components vanish. Moreover, the handedness relation \eqref{handness} for the tetrad reduces to 
\begin{equation}
    e^i_{(1)}e^j_{(2)}e^k_{(3)}\epsilon_{ijk}=1 \ ,
\end{equation}
 where $\epsilon_{ijk}=\epsilon_{0ijk}$. 
Expanding the gravitational perturbations in Fourier space and helicity basis (L = Left, R = Right), we obtain:
\color{black}
\begin{equation}
    h_{ij}(t,\vec{x})=\int\frac{d^3\vec{k}}{(2\pi)^{3/2}}\sum_{\lambda=L,R}e_{ij}^{(\lambda)}(\vec{k})h_{\lambda,\vec{k}}(t)e^{i\vec{k}\cdot\vec{x}} \, \ , \   [h_{\vec{k},\lambda}]= -2 \ ,
    \label{FourierTensor2}
\end{equation}
\color{black}
where $e_{ij}^{L,R}$ are defined as follows:
\begin{equation}
  \left [ e_{ij}^{(R)}(\vec{k})\right]=\frac{1}{\sqrt{2}}\left( \left[e^{(+)}_{ij}(\vec{k})\right]+ i \left[e^{(\times)}_{ij}(\vec{k}) \right] \right)=\left[ e_{ij}^{(L)}(\vec{k})\right]^\dagger
\label{helicitybasis}\,.
\end{equation}  
The cross and plus polarization tensors are given by:
\begin{align}
  &  e_{ij}^{(+)}(\vec{k})= [e_1(\vec{k})]_i[e_1(\vec{k})]_j-[e_2(\vec{k})]_i[e_2(\vec{k})]_j \,,\\
  &e_{ij}^{(\times)}(\vec{k})= [e_1(\vec{k})]_i[e_2(\vec{k})]_j+[e_1(\vec{k})]_j[e_2(\vec{k})]_i \,,
\end{align}
with $e_3(\vec{k})=\vec{k}/\vert\vec{k}\vert$.   By construction, the polarization tensors satisfy the following relations:
\begin{align}
   & e^{(L,R)}_{ij}(\vec{k})=e^{(L,R)}_{ij}(-\vec{k}) \ ,\label{prop1}\\
    &e^{(L)}_{ij}(\vec{k})\;e^{(L)\; ij}(\vec{k})=e^{(R)}_{ij}(\vec{k})\;e^{(R)\; ij}(\vec{k})=0 \ ,\label{prop2}\\
   & e^{(L)}_{ij}(\vec{k})e^{(R)\; ij}(\vec{k})=2 \ , 
   \label{prop3}
\end{align} 
while reality of $h_{ij}\in\mathbb{R}$ implies, $h_{L,\vec{k}}=h^\star_{R,-\vec{k}}$. Then, we can write the action with respect to the modes of GWs, 
\begin{equation}\label{GWact}
    S_{GW}=\frac{1}{4}\sum_{\lambda=L,R}\int dt \int d^3\vec{k} \,\left( 
      \vert \dot{h}_{\lambda,\vec{k}}\vert^2-k^2\vert h_{\lambda,\vec{k}}\vert^2 \ .
\right)
\end{equation}
Finally, the quantized GWs have the following free Hamiltonian,  
\begin{equation}
    \hat{\mathcal{H}}_{GW}^{(0)} = \int d^3\vec{k} \ \Omega_k\sum_{\lambda=L,R} \hat{\alpha}_{\lambda,\vec{k}}^\dagger\hat{\alpha}_{\lambda,\vec{k}}   .
  \label{GWsFree}
\end{equation}
with \color{black} $\Omega_k \equiv k$ denoting the frequency of the GW with momentum $\vec k$, and  \color{black} 
 \begin{equation}
    \hat{h}_{ij}(t,\vec{x})=\int \frac{d^3\vec{k}}{(2\pi)^{3/2}} \frac{1}{\sqrt{2\Omega_k}}\sum_{\lambda=L,R}\left[   e^{(\lambda)}_{ij,\vec{k}} \hat{\alpha}^\dagger_{\lambda,\vec{k}}\ e^{-ik\cdot x} + \text{h.c}  .       \right] \ ,
\label{mode_expansion_gravitons_Integral}
\end{equation}
where the creation and annihilation operators obey the following commutation relation:
\begin{equation}
\label{commutaiton_alpha_lambda}
    \left[\hat{\alpha}_{\lambda,\vec{k}} \ , \ \hat{\alpha}^\dagger_{\lambda^\prime , \vec{k}^\prime}\right] = \delta_{\lambda,\lambda^\prime} \ \delta^{(3)}\left(\vec{k}-\vec{k}^\prime \right) \ .
\end{equation}
 To avoid  infrared divergencies, we quantize in an arbitrary volume $V$, while also introducing a dimensionless pair of creation/annihilation operators using
\begin{equation}\label{intsum}
    \int \frac{d^{3}\vec{k}}{(2\pi)^{3/2}} \rightarrow \frac{1}{V}\sum_{\vec{k}} \ \ , \ \ \hat{\alpha}_{\lambda,\vec{k}}\rightarrow \sqrt{V} \ \hat{\alpha}_{\lambda,\vec{k}} \ .
\end{equation}
Then, we have, 
\begin{equation}
    \hat{h}_{ij}(t,\vec{x})=M_{\rm Pl}\sum_{\vec{k},\lambda} f_k\left[   e^{(\lambda)}_{ij}(\vec{k}) \ \hat{\alpha}^\dagger_{\lambda,\vec{k}}\ e^{-ik\cdot x} + \text{h.c}  .       \right]
    \label{mode_expansion_gravitons}
\end{equation}
where $\hat{\alpha}_{\lambda,\vec{k}}^\dagger \ ,\hat{\alpha}_{\lambda,\vec{k}}$  denote the dimensionless creation/annihilation operators from now on, and $f_k$ is given by: 
\begin{equation}
\label{strain}
    f_k\equiv \frac{\kappa}{\sqrt{2V\Omega_k}}\, , \quad [f_k]=0 \ ,
\end{equation}
which can be considered as the (dimensionless) single graviton strain.

We proceed to the derivation of the quantum Hamiltonians for the interactions \eqref{Interaction_Coherent_GR},\eqref{Interaction_second_order_GR} and \eqref{Interaction_CS}, keeping the axion field $b$ classical. This suffices for our purposes in this work, given that the $b$ field will be a condensate in the context of our analysis, as we shall show below. We start with the linear interaction \eqref{Interaction_Coherent_GR}, which, in the TT-gauge reads:
\begin{equation}
    \mathcal{H}^{(1)}_{I}(t)=-\frac{\kappa}{2}\int d^3\vec{x}\  T^{ij}(t,\vec{x})\ h_{ij}(t,\vec{x}) \ .
    \label{coherent_hamiltonian}
\end{equation}
Expanding to Fourier space and substituting the mode expansions for the gravitons \eqref{mode_expansion_gravitons}, we get:
\begin{equation}
\label{Hamiltonian_Quantum_Coherent}
   \hat{\mathcal{H}}^{(1)}_{I}(t)=-\sum_{\vec{k},\lambda}\left[ \beta_\lambda(t,\vec{k}) \hat{\alpha}^\dagger_{\lambda,\vec{k}} + \beta^\star_\lambda(t,\vec{k}) \hat{\alpha}_{\lambda,\vec{k}}  \right] 
\end{equation}
where, 
\begin{equation}
    \beta_\lambda(t,\vec{k}) = \frac{1}{2}f_k\  e^{i\Omega_k t}e^{(\lambda)}_{ij}(\vec{k})\ T^{ij}(t,-\vec{k})\ ,
\end{equation}
with, 
\begin{equation}
    T^{ij}(t,-\vec{k})=\int d^3\vec{x} \ e^{-i\vec{k}\cdot\vec{x}}\ T^{ij}(t,\vec{x})\ .
\end{equation}
Then, up to an irrelevant phase \cite{Skagerstam:2018jkw} (see also the Magnus expansion \eqref{Magnus_expansion} and \eqref{Magnus_Dyson}), we can define the scattering matrix for the process of graviton emission due to the interaction as, 
\begin{equation}
    \hat{S}^{(1)}=e^{-i\int_{-\infty}^{+\infty} dt\ \mathcal{H}_{I}(t)}= \exp\left[\sum_{\vec{k},\lambda} \left( \xi_{\lambda,\vec{k}}\hat{\alpha}_{\lambda,\vec{k}}^\dagger -\xi_{\lambda,\vec{k}}^\star\hat{\alpha}_{\lambda,\vec{k}}       \right)\right]=\prod_{\vec{k},\lambda}\hat{D}(\xi_{\lambda,\vec{k}})
\end{equation}
where 
\begin{equation}
   \hat{D}(\xi_{\lambda,\vec{k}})=\exp\left[  \xi_{\lambda,\vec{k}}\hat{\alpha}_{\lambda,\vec{k}}^\dagger-\xi_{\lambda,\vec{k}}^\star\hat{\alpha}_{\lambda,\vec{k}}       \right]\ , \
   \label{displacement_operator}
\end{equation}
i.e. the displacement operator for the graviton modes, $\vert\lambda,\vec{k}\rangle$, with the displacement parameter $\xi_{\lambda,\vec{k}}$ given by, 
\begin{equation}
    \xi_{\lambda,\vec{k}}=i\int_{-\infty}^{+\infty} dt\ \beta_{\lambda}(t,\vec{k})\ .
\end{equation}
One can see that the emitted gravitons due to the interaction \eqref{Interaction_Coherent_GR} are produced in coherent states, resembling their behavior as classical GWs. 

For the interaction of \eqref{Interaction_second_order_GR}, stemming from GR, the Hamiltonian in the TT-gauge reads:
\begin{equation}
    \mathcal{H}^{(2)}_{I}= \frac{\kappa^2}{2}\int d^{3}\vec{x}\ h_{im} h^{m}_{\ j} \ \partial^{i} b\,\partial^{j} b  \ ,
\end{equation}
which, after substitution of the mode expansions \eqref{mode_expansion_gravitons} for the gravitons reads:
\begin{align}
\label{Hamiltonian_Quantum_GR_second_order_General}
\nonumber
\hat{\mathcal{H}}^{(2)}_{I}=&\frac{1}{2}\sum_{\vec{k},\vec{k}^\prime}\sum_{\lambda,\lambda^\prime}f_k f_{k^\prime}\Bigg(\int d^3 \vec{x} \  \partial^{i} b \ \partial^{j}b \ e^{-i(\vec{k}+\vec{k}^\prime)\cdot \vec{x}} \Bigg)\Bigg(e^{(\lambda)}_{im}(\vec{k})e^{(\lambda^\prime)}_{mj}(\vec{k}^\prime) \ \hat{\alpha}^\dagger_{\lambda,\vec{k}}\hat{\alpha}^\dagger_{\lambda^\prime,\vec{k}^\prime} \ e^{i(\Omega_k+\Omega_{k^\prime})\cdot t}\Bigg) \ + \ \text{h.c.} \\ 
+&\frac{1}{2}\sum_{\vec{k},\vec{k}^\prime}\sum_{\lambda,\lambda^\prime}f_k f_{k^\prime}\Bigg(\int d^3 \vec{x} \  \partial^{i} b \ \partial^{j}b \ e^{-i(\vec{k}+\vec{k}^\prime)\cdot \vec{x}}\Bigg)\Bigg(e^{(\lambda)}_{im}(\vec{k})\left[e^{(\lambda^\prime)}_{mj}(\vec{k}^\prime)\right]^\dagger  \hat{\alpha}^\dagger_{\lambda,\vec{k}}\hat{\alpha}_{\lambda^\prime,\vec{k}^\prime} \ e^{i(\Omega_k-\Omega_{k^\prime})\cdot t}\Bigg) \ + \ \text{h.c.}   \ .
\end{align}
This interaction, as well as the CS interaction, are quadratic to the gravitational waves. Such Hamiltonians will be related, through the evolution operator, to the squeezing operator, in analogy to quantum mechanics, as we will demonstrate later on. The states produced from such interactions are purely quantum in origin and their strength will be in general highly suppressed through the factor related to the mass of the axion $\mu_b$ over the Planck mass $M_{\rm Pl}$. \par
Following the same procedure, we proceed to the CS interaction of \eqref{Interaction_CS}, which in the TT-gauge takes the following form:
\begin{equation}
    \mathcal{H}_{I,CS}(t)=A\kappa^2\int d^3\vec{x}\ b(x)\  \epsilon^{ijk}\left( \partial_l \dot{h}^{m}_{j} \partial_m \partial_i h_{k}^l - \partial_l \dot{h}_{jm}\partial^l \partial_i h_{k}^m + \ddot{h}_{jl}\partial_i \dot{h}^{\ 
 l}_{k} \right) \ ,
 \label{Chern-Simons_Hamiltonian_bR*R}
\end{equation}
where $\epsilon^{ijk}$, $i,j,k=1,2,3$, is the totally antisymmetric Levi-Civita symbol in the spatial indices.

After substitution of the mode expansions \eqref{mode_expansion_gravitons}, the above Hamiltonian becomes:
\begin{equation}
\label{H_ineteraction_help_1}
  \begin{aligned}
    & \hat{\mathcal{H}}_{I,CS}^{(2)}= \ A\int d^3\vec{x} \  b(x) \sum_{\vec{k},\vec{k}^\prime}\sum_{\lambda,\lambda^\prime}f_k  f_{k^\prime}\Bigg\{ \\
    &e^{-i(k+k^\prime)\cdot x}\left[ -\Omega_k k_l  k^\prime _m  \ e^{(\lambda)}_{mj}(\vec{k}) \underbrace{\epsilon^{ijk} k^\prime _i \ e^{(\lambda^\prime)}_{lk}(\vec{k}^\prime)} +\Omega_k k_l  k^\prime _l   \ e^{(\lambda)}_{mj}(\vec{k}) \  \underbrace{\epsilon^{ijk}  k^\prime _i     e^{(\lambda^\prime)}_{mk}(\vec{k}^\prime)} - \Omega^{2}_k \Omega_{k^\prime}   \ e^{(\lambda)}_{jl}(\vec{k}) \   \underbrace{\epsilon^{ijk}   k^\prime _i  e^{(\lambda^\prime)}_{lk}(\vec{k}^\prime) } \ \right]  \hat{\alpha}^\dagger_{\vec{k},\lambda}\hat{\alpha}^\dagger_{\vec{k}^\prime,\lambda^\prime}
    \\
    +\ & e^{-i(k-k^\prime)\cdot x}\Bigg[ -\Omega_k k_l  k^\prime _m  \ e^{(\lambda)}_{mj}(\vec{k}) \    \underbrace{\epsilon^{ijk}  k^\prime _i \left[e^{(\lambda^\prime)}_{lk}(\vec{k}^\prime) \right]^\dagger } +\Omega_k k_l  k^\prime _l   \ e^{(\lambda)}_{mj}(\vec{k}) \   \underbrace{ \epsilon^{ijk}  k^\prime _i    \left[e^{(\lambda^\prime)}_{mk}(\vec{k}^\prime)\right]^\dagger}  \\
    & \hspace{18mm}  -  \Omega^{2}_k \Omega_{k^\prime}  \ e^{(\lambda)}_{jl}(\vec{k}) \  \underbrace{\epsilon^{ijk}  k^\prime _i       \left[e^{(\lambda^\prime)}_{lk}(\vec{k}^\prime)\right]^\dagger }\  \Bigg]  \hat{\alpha}^\dagger_{\vec{k},\lambda}\hat{\alpha}_{\vec{k}^\prime,\lambda^\prime}  \Bigg\} 
    + \text{h.c.}  \ .
\end{aligned}
\end{equation}
The under-braced terms satisfy the following identity~\cite{Alexander:2004wk}, 
\begin{equation}
    \epsilon^{ijk}k_je^{(\lambda)}_{kl}(\vec{k)} = i l_{\vec{k}} l_{(\lambda)}  k \  e^{(\lambda)}_{il}(\vec{k}) \ ,
\end{equation}
where
\begin{equation}
 l_{\vec{k}}=  \left\{
\begin{array}{ll}
      +1, & \theta_k< \pi/2 \\
     -1, &  \theta_k> \pi/2 \\
\end{array} 
\right. \ \ \ ,
\label{lk_definition}
\end{equation}
with $\theta_k$ the polar angle of $\vec{k}$ while also $l_R=-l_L=1$. The above identity can also be applied to the hermitian conjugate polarization tensor. So, equation \eqref{H_ineteraction_help_1} becomes:
\begin{equation}
\label{H_ineteraction_help_2}
  \begin{aligned}
     \hat{\mathcal{H}}_{I,CS}^{(2)}=&A\int d^3\vec{x} \  b(x) \sum_{\vec{k},\vec{k}^\prime}\sum_{\lambda,\lambda^\prime}i l_{\vec{k}^\prime}l_{\lambda^\prime}f_k  f_{k^\prime} \Omega_k \Omega_{k^\prime}\Bigg\{ \\
    & e^{-i(k+k^\prime)\cdot x}\Bigg[ k^\prime _m  e^{(\lambda)}_{mj}(\vec{k}) k_l e^{(\lambda^\prime)}_{jl}(\vec{k}^\prime) -k_l k^\prime_{l}e^{(\lambda)}_{mj}(\vec{k})e^{(\lambda^\prime)}_{mj}(\vec{k}^\prime) + \Omega_k \Omega_{k^\prime}e^{(\lambda)}_{jl}(\vec{k})e^{(\lambda^\prime)}_{jl}\ \Bigg]  \hat{\alpha}^\dagger_{\vec{k},\lambda}\hat{\alpha}^\dagger_{\vec{k}^\prime,\lambda^\prime}
    \\
    +\ & e^{-i(k-k^\prime)\cdot x}\Bigg[-k^\prime _m  e^{(\lambda)}_{mj}(\vec{k}) k_l \left[e^{(\lambda^\prime)}_{jl}(\vec{k}^\prime)\right]^\dagger +k_l k^\prime_{l}e^{(\lambda)}_{mj}(\vec{k})\left[e^{(\lambda^\prime)}_{mj}(\vec{k}^\prime)\right]^\dagger - \Omega_k \Omega_{k^\prime}e^{(\lambda)}_{jl}(\vec{k})\left[e^{(\lambda^\prime)}_{jl}(\vec{k}^\prime)\right]^\dagger\  \Bigg]  \hat{\alpha}^\dagger_{\vec{k},\lambda}\hat{\alpha}_{\vec{k}^\prime,\lambda^\prime}  \Bigg\} 
    + \text{h.c.}  \ 
\end{aligned}
\end{equation}
We may express \eqref{H_ineteraction_help_2} in a more compact form, i.e.
\begin{align} 
\label{Hamiltonian_CS_General}
\nonumber
\hat{\mathcal{H}}_{I,CS}^{(2)}=&\sum_{\vec{k},\vec{k}^\prime}\sum_{\lambda,\lambda^\prime}e^{i(\Omega_k + \Omega_{k^\prime})\cdot t}\Bigg(\int d^3\vec{x} \  b(x)e^{-i(\vec{k}+\vec{k}^\prime)\cdot \vec{x}}\Bigg)f_{\lambda,\lambda^\prime}(\vec{k},\vec{k}^\prime) \ \hat{\alpha}^\dagger_{\vec{k},\lambda}\hat{\alpha}^\dagger_{\vec{k}^\prime,\lambda^\prime} + \text{h.c.} \\
+&\sum_{\vec{k},\vec{k}^\prime}\sum_{\lambda,\lambda^\prime}e^{i(\Omega_k + \Omega_{k^\prime})\cdot t}\Bigg(\int d^3\vec{x} \  b(x)e^{-i(\vec{k}-\vec{k}^\prime)\cdot \vec{x}}\Bigg)g_{\lambda,\lambda^\prime}(\vec{k},\vec{k}^\prime) \ \hat{\alpha}^\dagger_{\vec{k},\lambda} \hat{\alpha}_{\vec{k}^\prime,\lambda^\prime} + \text{h.c.} \ ,
\end{align}
where we defined the joint functions:
\begin{align}
 \label{F_joint_function}
 f_{\lambda,\lambda^\prime}(\vec{k},\vec{k}^\prime)&=   iAf_k f_{k^\prime} \Omega_k \Omega_{k^\prime} l_{\vec{k}^\prime}l_{\lambda^\prime} \Bigg(   k^\prime _m  \ e^{(\lambda)}_{mj}(\vec{k})\ k_l\  e^{(\lambda^\prime)}_{jl}(\vec{k}^\prime)  -    k_l  k^\prime _l   \ e^{(\lambda)}_{mj}(\vec{k}) \   e^{(\lambda^\prime)}_{mj}(\vec{k}^\prime)    + \Omega_k \Omega_{k^\prime}  \ e^{(\lambda)}_{jl}(\vec{k}) \  e^{(\lambda^\prime)}_{jl}(\vec{k}^\prime)  \     \Bigg) \ ,
     \\
 \label{G_joint_function}
 g_{\lambda,\lambda^\prime}(\vec{k},\vec{k}^\prime)&=iAf_k f_{k^\prime} \Omega_k \Omega_{k^\prime} l_{\vec{k}^\prime}l_{\lambda^\prime} \Bigg(-  k^\prime _m   e^{(\lambda)}_{mj}(\vec{k})\ k_l  \left[e^{(\lambda^\prime)}_{jl}(\vec{k}^\prime) \right]^\dagger +    k_l  k^\prime _l    e^{(\lambda)}_{mj}(\vec{k})   \left[ e^{(\lambda^\prime)}_{mj}(\vec{k}^\prime)\right]^\dagger  -  \Omega_k  \ \Omega_{k^\prime}   \ e^{(\lambda)}_{jl}(\vec{k})  \left[ e^{(\lambda^\prime)}_{jl}(\vec{k}^\prime)\right]^\dagger\   \Bigg) \ .
\end{align}
 Now in order to proceed, we need to calculate the contractions between the polarizations tensors and the wave vectors appearing in the definition of the joint functions \eqref{F_joint_function} and \eqref{G_joint_function}. In order to accomplish that, one has to calculate the following quantities,
\begin{equation}
    k^\prime_m e^{(\lambda)}_{mj}(\vec{k})e^{(\lambda^\prime)}_{jl}(\vec{k}^\prime)k_l\ \ \text{and}\ \ e^{(\lambda)}_{ij}(\vec{k})e^{(\lambda^\prime)}_{ij}(\vec{k}^\prime)\ ,
\end{equation}
for each combination of $\lambda$ and $\lambda^\prime$. In order to calculate these quantities, we define the triad for an arbitrary momentum, $\vec{k}$, as follows~\cite{Alexander:2004wk}, 
\begin{equation}
    \begin{aligned}
        &e_1(\vec{k})=(\sin\varphi_k , -\cos\varphi_k, 0    )\\
        & e_2(\vec{k})=l_{\vec{k}}(\cos\theta_k\cos\varphi_k,\cos\theta_k\sin\varphi_k,-\sin\theta_k)\\
        & e_3(\vec{k})= \vec{k}/k=(\sin\theta_k\cos\varphi_k,\sin\theta_k\sin\varphi_k,\cos\theta_k) \ ,
    \end{aligned}
\end{equation}
and similarly for $\vec{k}^\prime$, while $l_{\vec{k}}$ is defined in \eqref{lk_definition}. Then, choosing the vector to lie on the z-y plane, i.e. $\varphi_k=\varphi_{k^\prime}=\frac{\pi}{2}$, while also $\theta_k=\frac{\pi}{2}-\frac{\Delta\theta}{2}$ and $\theta_{k^\prime}=\frac{\pi}{2}+\frac{\Delta\theta}{2}$, with $\Delta\theta$ denoting their mutual angle, we obtain, 
\begin{equation}
\label{first_terms_of_f_lambda}
\begin{aligned}
    &k^\prime_m e^{(R)}_{mj}(\vec{k})e^{(R)}_{jl}(\vec{k}^\prime)k_l= -\frac{kk^\prime}{4}\csc^2\left(\frac{\Delta\theta}{2}\right)\sin^4\Delta\theta\\
    &k^\prime_m e^{(R)}_{mj}(\vec{k})e^{(L)}_{jl}(\vec{k}^\prime)k_l=kk^\prime\sin^2\left(\frac{\Delta\theta}{2}\right)\sin^2\Delta\theta\\
    &k^\prime_m e^{(L)}_{mj}(\vec{k})e^{(R)}_{jl}(\vec{k}^\prime)k_l= kk^\prime\sin^2\left(\frac{\Delta\theta}{2}\right)\sin^2\Delta\theta \\
    &k^\prime_m e^{(L)}_{mj}(\vec{k})e^{(L)}_{jl}(\vec{k}^\prime)k_l= -\frac{kk^\prime}{4}\csc^2\left(\frac{\Delta\theta}{2}\right)\sin^4\Delta\theta \ ,
    \end{aligned}
\end{equation}
which are (self-consistently) vanishing for $\Delta\theta=\pi$, while also:  
\begin{equation}
\label{second_terms_of_f_lambda}
\begin{aligned}
    &k^\prime_l e^{(R)}_{mj}(\vec{k})e^{(R)}_{mj}(\vec{k}^\prime)k_l= 2 k k^\prime\cos^4\left(\frac{\Delta\theta}{2}\right)\\
    &k^\prime_l e^{(R)}_{mj}(\vec{k})e^{(L)}_{mj}(\vec{k}^\prime)k_l=2 k k^\prime\sin^4\left(\frac{\Delta\theta}{2}\right)\\
    &k^\prime_l e^{(L)}_{mj}(\vec{k})e^{(R)}_{mj}(\vec{k}^\prime)k_l=2 k k^\prime\sin^4\left(\frac{\Delta\theta}{2}\right)\\
    &k^\prime_l e^{(L)}_{mj}(\vec{k})e^{(L)}_{mj}(\vec{k}^\prime)k_l=2 k k^\prime\cos^4\left(\frac{\Delta\theta}{2}\right)\\
    \end{aligned}
\end{equation}
following from:
\begin{equation}
\begin{aligned}
\label{formulas_polarizations_contractions}
    &e^{(R)}_{ij}(\vec{k})e^{(R)}_{ij}(\vec{k}^\prime)= 2\cos^2\left( 
\frac{\Delta\theta}{2}   \right)  \\
    &e^{(R)}_{ij}(\vec{k})e^{(L)}_{ij}(\vec{k}^\prime)=  2\sin^2\left( 
\frac{\Delta\theta}{2}   \right)\\
    &e^{(L)}_{ij}(\vec{k})e^{(R)}_{il}(\vec{k}^\prime)= 2\sin^2\left( 
\frac{\Delta\theta}{2}   \right) \\
    &e^{(L)}_{ij}(\vec{k})e^{(L)}_{ij}(\vec{k}^\prime)=  2\cos^2\left( 
\frac{\Delta\theta}{2}   \right) \\
    \end{aligned}
\end{equation}
and the fact that $k^\prime_l \ k_l = k k^\prime\cos\Delta\theta$. For $\Delta\theta=\pi$, in the last equations \eqref{formulas_polarizations_contractions} only the off-diagonal elements are non-trivial, as the normalization of the polarization tensors requires. 

\color{black}
 In the presence of quadratic interactions of the graviton, we will show that  the coherent states produced by the cloud ({\it cf.} \eqref{displacement_operator}) are squeezed, forming a squeezed-coherent state of the following form, 
\begin{equation}
    \vert\psi\rangle = \prod_{\lambda,\vec{k}}\hat{D}(\xi_{\lambda,\vec{k}})\hat{S}_{GR}^{(2)}\vert 0\rangle,  
    \label{Squeezed-Coherent}
\end{equation}
where $\vert 0 \rangle$ is the appropriate vacuum state, $\hat{D}(\xi_{\lambda,\vec{k}})$ is the displacement operator \eqref{displacement_operator} and $S^{(2)}_{GR}$ is the multi-mode squeezing operator of \eqref{Scattering_Multimode_Squeeze_GR}.\footnote{\color{black}We do not include here the corresponding squeezing operator produced via the chiral gCS anomaly, due to its large suppression as shown in section \ref{sec:cs}.\color{black}} 
\color{black}

\color{black}

\section{BH Superradiance and the Gravitational Atom}\label{sec:BHSuperradiance}

Superradiance in (quantum) optics~\cite{Gross:1982dkt} describes a scattering phenomenon,  entailing the enhancement of the amplitude of the reflected wave, as compared to that of the incident wave. The latter is due to a collective spontaneous emission of photons, and does not  imply any instabilities in the system. However, such instabilities do characterise the analogue effect of BH superradiance, which is of interest to our work here. 

BH superradiance \cite{BritoCardoso} is a phenomenon associated with radiation enhancement in a system that ordinarily dissipates energy. \color{black} BHs provide the arena of such dissipative systems, where energy and matter that passes beyond a certain point (the event horizon) cannot escape. As reviewed in Appendix \ref{app:BHSuperradiance}, the Klein-Gordon equation for a massive (pseudo)scalar field in a rotating (Kerr-type~\cite{Kerr:1963ud}) BH background admits quasibound states, which are labeled by integer numbers $\left(n,l,m\right)$.
In the non-relativistic regime \cite{Detweiller}, the angular part of the solution is described by spherical harmonics, while the effective radial equation reduces to a Coulomb-like problem, with solutions resembling the wavefunctions of the hydrogen atom. 
\color{black} Under such circumstances, the (pseudo)scalar-field-BH system exhibits an {\it instability}~\cite{Detweiller}, due to the presence of a positive imaginary part of the field's frequency ($\omega_I > 0$)  ({\it cf.} \eqref{imaginary_part_frequency_help0}, \eqref{imaginary_part_frequency}).  
The imaginary part of the frequency \color{black} denotes the strength of the superradiant instability, which grows exponentially like $e^{\omega_I t}$ \color{black} ({\it cf.} figure~\ref{growth_rate} and relevant discussion in Appendix \ref{app:BHSuperradiance}), \color{black} as long as the superradiance condition of eq.\eqref{superradiance_condition} $\omega<m\,  \Omega_{H}$ holds. Here, $\omega$ (the real part of the frequency) is given by~\cite{Detweiller,BritoCardoso}:

\begin{align}\label{realfreq}
\omega \approx \mu_b\left(1 - \frac{a_{\mu}^2}{2n^2}\right)\,,
\end{align}
where
\begin{equation}\label{amudef}
    a_\mu \equiv G \mathcal{M}\mu_b
\end{equation}
is the dimensionless "gravitational atom" coupling (in the non-relativistic regime $a_\mu \ll 1$). In this regime, the classical axion field is given by:
\begin{equation}
\label{Axion_Non_Relativistic_Solution}
    b(t,r,\theta,\varphi)=\sum_{nlm}e^{-i \omega_{nlm} t} \sqrt{\frac{N_{nlm}}{2\mu_b}}\Psi_{nlm} + \text{c.c} \ ,
\end{equation}
where $\omega_{nlm}$ is the complex frequency of the scalar field in a state characterised by quantum numbers $n, l, m$, and 
$N_{nlm}$ denotes the number of axions in the respective state, with $\Psi_{nlm}$ obeying the one particle wavefunction normalization $\int d^3x \vert \Psi_{nlm}\vert^2
=1$. As long as the superradiance condition is satisfied, the axionic-cloud will grow at a rate that must be faster than the relevant evolution timescale of the BH.
The most dominant mode corresponds to the "$2p$-axion state" $(n=2,l=m=1)$, 
\begin{equation}
\label{Psi_2p_Wavefunction}
    \Psi_{2p}(\vec{x})=\frac{1}{8\sqrt{\pi}}\frac{r}{r_0^{5/2}}e^{-\frac{r}{2r_0}} \ 
e^{i\varphi}\sin\theta \ ,
\end{equation}
where 
\begin{equation}
\label{bohr_radius}
    r_0=\left(a_\mu\mu_b\right)^{-1}
\end{equation}
denotes the "gravitational Bohr radius". The characteristic time scale of the superradiant instability for the $2p$-state scales as ({\it cf.} figure \ref{growth_rate}):
\begin{equation}
\label{timescale_superradiance}
\tau_s=\frac{1}{\omega_I (2p)}=24\ \mu_b^{-1}\left(\frac{\alpha}{G\mathcal{M}}\right)^{-1} a^{ -8}_\mu \ ,
\end{equation}
\color{black} where $\omega_I$ for this state is given in \eqref{2p_imaginary_omega}. \color{black}
The maximum number of axions occupying the dominant mode is given by \cite{arvanitaki3,arvanitaki4,Bernal_2022}:
\begin{equation}
\label{N_2p_Max}
    N_{2p}^{max}\sim 10^{74}\left(\frac{\Delta \alpha_\star}{0.1}\right)\left(\frac{\mathcal{M}}{M_{\odot}}\right)^{2} \ ,
\end{equation}
where $\alpha_\star=\alpha/(G\mathcal{M})$ and $\Delta \alpha_\star \sim \mathcal{O}(0.1)$ denotes the difference between the initial and final BH spin. The radii of the "$2p$-axionic cloud" is given by: 
\begin{equation}
\label{r_cloud_and_Delta_r}
    \langle r_{c} \rangle = 5 \ r_0  \quad \text{with} \quad \Delta r_{c} = \sqrt{5} \ r_0.
\end{equation}
In the non relativistic limit $\left(a_\mu \ll 1\right)$, $r_c$ is much larger than the dimensions of the BH $r_+ \sim G \mathcal{M}$. 
Therefore, we proceed to the approximation scheme of ignoring the curvature effects and quantize gravitational waves in the environment of the axionic cloud with a flat spacetime background.

It is well known that in Chern-Simons gravitational theory \cite{Jackiw:2003pm}, BH exhibit axionic hair \cite{Duncan:1992vz, Yagi:2012ya,Chatzifotis:2022mob}. The latter corresponds to stationary configurations that solve the gravitational equations of motion. On the other hand, the superradiant instability of the axion field interacting with the anomaly corresponds to the non-stationary solutions of the (pseudo-) scalar Klein-Gordon equation in the Kerr background sourced by $R_{CS}$.  In both cases, the effect of the higher curvature coupling with the axion field is highly suppressed by the BH mass over the Planck mass, and thus the effect is negligible, unless we are dealing with extremely small BHs, with masses comparable to the Planck mass. This means, that the process of superradiance, in the decoupling limit \cite{Teukolsky}, can be considered identical to the case of GR \cite{Richards:2023xsr}.   
 
\section{The Multi-Mode Squeezed States of Gravitons}\label{sec:Multimode}

From the local point of view, astrophysical sources of non-classical GWs are quite limited in the literature, being although of main importance, especially from the point of view of producing a large enough squeezing parameter. In Quantum Optics, non-classical states of photons are produced by non-linear interactions of the electromagnetic field in the presence of a medium. Spontaneous Parametric Down Conversion (SPDC) \cite{PhysRevA.31.2409}, is a physical process  that takes place in certain types of crystals~\cite{Wu:1986zz}. There, one high-energy (``pump’’) photon, coming from a laser, splits into two lower-energy ones. These two new photons are "entangled," which in layman terms means that their properties are deeply correlated, even if they are far apart. In contrast, Spontaneous Four Wave Mixing (SFWM) \cite{SFWM1,SFWM2,SFWM_3} is another non-linear process in quantum optics, where two ``pump’’ photons interact inside a nonlinear medium. Through quantum-mechanical interactions, the non-linearity of the medium generates two entangled photons. This leads to a quadratic dependence of the pump field, in contrast to the linear dependence of SPDC. In SFWM, all photons involved in the process generally exist in the same frequency range. The goal is to examine whether an astrophysical source is a viable option for producing squeezed graviton states with a large enough squeezing parameter. Outside a rotating BH (BH), superradiant instability leads to the formation of an axionic cloud/condensate around the BH; the so called "gravitational atom" \cite{arvanitaki,arvanitaki3,arvanitaki4,BritoCardoso,VicenteCardoso}. \par
\begin{figure}[ht]
    \centering
    \includegraphics[width=0.6\linewidth]{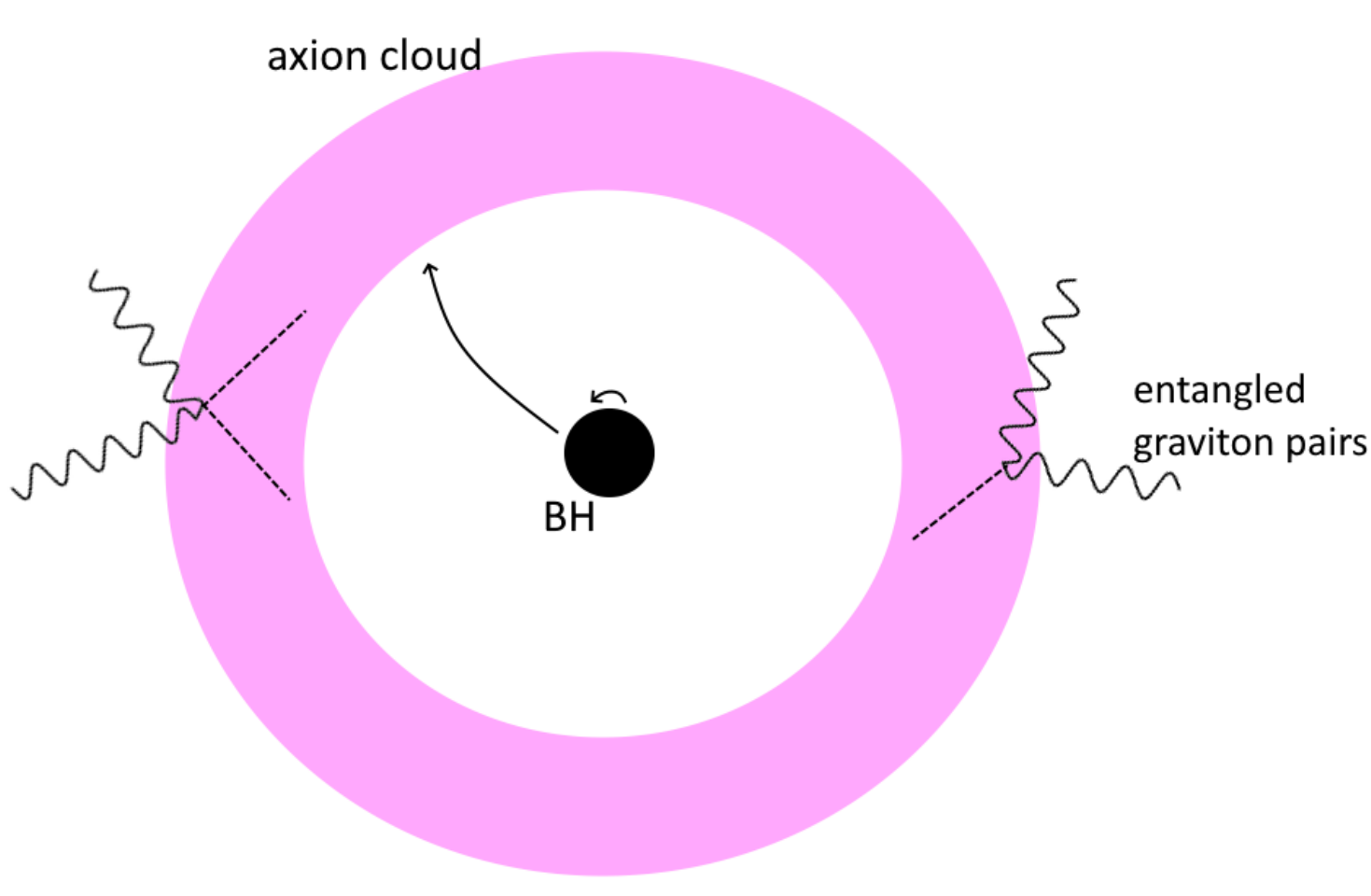}
    \caption{The superradiant axionic cloud around a rotating BH. The non-linear axion-graviton interactions are responsible for the production of pairs of entangled gravitons; the processes involved is the annihilation of two axions into two gravitons (in analogy to SFWM), and the axion decay into two gravitons (in analogy to SPDC). Picture taken from \cite{Dorlis:2025zzz}. }
    \label{axion_cloud_entangled_gravitons}
\end{figure}

 
The interaction linear to the graviton \eqref{Interaction_Coherent_GR} is responsible for the production of gravitational waves (coherent states of the graviton, see for example \cite{Skagerstam:2018jkw}). The remaining interactions correspond to processes including two gravitons. The first one \eqref{Interaction_second_order_GR}, stemming from General Relativity, involves two axions annihilating into two gravitons, in analogy to SFWM (the axion condensate-field plays the role of the ''pump" field), while the second one \eqref{Interaction_CS} is responsible for the axion decay into two gravitons, in analogy to SPDC ({\it cf.} figure \ref{axion_cloud_entangled_gravitons}). As we shall show below, the so-produced two-graviton states are entangled, given that they cannot be separated into single states~\cite{Law:2000hyw} (\color{black} {\it cf.} also discussion in Appendix \ref{appC:takagi}, in particular subsection~\ref{sec:appD}\color{black}). In what follows, we focus on the non-classical Hamiltonians  \eqref{Hamiltonian_Quantum_GR_second_order_General} and \eqref{Hamiltonian_CS_General}. \par

\subsection{\textbf{General Relativity Induced Squeezing Operator - Correlation Functions}}\label{sec:VIA}

We begin by substituting into the GR interaction Hamiltonian \eqref{Hamiltonian_Quantum_GR_second_order_General} the condensate axion field of \eqref{Axion_Non_Relativistic_Solution} stemming from the superradiant instability. Focusing only on the $"2p-$axions state", as this state is several orders of magnitude stronger than the others, the Hamiltonian \eqref{Hamiltonian_Quantum_GR_second_order_General} takes the following form:
\begin{align}
\label{Hamiltonian_Quantum_GR_second_order_Condensate_help}
\hat{\mathcal{H}}^{(2)}_{I}= \sum_{\vec{k},\vec{k}^\prime}\sum_{\lambda,\lambda^\prime}e^{i(\Omega_k+\Omega_{k^\prime} - 2\omega)\cdot t}f_k f_{k^\prime}  N_{2p}\Bigg(\int d^3 \vec{x} \  \frac{\partial^{i} \Psi_{2p} \ \partial^{j}\Psi_{2p}}{4\mu_b} \ e^{-i(\vec{k}+\vec{k}^\prime)\cdot \vec{x}}  \, e^{(\lambda)}_{im}(\vec{k})\, e^{(\lambda^\prime)}_{mj}(\vec{k}^\prime)  \ \hat{\alpha}^\dagger_{\lambda,\vec{k}}\hat{\alpha}^\dagger_{\lambda^\prime,\vec{k}^\prime} \Bigg) \ + \ \text{h.c.}  
\end{align}
where $\omega\equiv\omega_{2p}$ is the energy of $2p-$axions, which in the non-relativistic limit satisfies $\omega\approx \mu_b$ and $N_{2p}$ is the number of axions in the dominant $2p-$state of the cloud. In the above Hamiltonian, following the logic behind SFWM and SPDC, we ignored terms with mixed creation and annihilation operators for the graviton. These terms come with exponentials of the form $\left(\Omega_k+\Omega_{k^\prime} + 2\omega\right),\left(\Omega_k-\Omega_{k^\prime} + 2\omega\right),\left(\Omega_k-\Omega_{k^\prime} - 2\omega\right)$. Such interactions are ignored under the Rotating Wave Approximation (RWA), since under the assumption of a long enough interaction time, these terms oscillate rapidly and are subdominant. The evolution operator for this process takes the form of the multimode squeezing operator \cite{multimode}~\footnote{The reader is referred to Appendix \ref{appC:takagi} for a comprehensive introductory discussion on multimode squeezing via Takagi decomposition~\cite{Takagi1933,Houde:2024mkj} in the context of quantum optics, whose basic features are shared with our gravity case discussed here.}, and is given by:
\begin{equation}
\label{Scattering_Multimode_Squeeze_GR}
    \hat{S}^{(2)}_{GR}=\exp\left[ \frac{1}{2}\sum_{I,J}\mathcal{G}^{(GR)}_{IJ}\ \hat{\alpha}^\dagger_I\hat{\alpha}^\dagger_J -h.c. \right]  \ ,
\end{equation}
where we introduced the tuple index $I=(\lambda,\vec{k})$ denoting the graviton states, and we defined:
\begin{equation}
\label{G_GR_General}
    \mathcal{G}^{(GR)}_{IJ}=-2\ i\ \mathcal{F}^{(GR)}_{IJ} \  T \ \text{sinc}\left[\left(\Omega_k + \Omega_{k^\prime} - 2\omega \right) \frac{T}{2}\right] ,
\end{equation}
with
\begin{equation}
\label{F_GR_1}
    \mathcal{F}^{(GR)}_{(\vec{k},\lambda)(\vec{k}^\prime,\lambda^\prime)}=f_k f_{k^\prime}  N_{2p}\Bigg(\int d^3 \vec{x} \  \frac{\partial^{i} \Psi_{2p} \ \partial^{j}\Psi_{2p}}{4\mu_b} \ e^{-i(\vec{k}+\vec{k}^\prime)\cdot \vec{x}}  \, e^{(\lambda)}_{im}(\vec{k})\, e^{(\lambda^\prime)}_{mj}(\vec{k}^\prime)\Bigg) \  .
\end{equation}
The time integration of the Hamiltonian will yield a characteristic timescale $T$, which denotes the lifetime of the classical (coherent) source that drives the process; in this case, the axionic condensate around the BH. For clarity, note that if $\mathcal{G}_{IJ}\sim \delta_{IJ}$, the evolution operator  \eqref{Scattering_Multimode_Squeeze_GR} reduces to the case of one mode squeezing. In addition, energy conservation $\Omega_k + \Omega_{k^{\prime}} \approx 2\omega$, implies that 
\begin{equation}
\label{G_max_GR}
    \mathcal{G}^{(GR)}_{IJ}\approx-2i  \ T \mathcal{F}^{(GR)}_{IJ} \ .
\end{equation} 
An important part to clarify is that, in the definition of the evolution operator, the time ordering of Dyson's formula has been omitted under the assumption of nearly rare event rate. This becomes more evident when one considers the Magnus expansion for the evolution operator \cite{magnus1, magnus2}, which is an expansion of the exponent according to the commutators of the Hamiltonian at different times. Specifically, the Magnus expansion has the following form, 
\begin{equation}
    \hat{S}=\exp\left[\sum_nM_n\right]\  , 
    \label{Magnus_expansion}
\end{equation}
where the operators $M_n$ are related to the commutators of the Hamiltonian at different times. To clarify, consider the Dyson's expansion written as $\hat{S}=I+\sum_nP_n$. Then, the coefficients of each expansion are related through, 
\begin{equation}
\begin{aligned}
&M_1=P_1\\
&M_2=P_2-P_1^2\\
&M_3=P_3-\frac{1}{2}\left( P_1P_2+P_2P_1    \right)+\frac{1}{3}P_1^3\\
&M_4=\dots \ ,
\end{aligned}   
\label{Magnus_Dyson}
\end{equation}
from which it is clear that the Magnus expansion can be understood as a re-summation of the Dyson's expansion terms. In this sense, as is customary in SPDC and SFWM, we make the assumption that events of graviton production obey $[\mathcal{H}_{int}(x^\prime),\mathcal{H}_{int}(x)]=0$, that is, the emissions occur at spacelike separated events. This assumption, viewed in terms of the Magnus expansion \eqref{Magnus_expansion}, means that the produced quantum state corresponds to a bath of entangled particles produced at tree level, without including loop contributions.

\par

We shall proceed to the analytic calculation of each quantity involved inside \eqref{F_GR_1}. Then, we will be in place to make safe estimates regarding the strength of the interaction, captured inside \eqref{G_GR_General} and as a consequence, the strength of the squeezing of the gravitational waves. We re-write \eqref{F_GR_1} as follows:
\begin{equation}
\label{F_GR_correlations}
\mathcal{F}^{(GR)}_{(\vec{k},\lambda)(\vec{k}^\prime,\lambda^\prime)} = \frac{  f_k f_{k^\prime} N_{2p}}{4\mu_b}  e^{(\lambda)}_{im}(\vec{k}) \, I_{ij}(\vec{k},\vec{k}^\prime )\, e^{(\lambda^\prime)}_{mj}(\vec{k}^\prime) \ , 
\end{equation}
where $I_{ij}$, $i=1,2,3$, is the tensorial (under rotations)  structure denoting the influence of the source,
\begin{equation}
\label{I_ij_definition}
    I_{ij}(\vec{k},\vec{k}^\prime)=\int d^{3} \vec{x} \ (\partial_i \Psi_{2p}(\vec{x}))\, (\partial_j \Psi_{2p}(\vec{x}))\, e^{-i(\vec{k}+\vec{k}^\prime)\cdot \vec{x}} \ ,  
\end{equation}
with $\Psi_{2p}$ given by \eqref{Psi_2p_Wavefunction}. Performing the differentiation of \eqref{I_ij_definition}, we obtain:
\begin{equation}
\label{I_ij_General_form}
\begin{aligned}
     I_{ij}(\vec{k},\vec{k}^\prime)=&\frac{\left(\delta_{1i}+i\delta_{2i}\right)\left(\delta_{1j}+i\delta_{2j}\right)}{64\pi  r^{5}_0}\int d^{3}\vec{x} \ e^{i \vec{q}\cdot\vec{x}}  e^{-r/r_0}  - 
     \frac{\left(\delta_{1i}+i\delta_{2i}\right)}{128\pi  r^{6}_0}\int d^{3}\vec{x} \ e^{i \vec{q}\cdot\vec{x}} \ e^{-r/r_0} \ \sin\theta \ e^{i\varphi} x_j \ \\ &-
     \frac{\left(\delta_{1j}+i\delta_{2j}\right)}{128\pi  r^{6}_0}\int d^{3}\vec{x} \ e^{i \vec{q}\cdot\vec{x}} \ e^{-r/r_0} \ \sin\theta \ e^{i\varphi} x_i  \ \\
   &+ \frac{1}{256\pi  r^{7}_0}\int d^{3}\vec{x} \ e^{i \vec{q}\cdot\vec{x}} \ e^{-r/r_0} \ \sin^{2}\theta \ e^{2i\varphi} x_i x_j \ ,
\end{aligned}
\end{equation}
where $\vec{q}=-\vec{k}-\vec{k}^\prime$. We may now define the following integrals:
\begin{align}
    \label{I(1)} I^{(1)}&=   \int d^{3}\vec{x} \ e^{i \vec{q}\cdot\vec{x}} \ e^{-r/r_0} \ ,       \\
    \label{I(2)} I^{(2)}&=  \int d^{3}\vec{x} \ e^{i \vec{q}\cdot\vec{x}} \ e^{-r/r_0} \ \sin\theta \ e^{i\varphi} \ , \\
    \label{I(3)} I^{(3)}&=  \int d^{3}\vec{x} \ e^{i \vec{q}\cdot\vec{x}} \ e^{-r/r_0} \ \sin^{2}\theta \ e^{2i\varphi}  \ . 
\end{align}
One can show  that \eqref{I_ij_General_form} takes the following form:
\begin{align}
\label{I_ij_tensorial_structure}
     I_{ij}(\vec{k},\vec{k}^\prime)=\frac{\left(\delta_{1i}+i\delta_{2i}\right)\left(\delta_{1j}+i\delta_{2j}\right)}{64\pi  r^{5}_0} \ I^{(1)}   -   \Bigg(\frac{\left(\delta_{1i}+i\delta_{2i}\right)}{128\pi  r^{6}_0}I^{(2)}_j + i \leftrightarrow j \Bigg)  +
    \frac{1}{256\pi  r^{7}_0}I^{(3)}_{ij} \ .
\end{align}
where
\begin{equation}
    I^{(2)}_i=\left(-i\partial_{q_j}\right) \ I^{(2)} \ \ \ \text{and} \ \ \ I^{(3)}_{ij}=\left(-i\partial_{q_i}\right)\left(-i\partial_{q_j}\right) \ I^{(3)} \ .
\end{equation}
We can calculate the integrals \eqref{I(1)},\eqref{I(2)},\eqref{I(3)}, using the following formula, 
\begin{equation}
    \label{Integral_formula}
    I_{lm}=\int d^{3} \vec{x}\ e^{i \vec{q}\cdot\vec{x}} f(r)  \ Y_{lm}(\theta,\varphi) = 4\pi \, i^l\, Y_{lm}(\hat{q})\int_{0}^{\infty}dr\ r^2 f(r) \ j_{l}(qr) \ ,
\end{equation}
where $Y_{lm}(\theta,\varphi)$ are the spherical harmonics and $j_l(qr)$ the Bessel functions of the first kind. Performing the corresponding integrals, we arrive at the following results:
\begin{align}
    \label{I(1)RESULT}
    I^{(1)}&=\frac{8 \pi r^{3}_0}{\left(1+q^{2}r^{2}_0\right)^2} \ , \\
    \label{I(2)RESULT}
    I^{(2)}&= 8 \pi i \  r^{3}_0\frac{qr_0}{\left(1+q^2 r^{2}_0\right)^{2}}\sin \theta_q e^{i\varphi_q} \ , \\ 
    \label{I(3)RESULT}
    I^{(3)}&= -4\pi  r^{3}_0 \ e^{2i\varphi_q}\sin^{2} \theta_q
\left(\frac{3}{q^3 r^3_0}\arctan(qr_0) - \frac{2}{\left(1+q^{2}r^{2}_0\right)^2} - \frac{3}{q^2 r^{2}_0\left(1+q^{2}r^{2}_0\right)}\right) \ ,
\end{align}
where $\theta_q,\varphi_q$ denote the polar and azimuthal angle for the vector $\vec{q}=-\left(\vec{k} + \vec{k}^\prime\right)$.\par  With these results at hand, we can proceed to the calculation of \eqref{I_ij_tensorial_structure}, i.e. calculate the derivatives of \eqref{I(2)RESULT} and \eqref{I(3)RESULT}. By defining the dimensionless variable 
\begin{equation}
\label{dimensionless_Variable_Q}
    Q_i\equiv r_0 q_i \ ,
\end{equation}
and also making use of the fact that $\sin \theta_q e^{i\varphi_q}=\frac{Q_1}{Q}+i\frac{Q_2}{Q}$ and that $\partial_{q_i} = r_0 \partial_{Q_i}$, we obtain:
\begin{equation}
    I^{(2)}_j = 8\pi\frac{r^{4}_0}{\left(1+Q^2\right)^2}\Bigg(\delta_{1j}+i\delta_{2j} - 4\frac{Q_1 + i Q_2}{1+Q^2}Q_j\Bigg) \ . 
\end{equation}
The calculation of $I^{(3)}_{ij}$ is a quite complex problem, so we'll try to break it down into smaller, individual parts. As we mentioned, $I^{(3)}_{ij}$ will be attained after the double differentiation $\left(-\partial_{q_i}\partial_{q_j} \ I^{(3)}\right)=\left(-r^{2}_0\partial_{Q_i}\partial_{Q_j}I^{(3)}\right)$. We shall express $I^{(3)}$ in the following form:
\begin{equation}
    I^{(3)}=-4\pi r^{3}_0 \ A   B \ ,
\end{equation}
where
\begin{align}
    A&=\left(\frac{Q_1}{Q}+i\frac{Q_2}{Q}\right)^2 \ , \\
    B&=\frac{3}{Q^3}\arctan(Q) - \frac{2}{\left(1+Q^{2}\right)^2} - \frac{3}{Q^2\left(1+Q^2\right)} \ . 
\end{align}
The relative terms of $I^{(3)}_{ij}$ are given by:
\begin{equation}
    \label{I(3)_ij_help}
    I^{(3)}_{ij}=4\pi r^{5}_0\left(\partial_{Q_i}A \ \partial_{Q_j}B + \partial_{Q_i}B \ \partial_{Q_j}A +A \ \partial_{Q_i}\partial_{Q_j}B + B \ \partial_{Q_i}\partial_{Q_j}A\right) \ ,
\end{equation}
where these terms are presented below:
\begin{align}
    \partial_{Q_i}A =& \  2 \ \frac{Q_1 + iQ_2}{Q}\Bigg(\frac{\delta_{1i} + i\delta_{2i}}{Q}-\frac{Q_1 + iQ_2}{Q^{3}}Q_i\Bigg) \ , \\
    \partial_{Q_i} B =& \Bigg(\frac{9+24Q^{2} + 23Q^{4}}{Q^{4}\left(1+Q^{2}\right)^{3}}-\frac{9 \arctan(Q)}{Q^{5}} \Bigg)Q_i     \    \end{align}
and
\begin{align}
     \nonumber
     \partial_{Q_j}\partial_{Q_i}A =&  \frac{2}{Q^{2}}\left(\delta_{1i}+i\delta_{2i}\right)\left(\delta_{1j}+i\delta_{2j}\right)-4\frac{Q_1 + iQ_2}{Q^{4}}\Bigg(\left(\delta_{1i}+i\delta_{2i}\right) \ Q_j + i\leftrightarrow j\Bigg)   \\
     -&\ 2 \ \frac{\left(Q_1 + iQ_2\right)^{2}}{Q^{4}}\delta_{ij} + 8 \ \frac{\left(Q_1 + iQ_2\right)^{2}}{Q^{6}}Q_{i}Q_{j} \ ,
\end{align}
\begin{align}
     \nonumber
     \partial_{Q_j}\partial_{Q_i}B =&\Bigg(\frac{9+24Q^{2} + 23Q^{4}}{Q^{4}\left(1+Q^{2}\right)^{3}}-\frac{9 \arctan(Q)}{Q^{5}} \Bigg)\delta_{ij}  \ + \ 
     \Bigg(-\frac{45+165Q^2 + 219Q^4 + 147 Q^6}{Q^{6}\left(1+Q^{2}\right)^4}  \\ 
     &+45 \ \frac{\arctan(Q)}{Q^{7}}\Bigg)Q_i Q_j \ .
\end{align}

\begin{figure}
    \centering
    \includegraphics[width=0.5\linewidth]{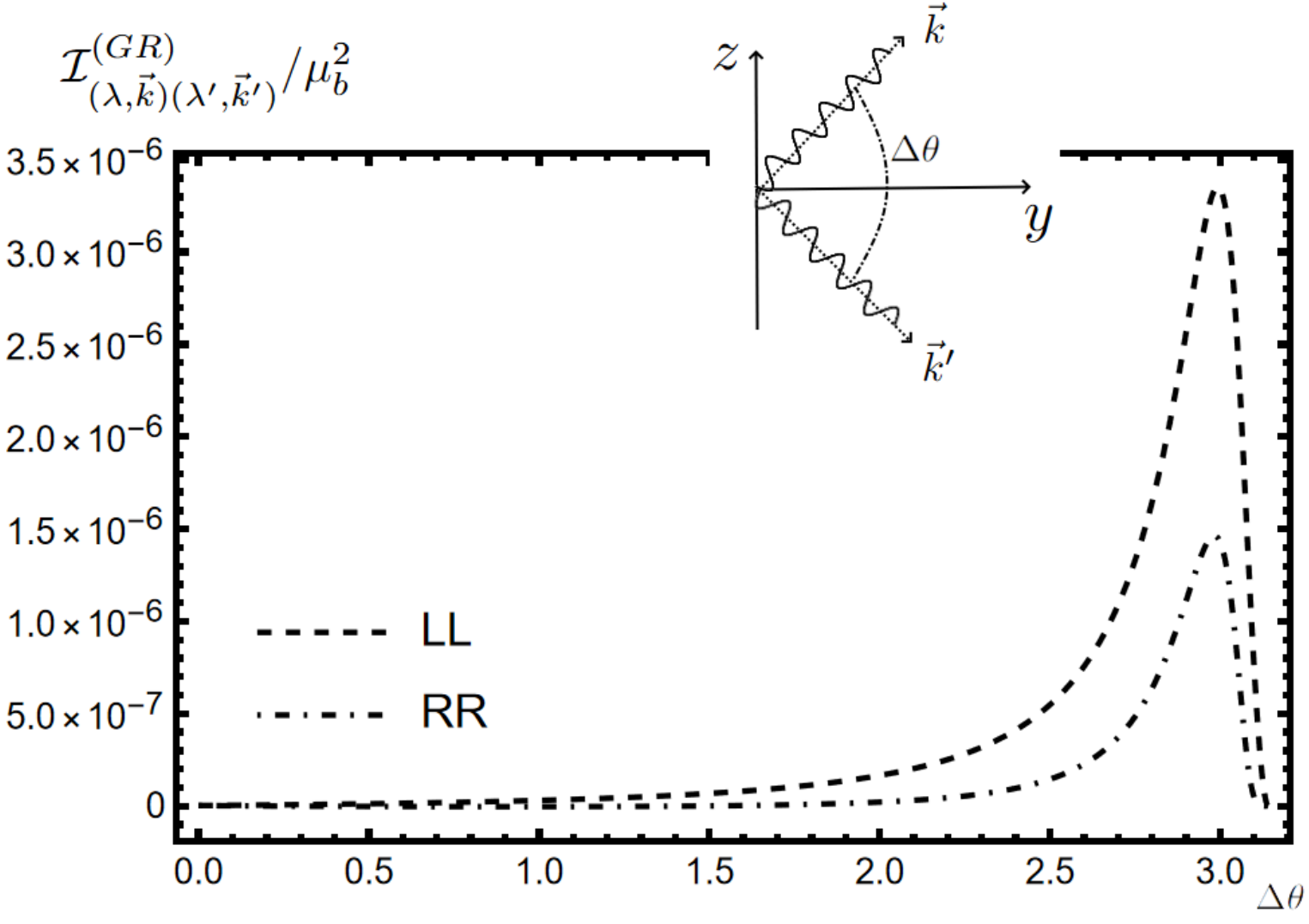}
    \caption{Polarization correlations from the GR induced interaction for LL and RR polarization pairs. The asymmetry is the result of the $2p$-state of the axionic cloud.  The plots have been made for $a_{\mu}=0.1$, with $\vec{k}+\vec{k}^\prime$ lying on the y-axis, \color{black} and $k=k^\prime=\mu_b$. Figure taken from \cite{Dorlis:2025zzz}.\color{black}}
    \label{GR_Correlations_LL_RR}
\end{figure}

\begin{figure}
    \centering
    \includegraphics[width=0.5\linewidth]{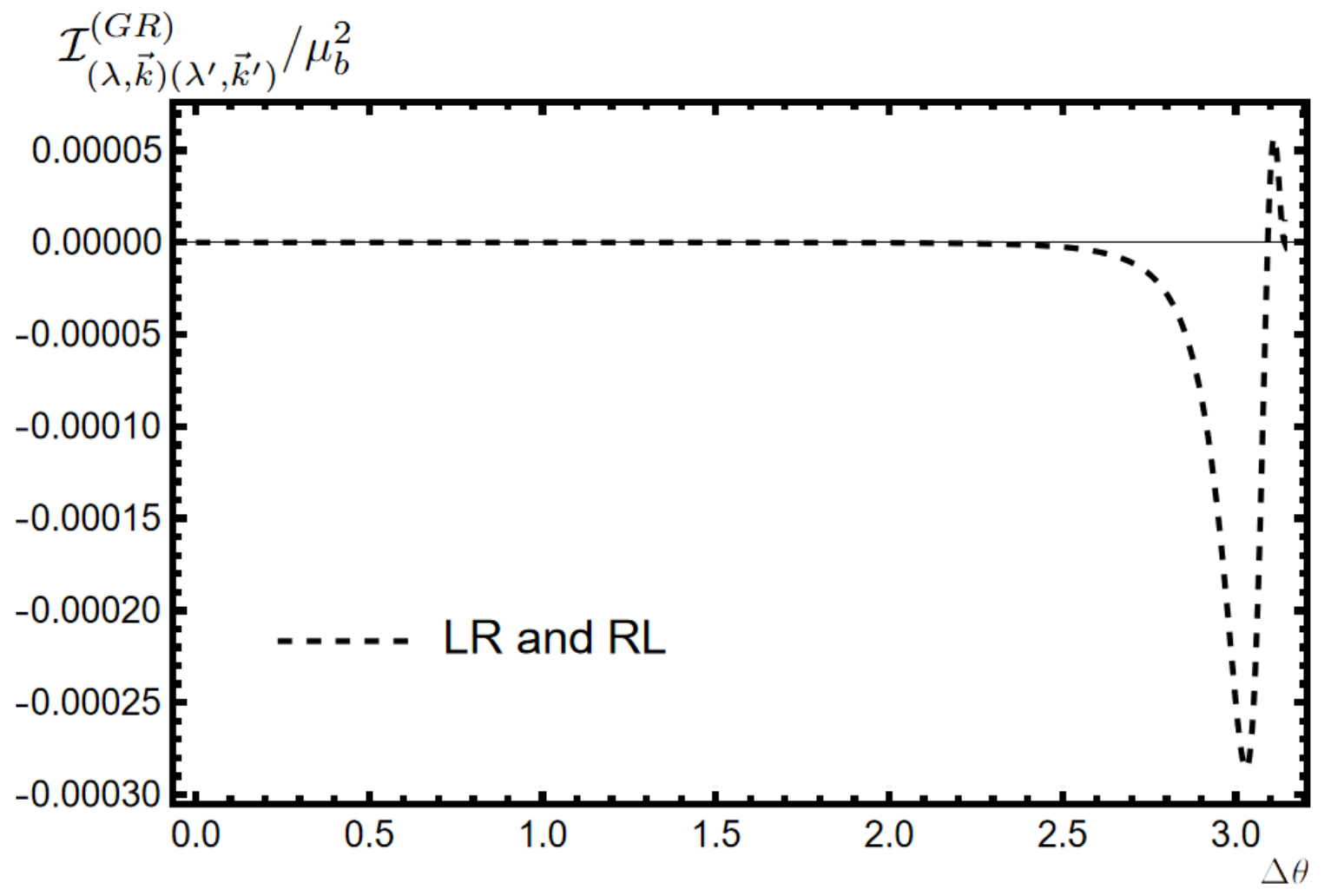}
    \caption{Polarization correlations from the GR induced interaction for LR and RL polarization pairs.  The plots have been made for $a_{\mu}=0.1$, with $\vec{k}+\vec{k}^\prime$ lying on the y-axis, \color{black} and $k=k^\prime=\mu_b$. \color{black} Figure taken from \cite{Dorlis:2025zzz}.\color{black} }
    \label{GR_Correlations_LR}
\end{figure}

\begin{figure}
    \centering
    \includegraphics[width=0.5\linewidth]{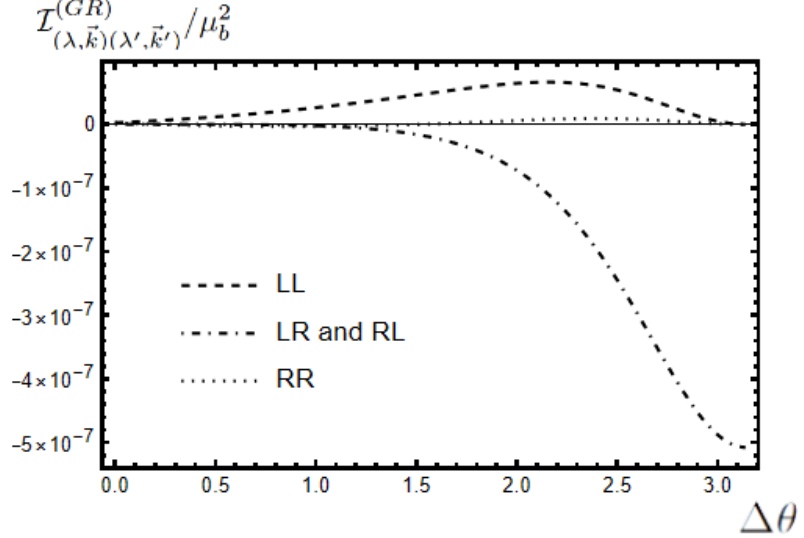}\hfill\includegraphics[width=0.5\linewidth]{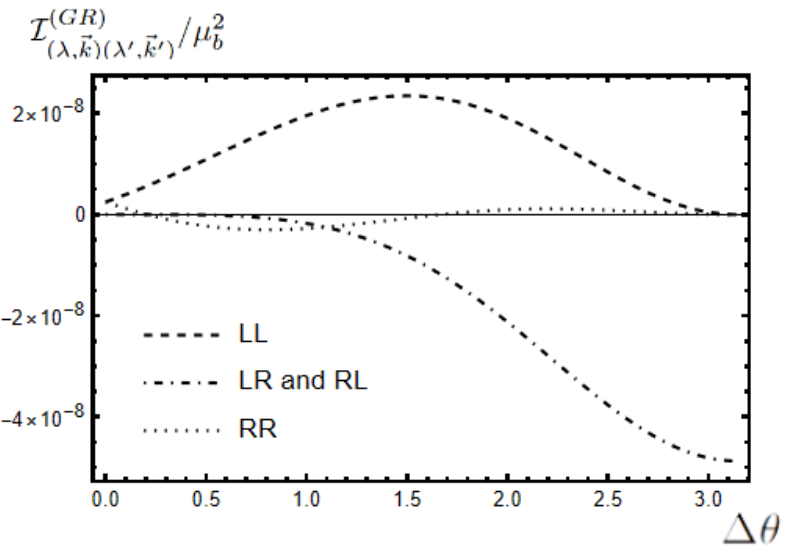}
    \caption{Polarization correlations from the GR induced interaction in the case that each graviton is emitted at different momenta ($k \neq k^\prime$) obeying though $k+k^\prime= 2\mu_b$.  The plots have been made for $a_{\mu}=0.1$, with $\vec{k}+\vec{k}^\prime$ lying on the y-axis. One can observe that the case of $k=k^\prime=\mu_b$ is the dominant one.  Left panel: $k=\frac{3}{2}\mu_b $ and $k^\prime = \frac{1}{2}\mu_b$. Right panel: $k=1.9\mu_b $ and $k^\prime = 0.1\mu_b$.  }
    \label{GR_Correlations_different_momenta}
\end{figure}

Hence, we can substitute all these results to the main equation \eqref{I_ij_tensorial_structure} for $I_{ij}$, and then estimate the maximum value of the following contracted quantity appearing in \eqref{G_GR_General}:
\begin{equation}
\label{I_GR_correlations}
 \mathcal{I}^{(GR)}_{(\lambda,\vec{k}),(\lambda^\prime,\vec{k}^\prime)}\equiv e^{(\lambda)}_{im}(\vec{k}) I_{ij}(\vec{k},\vec{k}^\prime \ )e^{(\lambda^\prime)}_{mj}(\vec{k}^\prime) \,.
\end{equation}
To this end, upon choosing the direction of the vectors $\vec{k},\vec{k}^\prime$ appropriately, so as to maximize the entanglement of the emitted gravitons, we can calculate explicitly the correlators \eqref{I_GR_correlations}. First, we can see that 
$Q$, defined in \eqref{dimensionless_Variable_Q}, 
is given by 
\begin{align}
  Q = r_0 \sqrt{k^2 + k^{\prime  2}+2kk^{\prime}cos(\Delta\theta)}\,,  
\end{align}
where $\Delta \theta$ denotes the relative angle between $\vec{k}$ and $\vec{k}^\prime$. Choosing the non - relativistic parameter to acquire the value:
\begin{equation}
    a_\mu\sim 0.1 \ ,
\end{equation}
to be consistent with the non-relativistic superradiance approximation \cite{BritoCardoso}, and also assuming that the order of magnitude of the vectors $\vec{k}$ and $\vec{k}^\prime$ to be $k\approx k^\prime \approx \mu_b$, since in this case the dominant contribution arises ({\it cf.} figure \ref{GR_Correlations_different_momenta}), we can plot the behavior of the contracted quantity \eqref{I_GR_correlations},which is shown in Figures \ref{GR_Correlations_LL_RR} and \ref{GR_Correlations_LR}.
The generated entangled graviton pairs acquire the following polarization states, 
\begin{equation}\label{EPRGR}
    \vert \Psi_{GR}\rangle=\frac{1}{2}\left(\mathcal{G}^{(GR)}_{(R,\vec{k})(L,\vec{k}^\prime)}\vert RL\rangle+\mathcal{G}^{(GR)}_{(L,\vec{k})(R,\vec{k}^\prime)}\vert LR\rangle+\mathcal{G}^{(GR)}_{(L,\vec{k})(L,\vec{k}^\prime)}\vert LL\rangle+\mathcal{G}^{(GR)}_{(R,\vec{k})(R,\vec{k}^\prime)}\vert RR\rangle        \right)\ ,
\end{equation}
with the opposite polarization correlations being enhanced, i.e.
\begin{equation}\label{EPRGR1}
    \mathcal{G}^{(GR)}_{(R,\vec{k})(L,\vec{k}^\prime)}, \,\, \mathcal{G}^{(GR)}_{(L,\vec{k})(R,\vec{k}^\prime)} \, \gg \, \mathcal{G}^{(GR)}_{(L,\vec{k})(L,\vec{k}^\prime)}, \,\, \mathcal{G}^{(GR)}_{(R,\vec{k})(R,\vec{k}^\prime)}\ .
\end{equation}

From figs.~\ref{GR_Correlations_LR} and \ref{GR_Correlations_different_momenta} we observe that 
\color{black} 
\begin{align}\label{GLeqGR}
\mathcal{G}^{(GR)}_{(R,\vec{k})(L,\vec{k}^\prime)} = \mathcal{G}^{(GR)}_{(L,\vec{k})(R,\vec{k}^\prime)} \,, 
\end{align}  which, in combination with 
\eqref{EPRGR1}, \color{black} implies the approximate form of the entangled graviton states in the framework of GR  ({\it i.e.} in the absence of gCS anomalies):
\begin{equation}\label{EPRGRsymm}
    \vert \Psi_{GR}\rangle \simeq \frac{1}{2}\, \mathcal{G}^{(GR)}_{(R,\vec{k})(L,\vec{k}^\prime)} \left(\vert RL\rangle+ \vert LR\rangle \right)\ ,
\end{equation}
that is, the cross-polarization correlations of the entangled graviton states (of Einstein-Podolsky-Rosen (EPR)~\cite{epr} type) are approximately symmetric under the interchange of $L \leftrightarrow R$.\footnote{This symmetry under the interchange 
$L \leftrightarrow R$ is only approximate, in view of fig.~\ref{GR_Correlations_LL_RR} and Eq.~\eqref{EPRGR1}.}


One can easily see that the vast suppression stemming from the gravitons strain, defined in \eqref{strain}, which comes in the form of $\sim \left(G \mu^{2}_b\right)$ is almost balanced from the number of axions occupying the $2p$-state of the axionic cloud. 
We can make use of \eqref{N_2p_Max}, and write it as follows (upon assuming that $\Delta \alpha_\star \sim \mathcal{O}(0.1)$ ):
\begin{equation}
\label{N_times_mu_over_planckmass}
     N_{2p}^{max}\left(\frac{ \mu_{b}}{M_{\rm Pl}}\right)^2\sim10^{-3}a^{2}_\mu=10^{-5} \ ,
\end{equation}
where we replaced in \eqref{N_2p_Max} the BH mass through $\mathcal{M}=a_\mu/(\mu_b G)$ and the sun mass $M_\odot \sim 10^{33}$gr. It is obvious from \eqref{N_times_mu_over_planckmass} that the vast suppression stemming from $ \big(\mu_b/M_{\rm Pl} \big)^2$ is almost balanced from the large number of axions occupying the most dominant superradiant mode.

\subsection{\textbf{Gravitational CS Anomaly Induced Squeezing Operator - Correlation Functions}}\label{sec:cs}
For the anomaly induced interaction of \eqref{Hamiltonian_CS_General}, we can follow the same steps as before in the context of the superradiance process. The Hamiltonian acquires the following form:
\begin{equation}
\hat{\mathcal{H}}_{I,CS}^{(2)}=\sum_{\vec{k},\vec{k}^\prime}\sum_{\lambda,\lambda^\prime}e^{i(\Omega_k + \Omega_{k^\prime}-\omega)\cdot t}\sqrt{N_{2p}}\Bigg(\int d^3\vec{x} \   \Psi_{2p}e^{-i(\vec{k}+\vec{k}^\prime)\cdot \vec{x}}\Bigg)\Bigg(\frac{f_{LR}(\vec{k},\vec{k}^\prime)}{\sqrt{2\mu_b}} \ \hat{\alpha}^\dagger_{\vec{k},L}\hat{\alpha}^\dagger_{\vec{k}^\prime,L}+\frac{f_{RL}(\vec{k},\vec{k}^\prime)}{\sqrt{2\mu_b}} \ \hat{\alpha}^\dagger_{\vec{k},R}\hat{\alpha}^\dagger_{\vec{k}^\prime,R}\Bigg) + \text{h.c.} \ ,
\end{equation}
where $f_{LR},f_{RL}$ are given by \eqref{F_joint_function}.
The structure of the evolution operator, using the same argumentation as in the case of GR, will take the following form:
\begin{equation}
\label{Scattering_Multimode_Squeeze_CS}
    \hat{S}^{(2)}_{CS}=\exp\left[ \frac{1}{2}\sum_{I,J}\mathcal{G}^{(CS)}_{IJ}\ \hat{\alpha}^\dagger_I\hat{\alpha}^\dagger_J -h.c. \right] \ ,
\end{equation}
where again 
\begin{equation}\label{CSkern}
    \mathcal{G}^{(CS)}_{IJ}\approx-2i  \ T \mathcal{F}^{(CS)}_{IJ} \ .
\end{equation} 
In this case, $\mathcal{F}^{(CS)}$ can be written as follows:
\begin{equation}\label{IijFour}
   \mathcal{F}^{(CS)}_{(\lambda,\vec{k})(\lambda^\prime,\vec{k}^\prime)}=   iA\sqrt{\frac{N_{2p}}{2\mu_b}}f_k f_{k^\prime} \ \Omega^{2}_k 
 \ \Omega^{2}_{k^\prime}\ \mathcal{I}^{(CS)}_{(\lambda,\vec{k})(\lambda^\prime,\vec{k}^\prime)}
\end{equation}
where  $\mathcal{I}^{(CS)}_{(\lambda,\vec{k})(\lambda^\prime,\vec{k}^\prime)}$ is the angular polarization correlation function of the anomaly interaction, 
\begin{align}
    \mathcal{I}^{(CS)}_{(\lambda,\vec{k})(\lambda^\prime,\vec{k}^\prime)}=&l_{\vec{k}^\prime}l_{\lambda^\prime}\widetilde{\Psi}_{2p}(\vec{k}+\vec{k}^\prime)\left(   [e^{(3)}(\vec{k}^\prime)] _m  \ e^{(\lambda)}_{mj}(\vec{k})\  e^{(\lambda^\prime)}_{jl}(\vec{k}^\prime) [e^{(3)}(\vec{k})] _l 
   \  +\right.\left. \left(1-   \cos\Delta\theta \right)  \ e^{(\lambda)}_{mj}(\vec{k}) \   e^{(\lambda^\prime)}_{mj}(\vec{k}^\prime)      \     \right) \ ,
\end{align}
with $\widetilde{\Psi}_{2p}(\vec{k})$ denoting the Fourier transform of $\Psi_{2p}(\vec{x})$,
\begin{equation}
     \widetilde{\Psi}_{2p}(\vec{q})=\frac{128 i  \sqrt{\pi }   \ q\ r_{0}^{5/2}}{\left(4 q^2   r_{0}^2+1\right)^3}\sin \theta_q e^{i \varphi_q} \ , 
     \label{2p_fourrier}
\end{equation}
where $\vec{q}=-(\vec{k}+\vec{k}^\prime)$. In this process,  $E^{CS}=\mu_b$, signifying that only one axion is involved to the microscopic process, which also appears through the proportionality to the square root of the axion occupation number  $\sim\sqrt{N_{2p}}$. 

\begin{figure}[ht]
    \centering
    \includegraphics[width=0.5\linewidth]{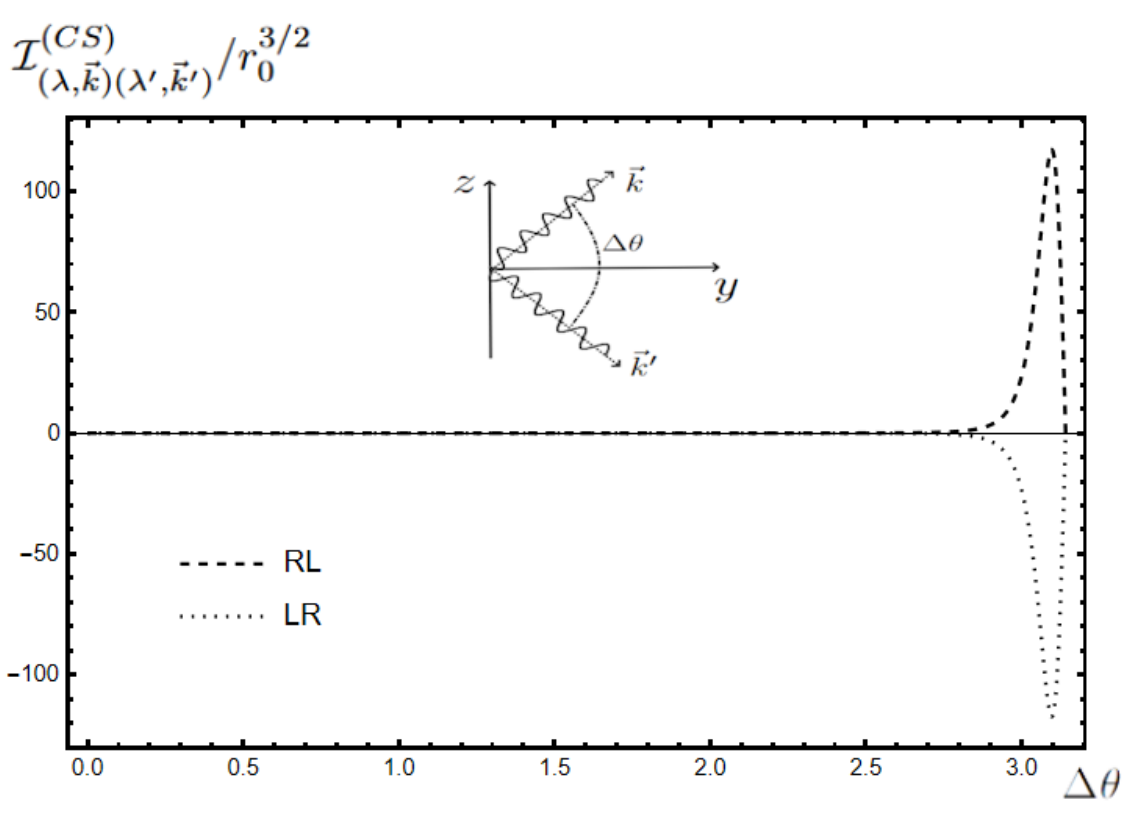}
    \caption{Polarization correlations from the gravitational CS anomaly induced interaction. Only pairs of opposite polarizations are produced; Maximal entanglement between the Left and Right polarizations. The plots have been made for $a_{\mu}=0.1$, with $\vec{k}+\vec{k}^\prime$ lying on the y-axis, \color{black}  and $k=k^\prime=\mu_b/2$. Figure taken from \cite{Dorlis:2025zzz}.\color{black}}  
    \label{CS_correlations}
\end{figure}

Using the contractions \eqref{first_terms_of_f_lambda}, \eqref{second_terms_of_f_lambda} and \eqref{formulas_polarizations_contractions}, we find, 
\begin{align}
\mathcal{I}^{(CS)}_{(L,\vec{k})(L,\vec{k}^\prime)}= \ &\mathcal{I}^{(CS)}_{(R,\vec{k})(R,\vec{k}^\prime)}=0\\
\mathcal{I}^{(CS)}_{(R,\vec{k})(L,\vec{k}^\prime)}=&-\mathcal{I}^{(CS)}_{(L,\vec{k})(R,\vec{k}^\prime)}=\frac{256\sqrt{2\pi}\ a_{\mu}^5\sqrt{1+\cos\Delta\theta}}{\left(2+a_{\mu}^2+2\cos\Delta\theta\right)^3}
    \sin^4\left(\frac{\Delta\theta}{2}\right)\ r_0^{3/2}\ ,
\end{align}
which are plotted in figure \ref{CS_correlations}. We assumed again that $k\approx k^\prime \approx\mu_b/2$, since this is when the main contribution arises as in the case of GR. Then, we observe a maximal entanglement state according the polarizations of the gravitons, in which only opposite helicity states are allowed to be produced, i.e. maximally entangled (Bell) states of the following form, 
\begin{equation}\label{EPRCS}
    \vert \Psi_{CS}\rangle = \frac{1}{2} \, \mathcal{G}^{(CS)}_{(R,\vec{k})(L,\vec{k}^\prime)}\Big( \vert LR\rangle-\vert RL\rangle     \Big)\, ,
\end{equation}
since $\mathcal{G}^{(CS)}_{(R,\vec{k})(L,\vec{k}^\prime)} = - \mathcal{G}^{(CS)}_{(L,\vec{k})(R,\vec{k^\prime})}$ ({\it cf.} fig.~\ref{CS_correlations}).
\color{black} We note that the antisymmetric nature of the entangled state \eqref{EPRCS} is due to the fact that the gCS anomaly terms \eqref{RCS} vanish identically for non-chiral GWs. 

\color{black}
If both gravitons have the same projection onto the z-axis (which is the axis of BH and cloud rotation in our setup), then only pairs of the same helicity are allowed. \color{black} When their sum is directed to the z-axis, indeed this is true, but it is limited by the distribution ({\it cf.} eq.\eqref{2p_fourrier}) and trivially vanishes, since $\theta_q=\pi$. When assuming $0<\theta_{\vec{k}+\vec{k}^\prime}<\pi/2$ ({\it cf.} figure \ref{CS_correlations}), where the distribution \eqref{2p_fourrier} is not trivial, we see exactly that only $LL$ and $RR$ pairs are allowed. As an illustrative example, consider the vector lying on the z-y plane with $\theta_{\vec{k}+\vec{k}^\prime}=\pi/4$. Then, the correlation functions read, 
\begin{align}
\mathcal{I}^{(CS)}_{(L,\vec{k})(L,\vec{k}^\prime)}= \ &-\mathcal{I}^{(CS)}_{(R,\vec{k})(R,\vec{k}^\prime)}=\frac{256\sqrt{\pi}\ a_{\mu}^5\sqrt{1+\cos\Delta\theta}}{\left(2+a_{\mu}^2+2\cos\Delta\theta\right)^3}
    \sin^4\left(\frac{\Delta\theta}{2}\right)\ r_0^{3/2}\\
\mathcal{I}^{(CS)}_{(R,\vec{k})(L,\vec{k}^\prime)}=&\mathcal{I}^{(CS)}_{(L,\vec{k})(R,\vec{k}^\prime)}=0\ ,
\end{align}
which is valid as long as $\Delta\theta<\pi/2$. In such a case the entangled single-pair state, takes the following form, 
\begin{equation}
    \vert\Psi_{CS}\rangle=\frac{1}{2}\mathcal{G}^{(CS)}_{(L,\vec{k})(L,\vec{k}^\prime)}(\vert LL\rangle-\vert RR\rangle)
\end{equation}
which is again an antisymmetric Bell state (as it is induced by the CS term). Furthermore, by opening up the relative angle $\Delta\theta>\pi/2$, the correlations are flipped, giving, 
\begin{align}
\mathcal{I}^{(CS)}_{(L,\vec{k})(L,\vec{k}^\prime)}= \ &\mathcal{I}^{(CS)}_{(R,\vec{k})(R,\vec{k}^\prime)}=0\\
\mathcal{I}^{(CS)}_{(L,\vec{k})(R,\vec{k}^\prime)}=&-\mathcal{I}^{(CS)}_{(L,\vec{k})(R,\vec{k}^\prime)}=\frac{256\sqrt{\pi}\ a_{\mu}^5\sqrt{1+\cos\Delta\theta}}{\left(2+a_{\mu}^2+2\cos\Delta\theta\right)^3}
    \sin^4\left(\frac{\Delta\theta}{2}\right)\ r_0^{3/2}\ .
\end{align}

\color{black}

One notices that both correlators \eqref{F_GR_correlations} and \eqref{IijFour} cannot be factorized into separable states of two gravitons. This demonstrates that the graviton pairs are indeed {\it entangled}~\cite{Law:2000hyw}. \color{black} This property also follows by computing the von Neumann entropy of these states, following methods in quantum optics (Takagi decomposition~\cite{Takagi1933,Houde:2024mkj}), as outlined in Appendix \ref{appC:takagi} for the case of relevant, albeit simplified, examples, {\it cf.} \eqref{entrexample2}, upon making the correspondence 
of $\vert \mathcal G_{IJ}\vert^2$ with the squeezing parameter $r^2$ of the pertinent squeezing operator \eqref{examplesq2}\color{black}.   Because of the {\it non-relativistic regime} of the axionic cloud in our analysis, the graviton pairs are emitted (approximately) in opposite directions (i.e. {\it almost anti-collinear})  as shown in figures \ref{GR_Correlations_LL_RR}, \ref{GR_Correlations_LR} and \ref{CS_correlations}. However, going beyond the non-relativistic regime, a larger volume of
the graviton phase space is occupied and the nearly anti-collinear
emission breaks down. This would naturally spread out the narrow peaks of figures \ref{GR_Correlations_LL_RR},  \ref{GR_Correlations_LR} and \ref{CS_correlations}, with contributions from different emission angles, as well.

The CS interaction differs from the GR case in two basic features. Firstly, the dependence on the axion field is linear, and hence the enhancement will be proportional to $\sim\sqrt{N_{2p}}$ instead of $\sim N_{2p}$, in relation to GR. Furthermore, the CS-part of the theory describes a higher curvature interaction of the (pseudo-)scalar field with the gravitational anomaly of \eqref{RCS}. Hence, the interaction will be highly suppressed through the coupling constant of the theory, as well as due to the higher derivative terms yielding extra suppression to the strength of the interaction, of the form $\mu_b/M_{\rm Pl}$. This shall become clear when we will calculate the average number of gravitons in the squeezed vacuum state in the following chapter. 

\subsection{\textbf{${\rm g}$CS anomaly and modification of the entangled-graviton-correlation symmetries: comments}}\label{sec:VIC}

Finally, before closing this section, it worths noticing an analogy of the r\^ole of the gCS anomalous term in 
``contaminating'' ({\it cf.} \eqref{EPRCS}) with a ``wrong symmetry'' (antisymmetry under the interchange $L \leftrightarrow R$) the Einstein-Podolsky-Rosen (EPR)-type~\cite{epr} correlators of the entangled graviton states of GR, \eqref{EPRGRsymm}, which are approximately symmetric under $L \leftrightarrow R$. This situation is somewhat reminiscent of the so-called $\omega$-effect~\cite{Bernabeu:2003ym,Bernabeu:2006av}, associated with the modification of the EPR correlators of entangled neutral meson states in neutral-meson factories, as a result of the ill-defined nature of the 
CPT operator in cases of QG-induced decoherence of the quantum particles due to a spacetime foam ``environment''.

Indeed, in such cases, if CPT is a well-defined operator, even if it does not commute with the Hamiltonian of the system (due, {\it e.g.} to Lorentz-symmetry breaking), the initial state $\vert i 
\rangle$ of the entangled neutral mesons, produced by an initial decay of another neutral meson ({\it e.g.} $\Phi$-meson in the case of neutral Kaons in $\Phi$-factories), is antisymmetric due to the imposed Bose statistics (implying the standard result that a state of two identical bosons vanishes in quantum theory).  In that case,  the (normalised) entangled state $\vert i\rangle$ of neutral meson-antimeson pairs, $\vert M^0\rangle, \vert \overline M^0 \rangle$ reads:
\begin{align}\label{noomega}
\vert i \rangle = \frac{1}{\sqrt{2}}\Big(\vert M^0(\vec k)\,,\, \overline{M}^0(-\vec k) \rangle  -  \overline{\vert M}^0(\vec k)\,,\, M^0(-\vec k)\rangle \Big)\,,
\end{align}
where $\vec k$ is the spatial momentum in the Laboratory frame of the decay of the initial neutral meson into the (entangled) pairs of the $M^0,\overline M^0$-mesons. 

However, if the system is in a QG-spacetime ``environment'', which involves microscopic, potentially singular, spacetime fluctuations (such as microscopic (Planck-size) quantum-fluctuating BHs), then the CPT operator is not a well-defined quantum mechanical generator of the CPT combination of symmetries~\cite{Wald:1980nm}, due to the existence of decoherence in the effective low-energy theory, which a local observer, who is unable to detect the aforementioned QG-BH fluctuations, lives on. In such a case, it was conjectured in \cite{Bernabeu:2003ym}, and demonstrated in 
some stochastic toy-models of spacetime foam in \cite{Bernabeu:2006av}, that there is a contamination of $\vert i \rangle$  with the ``wrong-symmetry'' state:
\begin{align}\label{omega}
\vert i \rangle = \frac{1}{\sqrt{2}}\Big(\vert M^0(\vec k)\,,\, \overline{M}^0(-\vec k) \rangle -  \overline{\vert M}^0(\vec k)\,,\, M^0(-\vec k)\rangle \Big) + \frac{\omega}{\sqrt{2}}\, \Big(\vert M^0(\vec k)\,,\, \overline{M}^0(-\vec k) \rangle +  \overline{\vert M}^0(\vec k)\,,\, M^0(-\vec k)\rangle \Big)\,, \,  \omega = \vert \omega \vert \, \exp(\varphi_\omega)\,,
\end{align}
where the complex parameter $\omega$ (not to be confused with the real part of the axion oscillation frequency) parametrizes the observer's ignorance on the details of the spacetime foam. 
In concrete examples this parameter can be calculated within weak quantum-gravity approaches~\cite{Bernabeu:2006av}.
This is the $\omega$-effect. 

In our case, from \eqref{EPRGR1} and \eqref{EPRCS} we observe that a somewhat similar, but qualitatively and quantitatively different, in both form, origin and order-of-magnitude phenomenon occurs, in which the r\^ole of the ill-defined CPT operator is played by the gCS anomaly. The latter contaminates the GR entangled-state of two gravitons \eqref{EPRGR1}, which is approximately symmetric under the interchange of $L \leftrightarrow R$ polarizations, by the antisymmetric maximally-entangled Bell-state \eqref{EPRCS}. It will be interesting to see whether such a situation in graviton entangled states plays a r\^ole in determining unique (``smoking-gun'') effects of QG, similar to what the $\omega$-effect does for EPR particle states.

\section{Number of Gravitons in the Squeezed Vacuum State}\label{sec:number_of_gravitons}

We may now proceed to the calculation of the number of gravitons in the squeezed vacuum state 
\begin{equation}
    \label{Squeezed_vacuum_state}
    \vert \psi \rangle = \hat{S}\vert0\rangle \ ,
\end{equation}
for both the GR and gCS induced interactions, where the index $I=(\lambda,\vec{k})$ describes the graviton states. Following the analysis of \cite{multimode}, and noting that the evolution operator \eqref{Scattering_Multimode_Squeeze_GR} has the form of a multimode squeezing operator, one can see that:
\begin{align}
\label{SaS_multimode}
    \hat{S^\dagger} \hat{\alpha}_{I} \hat{S} &=\sum_J \left( \mu_{IJ}\hat{\alpha}_J \ + \ \nu_{IJ}\hat{\alpha}^\dagger_{J} \right) \,,\\ 
\label{SadaggerS_multimode}    
    \hat{S^\dagger} \hat{\alpha}^\dagger_{I} \hat{S} &=\sum_J \left( \mu^\star_{IJ}\hat{\alpha}^\dagger_J \ + \ \nu^\star_{IJ}\hat{\alpha}_{J} \right) \ ,
\end{align}
with the transformation coefficients given by:
\begin{align}
\label{mu_IJ_expansion} 
\mu_{IJ}=\delta_{IJ} + \frac{1}{2!}\sum_M \mathcal{G}_{IM}\mathcal{G}^\star_{MJ} \ +\frac{1}{4!}\sum_{M,L,N}\mathcal{G}_{IM}\mathcal{G}^\star_{ML}\mathcal{G}_{LN}\mathcal{G}^\star_{NJ} \ + \ \dots \,,
\end{align}
\begin{align} 
\label{nu_IJ_expansion}
\nu_{IJ} = \mathcal{G}_{IJ} + \frac{1}{3!}\sum_{M,L} \mathcal{G}_{IM}\mathcal{G}^\star_{ML} \mathcal{G}_{LJ} \ + \ \frac{1}{5!}\sum_{M,L,N,P} \mathcal{G}_{IM}\mathcal{G}^\star_{ML} \mathcal{G}_{LN}\mathcal{G}^\star_{NP} \mathcal{G}_{PJ} \ + \ \dots \, 
\end{align}
Using these transformations, one can estimate the average number of gravitons $N_{gr}$ in the squeezed vacuum state $|\psi\rangle=\hat{S}\vert 0 \rangle $. Firstly, the number of gravitons at the state denoted by $I$ is given by:
\begin{align}
\label{number_of_gavitons_I}
   \nonumber
   \langle N^{gr}_{I} \rangle =& \ \langle \psi \vert \hat{\alpha}^\dagger_{I}\hat{\alpha}_I \vert \psi \rangle = \langle 0 \vert \hat{S}^\dagger \hat{\alpha}^\dagger_{I}\hat{\alpha}_I \hat{S}\vert 0 \rangle=\langle 0 \vert \left(\hat{S}^\dagger \hat{\alpha}^\dagger_{I}\hat{S}\right)\left(\hat{S}^\dagger\hat{\alpha}_I \hat{S}\right)\vert 0 \rangle \\
   \nonumber
   =& \ \langle 0 \vert \sum_J \left( \mu^\star_{IJ}\hat{\alpha}^\dagger_J \ + \ \nu^\star_{IJ}\hat{\alpha}_{J} \right)\sum_L \left( \mu_{IL}\hat{\alpha}_L \ + \ \nu_{IL}\hat{\alpha}^\dagger_{L} \right)\vert 0 \rangle \\
   =& \ \sum_{J,L} \nu^\star_{IJ}\nu_{IL}\langle 0 \vert \hat{\alpha}_{J} \hat{\alpha}^\dagger_{L}\vert 0 \rangle = \sum_{J}\vert \nu_{IJ} \vert ^{2} \ ,
\end{align}
where we substituted eqs. \eqref{SaS_multimode},\eqref{SadaggerS_multimode} and we used the commutation relation of eq.\eqref{commutaiton_alpha_lambda}.
Hence, the total number of gravitons in the squeezed vacuum state $\vert \psi \rangle = \hat{S}\vert 0 \rangle$ is given by:
\begin{align}\label{Number_total_gravitons}
\langle N_{gr} \rangle  = \sum_{I}\langle N^{gr}_{I}\rangle = \sum_{I, J} 
\Big|\nu_{IJ}\Big|^2 \ ,
\end{align}
Substituting \eqref{nu_IJ_expansion}, one can obtain the following bound:
\begin{align}\label{Nmax}
\langle N_{gr} \rangle  = \sum_{I, J} 
\Big|\nu_{IJ}\Big|^2\, \lesssim \sum_{I,J} \Big( \Big|{\mathcal G}_{IJ}\Big|^2 + \Big|
\frac{1}{3!}\sum_{M,L} \mathcal{G}_{IM}\mathcal{G}^\star_{ML} \mathcal{G}_{LJ} \Big|^2 \ + \dots \Big)\, \ .
\end{align}
We remind the reader that in the case of a single-mode squeezed vacuum, i.e $\mathcal{G}_{IJ}=r\ e^{i\varphi}\ \delta_{IJ}$ the expression reduces to the well known relation for one - mode squeezing, i.e. $\langle N_{gr}\rangle=\sinh^2 r$. From 
\eqref{Nmax}, one observes that in the case $\sum_{I,J} \Big|(\mathcal{G}_{IJ})\Big|\ll1$,
the upper bound in \eqref{Nmax} is saturated, {\it i.e.}
\begin{align}\label{Nmaxsat}
\langle N_{gr}\rangle \simeq\sum_{I,J} \Big|(\mathcal{G}_{IJ})\Big|^2\, ,
\end{align}
implying that in such a case the average number of gravitons in the squeezed-vacuum state is highly suppressed. On the other hand, if the first term of the above infinite series \eqref{Nmax} is of order one, $\sum_{I,J} \Big|(\mathcal{G}_{IJ})\Big|^2\sim\mathcal{O}(1)$, the number of squeezed gravitons is on the verge of a highly non-linear enhancement similar to the exponential enhancement of the single mode squeezed state.    \par 

\subsection{GR - induced interaction}

For the GR interaction, we substitute \eqref{G_max_GR} into the sum, arriving at the following expression for the number of squeezed gravitons per solid angles, 
\begin{align} 
    \frac{d^2}{d\Omega d\Omega^\prime}\sum_{I,J} \Big|(\mathcal{G}_{IJ})\Big|^2=\left(\frac{\mu_b}{M_{\rm Pl}}\right)^{4}\left(\frac{N^{2}_{2p}}{128 \pi^3}\right)\left(T \mu_b\right) \sum_{\lambda,\lambda^\prime}\int d\tilde{k}d\tilde{k}^\prime \ \tilde{k}\tilde{k}^\prime\delta\left(
    \tilde{k}+ \tilde{k}^\prime -2\right)\Big | \widetilde{\mathcal{I}}^{(GR)}_{(\lambda,\vec{k})(\lambda^\prime,\vec{k}^\prime)}\Big|^{2} \ ,
 \label{N_gr_per_Solid_angles}
\end{align}
where $\tilde{k}=k/\mu_b$ is the dimensionless re-defined integration variable and $\widetilde{\mathcal{I}}=\mathcal{I}/\mu^{2}_{b}$ is the dimensionless correlation function defined in \eqref{I_GR_correlations}. 

In order to estimate the number of gravitons in the squeezed vacuum state, we shall consider the vector $\vec{q}=-\left(\vec{k}+\vec{k}^\prime\right)$ lying on $\theta_{k+k^\prime}=\varphi_{k+k^\prime}=\pi/2$ ({\it cf.} figure \ref{GR_Correlations_LL_RR}, \ref{GR_Correlations_LR} and \ref{GR_Correlations_different_momenta}). Hence, we can perform the integrals of  \eqref{N_gr_per_Solid_angles}, where the main contribution comes when $k\approx k^\prime\approx\mu_b$ (or $\tilde{k}\approx \tilde{k}\approx 1$), and when the angle between $\vec{k}$ and $\vec{k}^\prime$ is $\Delta \theta=0.964\pi$ ({\it cf.} figure \ref{GR_Correlations_LL_RR} and \ref{GR_Correlations_LR} for the maximum contribution). The final result, should be multiplied by a factor stemming from the integrations over the solid angles, which will yield a maximum contribution of $\sim \left(4 \pi \right)^2$. Hence, we obtain the following upper limit,
\begin{equation}
    \label{N_gr_RESULT}
  \sum_{I,J}\vert\mathcal{G}^{(GR)}_{IJ}\vert^2 \lesssim 2.5\times10^{-15} \left(T \mu_b\right)  .
\end{equation}
Note here that the large suppression induced by the ratio $\big(\mu_b/M_{\rm Pl}\big)^2$ is compensated by the number of axions in the cloud, $N_{2p}$, 
\begin{equation}
    N_{2p}\left(\frac{ \ \mu_b}{M_{\rm Pl}}\right)^2\approx 10^{-3}a_\mu^2 =10^{-5}\ ,
    \label{Axion_Number_Compensation}
\end{equation}
which can be easily verified in view of \eqref{N_2p_Max}.

\subsection{${\rm g}$CS - induced interaction}
On the other hand this is not the case for the anomaly induced interaction. Following the same analysis, substitution of the correlation function \eqref{IijFour} into \eqref{Nmaxsat}, we get:
\begin{align} 
    \frac{d^2}{d\Omega d\Omega^\prime}\sum_{I,J} \Big|(\mathcal{G}^{(CS)}_{IJ})\Big|^2=\frac{N_{2p}}{16\pi^3}\left(\frac{\mu_b}{M_{\rm Pl}}\right)^{4}\left(A \mu_b \right)^2\left(T \mu_b\right)\sum_{\lambda,\lambda^\prime}\int d\tilde{k}d\tilde{k}^\prime \ \tilde{k}^{5}\tilde{k}^{\prime  5}\delta\left(
    \tilde{k}+ \tilde{k}^\prime -1\right)\Big | \widetilde{\mathcal{I}}^{(CS)}_{(\lambda,\vec{k})(\lambda^\prime,\vec{k}^\prime)}\Big|^{2} \ .
    \label{N_CS_per_Solid_angles_CS}
\end{align}
As such, in view of \eqref{Axion_Number_Compensation}, the large occupation number of axions in the condensate cannot compensate the suppression from axion mass over the Planck mass, since: 
\begin{equation}
    N_{2p}\left(\frac{ \ \mu_b}{M_{\rm Pl}}\right)^4\approx 10^{-3}a_\mu^2\left(\frac{ \ \mu_b}{M_{\rm Pl}}\right)^2=10^{-5}\left(\frac{ \ \mu_b}{M_{\rm Pl}}\right)^2\ .
\end{equation}
Moreover, seeing the CS-coupling to the axion in terms of an effective field theory, arising for example from string theory \cite{Duncan:1992vz},  the coupling constant has to be suppressed from the cut-off scale of the theory. In terms of this, the term $(A\mu_b)^2$ introduces further suppression to the  order of $(\mu_b/M_{s})^2$. If one considers the string inspired coupling \cite{Duncan:1992vz} $A\sim 10^{-2}M_{\rm Pl}/M_s$, with $M_s$ the string scale, the final result acquires the following upper limit,
\begin{equation}
\label{N_cs_RESULT}
    \sum_{I,J} \Big|(\mathcal{G}^{(CS)}_{IJ})\Big|^2<10^{-10}\left(\frac{\mu_b}{M_s}\right)^4(\mu_b T)\ ,
\end{equation}
where again, we used the fact that the dominant contribution comes when $k\approx k^\prime \approx
\mu_b/2$, with the relative $\Delta \theta = 0.986 \pi$ acquired from figure \ref{CS_correlations}.

\subsection{Long Lived Axionic Clouds Enhancing Squeezing Effects and Observational Bounds}

What becomes clear from \eqref{N_gr_RESULT} and \eqref{N_cs_RESULT}, is the fact that the lifetime of the axionic cloud constitutes the final enhancement parameter capable of producing appreciable squeezing effects. In contrast to the short lived Quasi-normal Modes (QNMs), which, as shown in \cite{BH_Squeezer}, are not capable of producing significant amount of squeezed scalar particles around a BH, the longevity of axionic clouds seems to overcome this limitation. Identifying $T$ with the lifetime of the cloud,
\begin{equation}\label{tcloud}
    T=\tau_{\text{cloud}} \ ,
\end{equation}
measured in terms of the superradiance instability timescale \eqref{timescale_superradiance}, i.e. 
\begin{equation}\label{factor}
    \tau_{\text{cloud}}=\gamma \tau_s \ ,
\end{equation}
one can easily obtain that $\mu_b  T \approx 10^9  \gamma $. As have been already discussed, the necessary condition for the formation of the axionic cloud is the separation of timescales, i.e. $\tau_{\text{cloud}}\gg \tau_s$,  implying that $\gamma\gg1$. Although the lifetime of the axionic condensate is not explicitly defined in the literature,  numerical simulations have been carried out in the literature \cite{BritoScales1,BritoScales2,Yoshino_2014_scales,porto_scales}, and all seem to agree that the lifetime of the axionic cloud is many orders of magnitude larger than the instability timescale \eqref{timescale_superradiance}. For example, in \cite{porto_scales}, the lifetime of the cloud is given by:
\begin{equation}
\label{tau_cloud}
    \tau_{\text{cloud}}\approx10^9 \ \text{years}\left(\frac{\mathcal{M}}{10^5 M_{\odot}}\right)\left(\frac{0.1}{a_\mu}\right)^{15} \ ,
\end{equation}
which is valid for $a_\mu \leq 0.1$, as stated in \cite{porto_scales}. \par
\color{black}Multiplying equation \eqref{tau_cloud} with $\mu_b$, \color{black} and making use of the following units-relations:
\begin{align*}
    1~\text{GeV}^{-1} \approx6.582 \times 10^{-25} \ \text{s} \ ,\quad 
    M_{\rm Pl} \approx 10^{-39} M_{\odot}  \ , \quad   \,
    1 \ \text{year} \approx 3.1  \times 10^7 \ \text{s} \ ,
\end{align*}
\color{black}one obtains the following order of magnitude estimate, \color{black}
\begin{equation}
    \mu_b \, \tau_{\text{cloud}}\sim 10^{16} \ .
\end{equation}
This results in a factor in \eqref{factor}
\begin{equation}\label{factor2}
    \gamma\sim10^7 \ ,
\end{equation}
since $\mu_b \tau_s\sim10^{9}$ ({\it cf.} eq.\eqref{timescale_superradiance}). Hence, we are led to the  
following estimate for the upper limit of the inequality
\eqref{N_gr_RESULT}: 
\begin{equation}
    \label{N_gr_RESULT_final}
  \left(\sum_{I,J}\vert\mathcal{G}^{(GR)}_{IJ}\vert^2\right)_{max} \sim \mathcal{O}(10-60) \ . 
\end{equation}
\color{black}
Note that, the result regarding the upper bound of this quantity is generic, since it does not change  for different values of \color{purple}$\mu_b$ and $\mathcal{M}$, \color{black} as long as they satisfy the non - relativistic superradiance condition $ \mu_b G\mathcal{M}=a_\mu=0.1$.\footnote{\color{black} Nonetheless, we note that the numerical coefficient, denoted by $\sim$, in \eqref{factor2}, is phenomenological. Even a value of order $1.25$ suffices to push the limit in \eqref{N_gr_RESULT_final} to 75, which leads to uncertainties in the order of the produced squeezed gravitons, {\it cf.} \eqref{Ngrfinal}.\color{black}} \par 

In such a case, the total number of gravitons in the squeezed vacuum state defined in \eqref{Nmax} will be exponentially enhanced, resembling the $\sim \sinh^{2}r$ behavior of the one - mode case \cite{Scully_Zubairy_1997} (see also discussion in section \ref{appC:takagi}). \color{black} We thus have 
\begin{align}\label{Ngrfinal}
\langle N_{gr} \rangle \lesssim \mathcal O(10^6-10^7)\,.
\end{align}
\color{black} In such cases, the expectations of~\cite{Guerreiro:2019vbq,Coradeschi:2021szx,Guerreiro:2021qgk} for large values of the squeezing parameter for gravitons in GW, which would allow for the possibility of observational signatures of squeezed graviton entangled states, might be realized. 

Our estimates \eqref{N_gr_RESULT_final} for the GW squeezing parameter from Kerr BHs are compatible with the observational bound on the single-mode squeezing parameter $r < 41$ of \cite{Hertzberg:2021rbl}, inferred by the non-observation of squeezing effects in the GW signals of the current LIGO-Virgo interferometric data~\cite{LIGOScientific:2016aoc,McCuller:2021mbn}, \color{black} provided, of course, that the methods in \cite{Scully_Zubairy_1997,Hertzberg:2021rbl} for single-mode squeezing are valid in our multi-mode case. The latter aspect  needs further investigation, which we postpone to a forthcoming publication. 
Nonetheless, on accepting this assumption, and saturating the upper bound of \eqref{N_gr_RESULT}, corresponding to GR-type squeezing effects, and 
representing $\vert \mathcal G_{IJ}\vert$ as the squeezing parameter $r_{\rm multimode}$, 
we can place an upper bound on the 
allowed axion-cloud life times. Indeed, for superradiance time scales of order \eqref{timescale_superradiance}, i.e. $\mu_b \, \tau_s \sim 2.4 \times 10^9$, and setting $r^{2}_{\rm multimode} < 41$ and $T = \gamma \, t_s$, 
we do obtain, the bound for the allowed regions of the axionic-clouds lifetimes:
\begin{align}\label{allowedclouds}
(41)^2 > r_{\rm multimode}^2 \simeq 6 \times 10^{-6} \,  \gamma \, \Rightarrow \, 
\gamma < 2.8 \times 10^{8} \,,
\end{align}
that is, models predicting 
\color{black}
\be\label{exclT}
T < 2.8 \times 10^{8} \, \tau_s \ee\color{black}
are excluded already, based on the non-observation of squeezing effects in current LIGO-Virgo data.\footnote{\color{black}The reader is reminded at this stage that such lifetimes are only an order of magnitude longer than the lifetime \eqref{factor2}, 
characterizing models discussed in \cite{porto_scales},
which corresponds to 
the upper bound of the range exhibited on the right-hand side of \eqref{N_gr_RESULT_final}\color{black}}
 \color{black} By contrast, even with such long lifetimes for the axionic cloud, the axion decay process induced by the anomaly \eqref{Chern-Simons_Hamiltonian_bR*R} is still highly suppressed. At these low energy scales, characterized by the axion mass, $\mu_b$, the above results are consistent with the EFT approach of the graviton, in which the higher curvature interactions have to be subdominant with respect to the GR induced interactions \cite{Donoghue:1994dn}.

The analysis presented here is constrained to the flat-background approximation and is highly reliant on the longevity of the axionic clouds under consideration. Nevertheless, such an approach can be extended in order to consider curvature effects around the rotating BH, which are in place to provide an additional enhancement to the squeezing process. Further numerical simulations are essential with the intention of acquiring more accurate results. 

\color{black}

\section{Multimode squeezing and Takagi-Autonne-Schmidt modes}\label{sec:Takagi}

\color{black} In this section \color{black}
\cmaa
we elaborate on the nature of  squeezed  GW states found in \eqref{Scattering_Multimode_Squeeze_GR}, \color{black} and comment on  their observational prospects\color{black}\cbl \cmaa. This state will be characterised rigorously in terms of independent squeezed modes using a reformulation based on the methods of Takagi \cite{Houde:2024mkj}. This is advantageous conceptually in analysis  of both the squeezing and its detection in an idealised GW detector, such as a Fabry-Perot interferometer. The treatment of the Fabry-Perot interferometer as a device for investigating the quantum nature of GWs is fairly new \cite{Pang:2018eec}. It relies on earlier works on system-bath interactions \cite{GardinerCollett1985,CaldeiraLeggett1983}. Realistic detectors will require a great deal of numerical modeling based on the same principles that we use. 

The state in \eqref{Scattering_Multimode_Squeeze_GR} has simple properties which makes it a member  of a class of states known as Gaussian states widely studied in quantum information and quantum optics \cite{RevModPhys.84.621,serafini2017quantum}. The   general zero mean multimode Gaussian state is given by the following squeezing operator $\mathcal S$ acting on a vacuum state
\begin{equation}\label{SO}
    \hat{\mathcal{S}}= \exp\left[  -\sum_{ij} A_{ij} \hat{a}^\dagger_i\hat{a}_j-\frac{1}{2}\sum_{ij}\left(B_{ij}\hat{a}_i^\dagger\hat{a}_j^\dagger  - B_{ji}^* \hat{a}_i\hat{a}_j  \right)   \right]
\end{equation}
\color{black}{where $\{\hat{a}_i\}$  are the components of a finite dimensional row of annihilation operators, and} \cmaa 
$A$, $B$ are finite dimensional matrices, with $A=A^{\dag}$ and $B=B^T$, with $T$ denoting transpose.\footnote{\cma Continuous variable generalisations are possible in terms of antilinear eigenvalue problems for Fredholm equations. \cbl} It is easy to prove that $\langle \hat{a}_{j} \rangle =0$ for all $j$. If $B=0$ we have $\langle\hat{a}_{i}\hat{a}_{j} \rangle=\langle\hat{a}_{i}^{\dagger}\hat{a}_{j}  \rangle = 0$. All cumulants higher than $2$ vanish, which is a characteristic of multi-dimensional Gaussian integrals. Such states typically arise necessarily in quantum systems where the effective Hamiltonian is quadratic in second quantised creation and annihilation operators.  If $B\neq0$, then we have squeezing and $\langle\hat{a}_{i}\hat{a}_{j} \rangle\neq 0,\langle\hat{a}_{i}^{\dagger}\hat{a}_{j}  \rangle \neq 0$. This is the situation adopted in  \eqref{Scattering_Multimode_Squeeze_GR}.\par 

There is a theorem of linear algebra\color{black}, proved by Autonne~\cite{autonne} and, independently, by Takagi~\cite{Takagi1933}, stating \cmaa that, for a complex symmetric matrix $\mathcal G$ (using the notation of 
\eqref{Scattering_Multimode_Squeeze_GR}), it is possible to find a unitary matrix $W$ such that ({\it cf.} \eqref{takagidec} of Appendix \ref{appC:takagi}):
\be\label{takagidec2}
\mathcal G= W\, \Sigma W^{T}\,,
\ee
where $\Sigma =diag\left( \sigma_{1} ,\sigma_{2} ,\ldots ,\sigma_{n} \right)$ is a non-negative matrix, {\it i.e.} $\sigma_{i} \ge 0, i=1, \dots n$. On defining \color{black} annihilation operators for the 
Takagi-Autonne-Schmidt (TAS)  modes 
\cmaa $\hat{b}_{\widetilde{I}}=W^\star_{I\widetilde{I}}\hat{a}_I$ ({\it cf.} \eqref{bmodes} of Appendix \ref{appC:takagi}), we have~\cite{autonne,Takagi1933}:
\begin{equation} 
\sum_{I,J} \mathcal G_{IJ}a_{I}^{\dagger}a_{J}^{\dagger}=\sum_{\widetilde{I}} \sigma_{\widetilde{I}} b_{\widetilde{I}}^{\dagger 2}, 
\end{equation}
where henceforth we denote indices pertaining to the $\hat b_{\tilde I}$ modes with $\widetilde{I}$. It is straightforward to show that $[b_{\widetilde{I}},b_{\widetilde{J}}]=[b_{\widetilde{I}}^{\dag},b_{\widetilde{J}}^{\dag}]=0$ and $[ b_{\widetilde{I}},b_{\widetilde{J}}^{\dag} ] =\delta_{\widetilde{I}\widetilde{J}}$. The TAS modes\footnote{\cmaa Apart from Autonne and Takagi, others such as Bloch-Messiah and Schmidt are also associated with this and similar decompositions~\cite{Houde:2024mkj}. Schmidt, in the  earliest work~\cite{Schmidt1907}, diagonalised a vector in a tensor product space, a method used for analysing bipartite pure states.\cbl} are characterised by squeezing parameters $\sigma_{\widetilde{I}} $ (see Appendix \ref{appC:takagi} for more details). So the density matrix in the original basis becomes a product of independent single mode density matrices in the new basis. This is a formal argument and, even for a finite set of operators, it is not generally possible to obtain the $W$ analytically.  When we have an infinite number of oscillators and we wish to avoid truncation of  the basis set (in a numerical treatment), we are led to a continuum description, as in our case ({\it cf.} \eqref{F_GR_correlations} and \eqref{IijFour}). There is a simple continuum version of the Takagi-Autonne method~\cite{autonne,Takagi1933},   which has been extensively developed mathematically \cite{GarciaMashreghiRoss2016ICSO,GarciaPutinar2006ComplexSymmetric} and there is related work in quantum optics \cite{CavesSchumaker1985TwoPhotonII}. We frame the discussion in the function space of complex square integrable functions, which is a Hilbert space. If we consider the index $I$ as $(\lambda, \vec{k})$,  we might wish  to associate $\mathcal G_{IJ}$ with a hermitian operator $\mathbb G$:
\begin{equation}
    \label{takagialter}(\mathbb{G} f)(I)=\int \mathcal{G}_{IJ}\  \overline{f}(J) d\mu \left( J \right)
\end{equation}
where the measure $d\mu (I)=\sum_{\lambda} d^{3}k$. Then, the Takagi decomposition~\eqref{takagidec2} resembles the following eigenvalue equation of the integral operator~\eqref{takagialter}, 
\begin{equation}
\mathbb G u_{\widetilde{I}}=\sigma_{\widetilde{I}} u_{\widetilde{I}}\,,
\end{equation}
which has the full form
\begin{equation}
\int \mathcal G_{IJ} \overline{u}_{\widetilde{I}} \left( J \right) d\mu \left( J \right) =\sigma_{\widetilde{I}} u_{\widetilde{I}}\left( I \right).
\label{eigenvalue_problem}
\end{equation}
The solution of these equations cannot usually be found analytically for \eqref{Scattering_Multimode_Squeeze_GR}. In terms of the $u_{\widetilde{I}}$ and $\sigma_{\widetilde{I}}$,
\begin{equation}
    \mathcal G_{IJ} =\sum_{\widetilde{I}} \sigma_{\widetilde{I}} u_{\widetilde{I}}(I)\  u_{\widetilde{I}}(J)
    \label{G_decomposition}
\end{equation}
and the TAS mode operators are
\begin{equation}
\hat{b}_{\widetilde{I}} = \sum_I \bar{u}_{\widetilde{I}}(I)\  \hat{a}_{I}.
\end{equation}
As in the finite dimensional case, the squeezing operator  $\mathcal S$ becomes (ignoring $A_{ij}$ in~\eqref{SO})
\begin{equation}
\mathcal S =\exp \left[ \frac{1}{2} \sum_{\widetilde{I}} \sigma_{\widetilde{I}}\, \hat{b}_{\widetilde{I}}^{2} -h.c. \right].
\end{equation}
With realistic hydrogenic wavefunctions $\psi_{nlm}$ \color{black} (see discussion in section \ref{sec:BHSuperradiance} and Appendix \ref{app:BHSuperradiance})\cmaa, the overlap $I_{ij}( \vec k,\vec k^{\prime} )$ ({\it cf.} \eqref{F_GR_correlations} and \eqref{IijFour}) contracted with TT polarisation tensors has a full $3-D$ angular structure, and a non-separable dependence on $(\lambda,\vec{k})$ and $(\lambda^\prime,\vec{k'})$ (see \eqref{I_ij_tensorial_structure}). The finite time gate \color{black} involving the sinc function  \eqref{G_GR_General} \cmaa has an argument which is the sum of frequencies, another contribution to non-separability of $I_{ij}( \vec k,\vec k^{\prime} )$.

We now introduce some physical intuition to simplify the eigenvalue problem~\eqref{eigenvalue_problem} so that we can make useful approximations for $I_{ij}$, which lead to analytically soluble TAS modes. The ${\rm{sinc}}(x)$ function in \eqref{G_GR_General} gives approximate energy conservation. It is peaked at $x=0$ and its first zeros are at $x=\pm \pi$. In our case the peak has a width given by:
\begin{equation}
\label{omegarange}
-\frac{2\pi}{T} <\left( \Omega_{k} +\Omega_{k^{\prime}} -2\omega \right) <\frac{2\pi}{T},
\end{equation}
where $T$ is a time scale related to astrophysical evolution ({\it i.e.} cloud-growth/depletion), which in our approach we identified with the lifetime of the axionic cloud ({\it cf.} \eqref{tcloud}). The quantity $\omega$ is the angular oscillation frequency of the bound (pseudo)scalar cloud that pumps the pair production. For $n=2$ it is approximately given by \color{black} $\omega \sim \mu_{b} ( 1-\frac{a_\mu^{2}}{8})$, ({\it cf.} \eqref{realfreq}, and  discussion in  Appendix \ref{app:BHSuperradiance})\cmaa.

 The graviton pair is radiated to the far zone and the dominant contributions to the $x$ integration comes from the stationary phase regions. Indeed, 
\be
 I_{IJ}\left( \vec{q} \right) =\int d^{3}x\  S_{IJ}\left( \vec{x} \right) \exp \left( -i\vec{q} .\vec{x} \right)
 \ee
 where $\vec q=\vec k+\vec k'$ and $S_{IJ}$ is a smooth localised function.
 Because the source size $r_c$  is large compared to \color{black} the  wavelength of the GWs \cmaa ($\lambda_{GW} \sim \frac{\pi}{\Omega_{0}}$ ), the phase in \eqref{I_ij_definition} is stationary giving $\vec k^{\prime}\sim -\vec k$. Of course, detailed numerical calculations also lead to this almost anti-collinear emission, \color{black} as discussed in section \ref{sec:Multimode}\cmaa. The polarisation tensors $e_{ij}^{(\lambda )}$, \color{black} appearing in the Fourier mode expansion of the GW tensor perturbations \eqref{FourierTensor2}\cmaa, 
 enforce helicity selection rules: same helicity dominance for the GR case and opposite-helicity correlations for CS.

We respect two universal structures in our soluble models: \textbf{(i)} there is an energy ridge $\Omega_{k} +\Omega_{k^{\prime}} \approx E$ with $E=2\mu_{b}$ for GR case and $E=\mu_{b}$ for the CS case. The width $\Delta$ of the ridge is $\Delta = 4 \pi /T$ \color{black}, ({\it cf. \eqref{omegarange}}), and \textbf{(ii)} anti-collinear emission peaked at $\vec{q}=\vec{k}+\vec{k'} \approx 0$ with angular width $\Delta \theta$ of order $1/kR$, $R$ being the cloud size.

These two features,  together with the helicity tensor, control the multimode correlation relevant for Takagi analysis~\cite{Takagi1933} ({\it cf.} Appendix \ref{appC:takagi}). We focus on the non-relativistic cloud and our soluble $\mathcal{G}$, replacing for our purposes here \eqref{G_GR_General} and 
\eqref{CSkern}, 
with
\begin{align}
\label{SolubleModel}
\mathcal{G}_{(\vec{k},\lambda),(\vec{k}^\prime,\lambda^\prime)}  =\mathcal{A}\exp \left( -\frac{\left( k +k^{\prime} -E \right)^{2}}{2\Delta^{2}} \right) \exp \left[ -\frac{\left(\vec{k}+\vec{k}^{\prime}\right)^{2}}{2\sigma_{+}^{2}} -\frac{\left( \vec{k}-\vec{k}^{\prime} \right)^{2}}{2\sigma_{-}^{2}} \right]\Pi_{\lambda \lambda^{\prime}}\ ,
\end{align}
where the proportionality constant $\mathcal{A}$ sets the order of magnitude of $\mathcal G$ and is different for the GR and CS cases, as discussed in section~\ref{sec:Multimode}. Its order of magnitude is not relevant for our purposes below. The quantity  $\Delta$ denotes the width of the Gaussian distribution for the energies, while  $\sigma_{\pm}$ denote the widths of the associated Gaussian distributions for the momenta. Depending on the relative magnitude of these widths, we may have collinear ($\sigma_- \ll \sigma_+$) or anti-collinear ($\sigma_- \gg \sigma_+$) situations. In our case, it is the latter that approximately characterises the non-relativistic BH-axion system, as discussed in section \ref{sec:Multimode}. The simplification~\eqref{SolubleModel} in the anti-collinear situation corresponds to $\Delta\theta=\pi$ in figures~\ref{GR_Correlations_LL_RR},~\ref{GR_Correlations_LR}  and~\ref{CS_correlations}. 

This model (despite its simplifications) gives: \textbf{(i)} a useful functional form of the squeezing modes in momentum space, and \textbf{(ii)} a scaling behaviour for the squeezing parameters, $\sigma_{\widetilde{I}}$. The simplifications of the model include the squeezing kernel ({\it cf. \eqref{momenta_kernel}} below) being real rather than complex; the helicity structure can be entangled with  momentum ({\it i.e} the role of helicity combinations LL, LR , RR and RL for the entangled gravitons). LR coupling, for example, can be  introduced by choosing  a structure of $\Pi_{\lambda \lambda^{\prime}}$ such as
\be\label{pimatrix}
\Pi =\left( \begin{matrix}0&1\\ 1&0\end{matrix} \right).\ee 
Generalisations become clear in the course of  the Takagi analysis for the spectrum associated with the kernel \eqref{SolubleModel}. Our analysis with $\Pi_{\lambda \lambda^{\prime}} =\delta_{\lambda \lambda^{\prime}}$ is readily modified to $\Pi_{\lambda \lambda^{\prime}}$ for  an arbitrary complex symmetric matrix and allows a correlation between momentum and helicity. Working in the specific anti-collinear case, we write:\footnote{In the actual situation encountered in section \ref{sec:Multimode}, the momenta $\vec k, \vec k^\prime$ lie in a 3-dimensional cone. This leads to enhancement of the squeezing effects, by a factor of $\mathcal O(10)$ compared to the oversimplified anti-collinear situation discussed here.}

\begin{equation}
\mathcal{G}_{(\vec{k},\lambda),(\vec{k}^\prime,\lambda^\prime)} =\mathcal{A}\ K(k,k^\prime)\  \Pi_{\lambda\lambda^\prime}\,,
\label{soluble_kernel_1}
\end{equation}
where
\begin{equation}
    K({k},{k}^\prime)=\exp \left[ -\frac{\left( k^{\prime}+k-E \right)^{2}}{2\Delta^{2}} \right] \exp \left[ -\frac{(k-k^\prime)^2 }{2\sigma_{+}^{2}} \right]\ .
    \label{momenta_kernel}
\end{equation}
In the above treatment, we made the simplification of writing the correlator $\mathcal{G}_{(\lambda,\vec{k})(\lambda^\prime,\vec{k}^\prime)}$ in a product form with respect to the momenta and the helicity modes of the graviton. In such a case, one can obtain the Takagi modes separately and then determine the full decomposition of the full graviton state (momenta and helicity). Starting with~\eqref{momenta_kernel}, we first define the new variable  $\widetilde{k}=k/\sqrt{\gamma}-E/2$, and similarly for $k^\prime$, to obtain, 
\begin{equation}
    K({k},{k}^\prime)= \exp\left[  -(\widetilde{k}+\widetilde{k}^\prime)^2 +\frac{\beta}{\gamma} \widetilde{k}\, \widetilde{k}^\prime           \right],
    \label{K}
\end{equation}
where 
\begin{align}\label{bg}
   \gamma=\frac{1}{2}\frac{\sigma_+^2+\Delta^2}{\Delta^2\sigma_+^2}\,, \qquad  
   \beta=\frac{\Delta^2-\sigma_+^2}{\Delta^2\sigma_+^2}\ . 
\end{align}
The above is a symmetric Gaussian on $\widetilde{k}$ and $\widetilde{k}^\prime$. Thus, one can obtain the Takagi decomposition by using Mehler's formula~\cite{Bateman:100233}, 
\begin{equation}
  \frac{1}{\sqrt{1-\rho^2}}\exp\left[  -\frac{\rho^2}{1-\rho^2}(x^2+y^2)+\frac{2\rho}{1-\rho^2}xy\right]=\sum_{n=0}^\infty\frac{(\rho/2)^n}{n!}H_n(x)H_n(y)\ , \quad 0 < \rho^2 < 1\,,
    \label{Mehler's_Formula}
\end{equation}
where $H_n(x)$ are the Hermite polynomials, that are orthogonal and complete,
\begin{align}
   & \int_{-\infty}^{+\infty}dx\ e^{-x^2}H_n(x)H_m(x)=2^n n! \sqrt{\pi}\ \delta_{mn} \ , \label{Hermite_orthogonality}\\
   & \sum_{n=0}^{+\infty}\frac{H_n(x)H_n(y)}{2^n \ n!}= \sqrt{\pi}\ e^{\frac{1}{2}(x^2+y^2)}\delta(x-y) \label{Hermite_completeness}\ . 
\end{align}
Multiplying \eqref{Mehler's_Formula} with $\sqrt{1-\rho^2}\, \exp(-x^2/2)\exp(-y^2/2)$, 
keeping $\rho^2 < 1$ strictly, 
we arrive at,
\begin{equation}
    \exp\left[ -\frac{1+\rho^2}{2(1-\rho^2)}(x^2+y^2)+\frac{2\rho}{1-\rho^2}xy \right]=   \sum_{n=0}^\infty s_n\chi_n(x)\chi_n(y), 
    \label{Mehler's_Formula1}
\end{equation}
where: 
\begin{align}
    &\chi_n(x)=\frac{1}{\sqrt{2^nn!\sqrt{\pi}}}e^{-\frac{1}{2}x^2}H_n(x),\\
    &s_n=\sqrt{\pi(1-\rho^2)}\rho^n\ .
\end{align}
Note that, on taking into account the orthogonality and completeness relations,  \eqref{Hermite_orthogonality} and \eqref{Hermite_completeness}, respectively, we see that $\chi_n(x)$ constitutes an orthonormal and complete set of functions, 
\begin{align}
    &\int_{-\infty}^{+\infty} dx\ \chi_n(x)\chi_m(x)=\delta_{nm}\,,\\
   & \sum_{n=0}^{+\infty}\chi_n(x)\chi_n(y)=\delta(x-y)\ .
\end{align}
In this sense, the right hand side of \eqref{Mehler's_Formula1} constitutes the Takagi decomposition for the Gaussian of the left hand side, with eigenvalues $s_n$. Thus, to find the Takagi decomposition of \eqref{K}, we have to bring the left hand side of \eqref{Mehler's_Formula1}, in the form of \eqref{K}. In order to achieve this, one has to perform the transformations  $x\rightarrow \sqrt{2(1-\rho^2)/(1+\rho^2)}\ x$ and $y\rightarrow \sqrt{2(1-\rho^2)/(1+\rho^2)}\ y$, to arrive at the following expression,
\begin{equation}
    \exp\left[-(x^2+y^2)+\frac{4\rho}{1+\rho^2}xy\right] = \sum_{n=0}^{+\infty} s_n\chi_n\left(\sqrt{\frac{2(1-\rho^2)}{1+\rho^2}}x\right)\chi_n\left(\sqrt{\frac{2(1-\rho^2)}{1+\rho^2}}y\right)\, ,
\end{equation}
which is exactly the left hand side of \eqref{K}, with 
\begin{equation}\label{rho}
    \rho=\frac{2\gamma}{\beta}\left(   1-\sqrt{1-\frac{\beta^2}{4\gamma^2}}   \right)\,,
\end{equation}
where 
we took the minus relative sign between the terms in the parenthesis on the right hand side of the equality in 
\eqref{rho}, due to 
the condition  $|\rho| <1$ ({\it cf.} \eqref{Mehler's_Formula}, which in our case is guaranteed by definition, since  
$|\beta|<2\gamma$ ({\it cf.} \eqref{bg}). Thus, the Takagi decomposition of \eqref{K} reads, 
\begin{equation}
    K(k,k^\prime)=\sum_{n=0}^{+\infty}\kappa_n U_n(k)U_n(k^\prime)\ , 
    \label{K_Takagi}
\end{equation}
where 
\begin{align}\label{Udef}
    U_n(k)= \left(\frac{\gamma(1+\rho^2)}{2(1-\rho^2)}\right)^{1/4} \chi_n\left(\sqrt{\frac{2(1-\rho^2)}{1+\rho^2}}\left(\frac{k}{\sqrt{\gamma}}-\frac{E}{2}\right)\right)  \ ,    \end{align}  and the squeezing parameters are:
    \begin{align}\label{squeeze}
    \kappa_n= \sqrt{\frac{2\pi}{\gamma}}\frac{1-\rho^2}{\sqrt{1+\rho^2}}\rho^n\ .
\end{align}

In view of the previous analysis for the cases of the non-relativistic cloud, as depicted in figs.~in~\ref{GR_Correlations_LL_RR},~\ref{GR_Correlations_LR} and~\ref{GR_Correlations_different_momenta}, we see that pairs of different helicity are dominant over the pairs of the same helicity. This forces us to assume for the $\Pi$ matrix the form given by~\eqref{pimatrix}. For the latter, the Takagi decomposition can be trivially obtained, 
\begin{equation}
   \Pi_{\lambda\lambda^\prime}=\sum_{m=1,2}w_{\lambda m}w_{m\lambda^\prime}\ , 
\end{equation}
where 
\begin{equation}
    w=\frac{1}{\sqrt{2}}\begin{pmatrix}
        1 & i\\
        1 & -i
    \end{pmatrix}\ .
\end{equation}
Thus, for the TAS modes, the notation for its indices reads $\widetilde{I}=(m,n)$, followed by $\lambda\rightarrow m$ and $\vec{k}\rightarrow n$, implied by each decomposition. Then, \eqref{soluble_kernel_1} acquires the TAS decomposition~\eqref{G_decomposition}, with the squeezing parameters given by:
\begin{align}\label{sigdef}
    \sigma_{\widetilde{I}}= \sigma_{(n,m)}=\mathcal{A}\, \kappa_n \,,
    \end{align}
    and the relevant eigenvectors by:
    \begin{align}\label{udef}
    u_{\widetilde{I}}(I)=u_{(n,m)(\lambda,\vec{k})} = U_n(k)\, w_{\lambda m}\ .
\end{align}    
The reader's attention is drawn to the fact that the above results pertain to finite $\Delta \ne 0$, $\sigma_+\ne 0$, and thus finite axion-cloud lifetimes $T < \infty$. The reader should recall from \eqref{Mehler's_Formula} that the limit $\rho \to 1$ corresponds to a $\delta$ function distribution on the left-hand side, which in our case translates to energy-momentum conservation in our anticollinear case. This is a singular limit, where the Autonne-Takagi decomposition cannot apply without proper regularization, which is what we have done above, in the context of our oversimplified approximation \eqref{momenta_kernel}, by keeping $\Delta, \sigma_+$ finite. In the more realistic situation encountered in our GW system in section \ref{sec:Multimode}, the quantities $\mathcal A$, $\sigma_+$, $\Delta$, are related to  each other and determined by the gravitational coupling $a_\mu$ ({\it cf.} \eqref{amudef}). The latter defines the (non-)relativistic  regime of the BH-ALP system, and thus the amount of squeezing. As such, the determination of these relations is crucial in order to draw conclusions about the values of the squeezing parameters, $\sigma_{\widetilde{I}}$.

\color{black}We postpone such a complete study for a future publication, where we shall also discuss detection prospects. As we shall show there, squeezing in the GW sector can appear as squeezing in the optical output of the detector. For our purposes here we only mention that, 
because the squeezed-quantum-graviton states produced by the ALP-gravity interactions in the BH cloud exhibit cross-polarisation correlations,
they can, in principle, be disentangled from classical stochastic backgrounds through an analysis of 
the cross-covariances between the interferometer outputs corresponding to orthogonal detection modes.
In practice, this corresponds to looking for off-diagonal structure in the detector-output matrix 
\be\label{Mdet}
M_{\mathrm{det}} = \mathbf H (\Omega) \, M_{\mathrm in}\, \mathbf H(\Omega)^T\,,
\ee
where 
$\mathbf H (\Omega)$ is the real transfer matrix built from the (resonant) cavity, of frequency $\Omega$, and mechanical susceptibilities of the interferometric detector, and 
${\langle M_{\mathrm{in}}\rangle}_{IJ} = \langle \mathbf{n}_{{\rm in}\, I} \, 
\mathbf{n}^T_{{\rm in}\,J} \rangle$,
with
\be\label{eta_in}
\mathbf{n}_{\rm in}(\Omega)
   =
   \begin{bmatrix}
     \widehat X_{\mathrm{in}}(\Omega) &
     \widehat Y_{\mathrm{in}}(\Omega) &
     \widehat \xi_{\mathrm{th}}(\Omega) &
     \widehat h_{1}(\Omega) &
     \widehat h_{2}(\Omega) &
     \cdots
   \end{bmatrix}^{\!T}\,,\ee
 (where $T$ denotes matrix transposition)  
 a vector of quantum-operator quantities. The latter    
represent
the GW input in the interferometer, along with the cavity optical mode, of frequency $\Omega$, in the resonant cavity regime, the input optical vacuum quadrature operators, 
$\widehat X_{\mathrm{in}}$ and $\widehat Y_{\mathrm{in}}$, and the operator pertaining to the thermal Langevin force,
$\widehat \xi_{\mathrm{th}}$. The quantities 
$\widehat h_m$, $m=1,2,$ in \eqref{eta_in} 
denote the standard GW-mode quantum operators, pertaining to the $\cross$ and $+$ polarisation, while the $\dots$ denote other inputs, associated with potential environmental (non gravitational) noise of the detector. Within the weak-graviton approximation which we are adopting, 
we assume linear response~\cite{Pang:2018eec}, 
\begin{align}\label{8.2}
\mathbf{b_{\rm det}} = \mathbf{H}(\Omega)\, \mathbf{n_{\rm in}},
\end{align}
where $\mathbf{b_{\rm det}}$ stands for the detected optical quadrature at the output port of the interferometer,
 
The Autonne–Takagi decomposition~\cite{autonne,Takagi1933} on $M_{\mathrm{det}}$ \eqref{Mdet}, therefore, identifies the \emph{detector supermodes}---the observable combinations of frequencies and polarizations that carry the 
gravitational squeezing into measurable optical correlations, 
\cbl \color{black} each corresponding to an independent squeezed spectral component of the output light.
This construction provides a direct operational link between the theoretically predicted multimode squeezed
graviton field and the experimentally accessible optical correlations.

\color{black}

\section{Conclusions and Outlook}\label{sec:Conclusions}

In this work, we have extended  and studied in detail a potential scenario for the production of squeezed entangled graviton states, proposed in~\cite{Dorlis:2025zzz}. If realised in Nature, this scenario could provide a means of experimentally proving the quantum nature of the gravitational interaction, through the detection of quantum squeezed graviton states. The scenario involves massive axion-like fields which form a kind of condensate (``axionic cloud'') in the exterior region of a rotating astrophysical BH (Kerr-type), and lead to the phenomenon of superradiance. These axions may : (i) have a geometrical origin, being associated with torsional degrees of freedom in Einstein-Cartan theories, (ii)  may be the model-independent (Kalb-Ramond) string axion, being dual to the spin-one antisymmetric tensor field of the (bosonic) massless gravitational multiplet of the string in (3+1)-dimensions (after string compactification) or (iii) may arise from specific compactification schemes ({\it compactification axions}) of string theory models, dependent on the specific model of string  compactification ~\cite{Svrcek:2006yi}. Nonetheless, provided the axion is massive, our main result, for the production of squeezed entangled graviton states from axions in clouds around the Kerr BH, is independent of the specific microscopic origin of the axion field.

In all of the above cases there are CS gravitationally-anomalous interactions coupled to the axions. Such terms are non-trivial in the presence of Kerr BH or chiral GW backgrounds. In our analysis above we have demonstrated that the anomaly-free (GR-type) terms in the respective gravitational effective actions (CS gravity models) lead to graviton-squeezing quantum effects that dominate, by many orders of magnitude, those induced by the CS gravitationally-anomalous interactions. It is found that, for sufficiently long-lived axionic clouds, the macroscopically large number of axions in them enhances, as a result of the superradiance,  the squeezing effects of QG to a point that the number of squeezed graviton states might be observable by (combinations of) future interferometers.

The different effects of GR vs CS terms in the CS-gravity effective action on the form of the entangled two-mode graviton states found by us explicitly, is worthy of further study. The effects, although qualitatively and quantitatively different, are reminiscent of the effects of quantum space-time foam inducing decoherence on entangled neutral-meson particle states produced in meson factories. For spacetime foam it is the ill-defined nature of the CPT operator that leads to a contamination of the EPR correlator with the ``wrong-symmetry'' state. In the gravitational systems discussed here, it is the CS gravitational anomaly, which leads to a  non-conservation of the axion stress-tensor, that modifies the EPR correlators of the entangled (non-separable) two-quantum-graviton polarization (Left, Right) states.

In our approach here we have considered the axion system sufficiently far from the BH horizon, so that the weak quantum-graviton approach (obtained by expanding around an approximately Minkowski spacetime background) suffices. This implies that, going beyond the two-graviton order in a perturbative expansion of weak graviton perturbations, is unimportant. However, such approximations may fail when one considers regions closer to the horizon of the BH. In such cases, the inclusion of higher-order quantum graviton fluctuations in the analysis may lead to novel results  for  multigraviton entangled states in the respective scattering or decay processes with the axions. This would be analogous to multiphoton, or more general, multipartite, entanglement states in quantum optics and quantum information~\cite{Pan:2012svn,Toth:2012lpv,Hiesmayr:2017xgx,Hiesmayr:2018rcm,Bala:2025gmv}. The same can be said for the inclusion of back-reaction effects of the axion clouds on the BH geometry itself, which becomes significant close to the horizon. These are important open issues which need to be studied. 
Nonetheless we believe that, through the current approximate study, we have provided convincing arguments to look for novel astrophysical sources of graviton squeezing, which could make the detection of quantum graviton fields (the fundamental carriers of the gravitational interaction) a realistic possibility, due to the enhancing effects of the superradiant axion-cloud. 

\color{black} We conclude by commenting briefly on observational prospects for the polarisation-entangled graviton states, produced in the axionic clouds surrounding rotating BHs. Due to the entangled nature of the gravitons, which is an inherent quantum property, it should be possible to disentangle potential signals from those of classical GWs, which, as discussed in \cite{Carney:2023nzz}, can mimic detector ``clicks'' that are expected to be produced in detection apparatuses of single gravitons~\cite{singlegrav}. Below we explain why in our proposed method such ambiguities can be avoided.

Following the reasoning in~\cite{Carney:2023nzz,Carney:2024dsj}, one has to carefully select the kind of measurement which  has to be performed in order to probe quantum properties of the signal. For example, counting statistics of the detector's ``clicks'' is in principle a possibility, since sub-Poissonian statistics for the number of events $N$ (``clicks''), {\it i.e.} $(\Delta N^2<\langle N\rangle )$, can be explained only via singular distributions in the space of the overcomplete basis formed by coherent states, and thus admit no classical analogue. Squeezed-coherent states  constitute such   states, that include quantum deviations from the classical-coherent graviton states, and arise naturally in our model  ({\it cf. }\eqref{Squeezed-Coherent}). 
In our scenario, significant squeezing parameters can be achieved, due to the large number of ALPs and the cloud's long lifetime. 
However, even if the amount of squeezing can be significantly enhanced, one might still argue that a quantum signature of gravity is still non-achievable through this process {\it alone}. Indeed, despite the fact that a sufficiently large squeezing parameter can induce a significant amount of noise, its domain for a non-classical imprint on the detector ({\it e.g.} in a strain detector like LIGO~\cite{LIGOScientific:2016aoc}),   is highly suppressed (sub-vacuum) and thus challenging to observe~\cite{Carney:2024dsj}.\footnote{\color{black} We stress once again that such challenges have been demonstrated explicitly to characterize interferometric detectors~\cite{Carney:2024wnp} within the context of effective field theory approaches to quantum gravity, which utilize weak-graviton excitations of the appropriate vacuum. Stronger QG signals, appearing as noise in the interferometers, might characterize non-effective field theory approaches to QG~\cite{Amelino-Camelia:1997pfm,Hogan:2007pk,Verlinde:2019xfb,Zurek:2020ukz,Li:2022mvy,Bub:2023bfi}. The detection of such large noise in current interferometers, therefore, if not attributed to conventional sources, would constitute evidence of QG beyond the standard framework of effective field theories.\color{black}}
 
One might argue that such a large noise domain (super-vacuum) could be explained by  classical noise of GWs. Having said that, though it seems possible, as explained in the text, to place upper bounds on single-mode squeezing~\cite{Hertzberg:2021rbl} via appropriate methods, such as treating these modes as Gaussian wave packets, this still does not constitute direct evidence of the quantum nature of gravity. 

However, all of the above considerations are limited to the case of a single graviton mode interacting with the detector and, hence, the relevant searches pertain to single-mode squeezing. By contrast, in our scenario, interactions of massive ALPs in the cloud produce {\it multimode entangled} (hence purely quantum) states of gravitons. What is important, in this respect, is the fact that the multi-mode squeezed state correlates different directions (modes) through such entanglement. This implies that noise measurements of differently arranged interferometers - interacting with different directions of emission -  might reveal correlations in the potentially observable noise domain (super-vacuum), which are  purely of quantum origin and thus overcome the quantum-classical degeneracy of the  detector's response. 

The Takagi decomposition method, discussed in this work, may prove invaluable to such studies, by providing a formal link between the entangled squeezed graviton states and the spectral modes accessible in the corresponding interferometric detection. \color{black}
 The framework developed in this work, section \ref{sec:Takagi} and Appendix \ref{appC:takagi}, unifies the quantum-field description of axion-induced superradiant
graviton emission with the linear detection theory of interferometers.  
The polarisation-entangled graviton states produced in axionic clouds present a qualitatively different
observable signature compared to classical gravitational waves, due to their 
cross-polarisation correlations, which distinguish them from classical stochastic backgrounds.

As argued in \cite{Carney:2023nzz},
demonstrating the quantum nature of such signals requires measurements that are sensitive to
nonclassical counting statistics---for example, sub-Poissonian fluctuations of the detected photon number,
$\Delta N^2 < \langle N\rangle$, as mentioned earlier.
As we have noted, squeezed-coherent graviton states, which arise naturally in our model,
are precisely such nonclassical states.
Although the direct observation of sub-vacuum noise in strain measurements (as in LIGO~\cite{LIGOScientific:2016aoc})
remains extremely challenging (see also related comments on \cite{Carney:2024wnp}, where such challenges on single-graviton detection are demonstrated explicitly to characterise the conventional weak-graviton effective field theory approach to gravity,  also adopted here),
our analysis shows that multimode squeezing correlating distinct emission directions
can, in principle, produce observable correlations between different detectors or interferometer arms,
revealing quantum signatures beyond the reach of single-mode analyses.

Moreover, in the present scenario the degree of entanglement resides in the graviton polarisation.
In principle, one could perform polarisation-resolved measurements of the interferometer outputs
to search for Einstein–Podolsky–Rosen-type correlations.
While current technology does not allow such polarisation-sensitive quantum measurements for gravitational fields,
the rapid progress in quantum metrology and optomechanical sensing may render such schemes plausible in the future.

Finally, we note that the same formalism applies to squeezed quantum-graviton states of
cosmological origin, such as those produced during inflation~\cite{Mukhanov:2007zz,Kanno:2018cuk,Kanno:2019gqw}.
The entanglement-induced decoherence of macroscopic interferometer mirrors
\cite{Kanno:2021gpt}
could, in principle, provide indirect evidence for the quantum character of such primordial states.
Furthermore, if populations of primordial rotating black holes were present during inflation,
the multimode superradiant mechanism discussed here could enhance
the primordial squeezing of the gravitational wave background.
Thus, the transfer-matrix and Takagi-mode framework developed in this work
may ultimately serve as a unified language for both astrophysical and cosmological
quantum-graviton observables.

\color{black}

 \color{black}

\begin{acknowledgments}
Results from this work have been presented in the {\it Wilzcek Quantum Centre (WQC) workshop on entanglement of high energy particles} (http://www.wilczekqc.net/show/174), 19-23 July 2025, WQC, USTC Shanghai Institute for Advanced studies, Shanghai (China). NEM wishes to thank Prof. Yu Shi for the invitation to give a plenary talk in the workshop, and for organising such an excellent, thought provoking event, with a lot of stimulating discussions. NEM also wishes to thank Beatrix Hiesmayr for stimulating discussions on multipartite entanglement. NEM thanks the University of Valencia and its Theoretical Physics Department for a visiting Research Professorship
supported by the programme  \emph{Atracci\'on de Talento}
INV25-01-15. 
The work of P.D. is supported by a graduate scholarship from the National Technical University of Athens (Greece).
The work of NEM and SS is supported in part by the UK Science and Technology Facilities research Council (STFC) under the research grant  ST/X000753/1, and by the UK Engineering and Physical Sciences Research Council (EPSRC) under the research grant No. EP/V002821/1. 
The work of S.-N.V. is supported by the Hellenic Foundation for Research and Innovation
(H.F.R.I. (EL.ID.EK.)) under the “5th Call for H.F.R.I. Scholarships to PhD Candidates” (Scholarship Number:
20572). NEM also acknowledges participation in the COST Association Actions CA21136 “Addressing observational
tensions in cosmology with systematics and fundamental physics (CosmoVerse)” and CA23130 ”Bridging high and
low energies in search of Quantum Gravity (BridgeQG)”.

\end{acknowledgments}

\appendix

\color{black}

\section{Axion fields in Various contexts}\label{app:axion}

In this Appendix we discuss the microscopic origin of axion-like fields in various gravitational frameworks of interest to us in this work, which contain rotating black holes.

\subsection{Torsion and Dynamical Axions in Einstein-Cartan Theories}

 We commence  by reviewing  the Einstein-Cartan theory of fermions in a contorted geometry~\cite{Cartan:1938ph,hehl}, following the treatment in \cite{Duncan:1992vz}. Fermion interactions with gauge fields do not affect the main conclusions,  and so  we only consider free Dirac fermions in a contorted geometry. It suffices for our purposes to consider one flavour ($N_f=1$)  of Dirac fermions. The corresponding action reads:
\begin{align}\label{axial}
   & \mathcal S_{\rm Torsfermi} = \frac{i}{2} \int d^4x \, \sqrt{-g}\, \Big[ \overline \psi (x) \, \gamma^{\mu} \, \mathcal{D}_\mu (\omega) \, \psi (x) - \overline{\mathcal D_\mu(\omega) \psi(x)} \gamma^{\mu} \, \psi(x) \Big]\, = S_{\rm Dirac-curved}(\mathring{\omega})  \nonumber \\ &+ \frac{1}{8} \, \int d^4x \sqrt{-g} \,  \overline{\psi} (x) \{\gamma^{c}\,, \,
\sigma^{ab}\}\, \mathcal{K}_{abc} \, \psi (x) = S_{\rm Dirac-curved}(\mathring{\omega}) - \frac{3}{4} \int d^4 x\, \sqrt{-g} \, S_\mu \, \overline \psi \, \gamma^{\mu}\, \gamma^{5} \, \psi 
\,,
\end{align}
where $S_{\rm Dirac-curve}(\mathring{\omega})$ is the free Dirac Lagrangian in the torsion-free curved spacetime with spin-connection $\mathring{\omega}$, 
$\mathcal{D}_\mu (\omega)$  is the gravitational covariant derivative in this spacetime, ${\mathcal D}_\mu(\omega) \, = \, \partial_\mu + i \, \omega^a_{\,\,\,b\,\mu} \, \sigma{^{b}_{a}}\; , \,\sigma^{ab} \equiv \frac{i}{4} [\gamma^{a}\,, \, \gamma^{b}]$, and the quantity $\omega^a_{\quad \mu \, b} = {\mathring{\omega}}^a_{\quad \mu\, b} + {\mathcal K}^a_{\,\,\,\mu\,b}$ is the contorted spin connection, with $\mathcal K$ the contorsion tensor~\cite{hehl,Duncan:1992vz},
$\mathcal{K}_{abc} =- \frac{1}{2}( T_{cab} - T_{abc} - T_{bca})$, and $T^\mu_{\,\,\,\nu\rho} = - T^\mu_{\,\,\,\rho\nu}$ the torsion.\footnote{Greek indices denote world-indices of the spacetime manifold, $\mathcal M$, while Latin indices $a,b,c, \dots$, are local Lorentz-frame indices, taking values in the Lie algebra $so(1,3)$ which pertain to the tangent space of $\mathcal M$ at a spacetime point $x$.} 
In form language~\cite{eguchi},  on considering the vielbein 1-form, $e^a$, and the spin connection 1-form, $\omega_{\  \  b}^{a}$, the torsion 2-from $T^a$ (of mass dimension $1$)~\cite{hehl} 
is defined as
\begin{align}
  T^{a}&=de^{a}+\omega_{\  \  b}^{a} \wedge e^{b}.
  \end{align}
The (generalised) curvature two form is given by 
$
  R_{\  b}^{a}=d\omega_{\  b}^{a} +\omega_{\  \  c}^{a} \wedge \omega_{\  b}^{c}$,
where $\wedge$ is the exterior product of forms~\cite{eguchi}. 
In \eqref{axial} we use the property of the (Dirac) $\gamma$-matrices $\{\gamma^{c}\,, \sigma^{ab}\} = 2 \epsilon^{abc}_{\quad \,d}\, \gamma^{d}\, \gamma^{5}$, 
and 
\begin{align}\label{Sdef}
S_d \equiv \frac{1}{3\!} \epsilon^{abc}_{\quad \, d} \, T_{abc}\,, \qquad \,\,({\rm or~in~form~language} \quad S \equiv \star \,T), 
\end{align}
which is a pseudovector dual to the totally antisymmetric part of the torsion $T^\mu_{\,\,\,\nu\rho}$. From \eqref{axial}, this is 
the only torsion component that couples to the axial fermionic current $J^{5\mu}= \overline \psi \,\gamma^{\mu} \, \gamma^{5} \, \psi$. Next, we add the action \eqref{axial} to the purely gravitational (generalised Einstein-Hilbert) part of the action in the presence of torsion:
\begin{align}\label{gravact}
S_{\rm grav} \equiv \frac{1}{2\kappa^2} \int d^4 x \, \sqrt{-g} \Big( R + \widehat \Delta \Big) + \frac{3}{4\kappa^2} \int \, S \, \wedge \, \star \,S \,,
\end{align}
with the term 
$\widehat \Delta$ containing quadratic forms of the remaining  components of the torsion, which are not totally antisymmetric ~\cite{Duncan:1992vz}. In this way we form the total action: 
\begin{align}\label{stotal}
\mathcal S_{\rm total}  = \mathcal S_{\rm Torsfermi} + S_{\rm grav} \,.
\end{align} 
The classical equations of motion with respect to the (non-propagating) field $S^\mu$, stemming from the action \eqref{stotal}, imply:
\begin{align}\label{SJ5}
 S_\mu = \frac{\kappa^2}{2} J_\mu^5 \,.  
\end{align}
Taking into account that the contribution of the symmetric Christoffel symbol to the conservation law of $S^\mu$ vanishes, due to \eqref{Sdef}, we observe from \eqref{SJ5}, that, if the axial fermion current is conserved, then 
there is conservation of the totally-antisymmetric-torsion current
\begin{align}\label{dS}
   \partial_\mu \, S^\mu = \frac{\kappa^2}{2} \nabla_\mu J^{5\, \mu} = 0\,,
\end{align}
and  of the corresponding charge $Q_S=\int d^3x S^0$, where $\nabla_\mu$ denotes the torsion-free gravitational covariant derivative.

In the presence of chiral fermions, there is a gravitational anomaly~\cite{Alvarez-Gaume:1983ihn}. Indeed, evaluating a one-loop Feynman graph of chiral fermions, in which the axial current vertex sits at one corner of a triangle and the other two corners are insertions of the energy momentum tensor (or equivalently insertion of two gravitons), one obtains 
\be
\label{axialanom}
\nabla _{\mu }J_{5}^{\mu }=\dfrac{1}{384\pi ^{2}}R_{\mu \nu \rho \sigma }\widetilde{R}^{\mu \nu \rho \sigma}\,,
\ee
with $\widetilde{R}^{\mu \nu \rho \sigma}$ the dual Riemann tensor (defined in \eqref{dualriem2} in the next section \ref{sec:CST}). This is an anomalous contribution of gravitational origin (exactly analogous to the gauge field case, which contributes to the right-hand-side of \eqref{axialanom} a term proportional to~\cite{Adler:1969gk}: $-\frac{1}{16\pi^2} \mathbf F_{\mu\nu} \, \widetilde{\mathbf F}^{\mu\nu}$, with $\mathbf F_{\mu\nu}$ the (non-Abelian in general) gauge field strength).
This new term is however locally Lorentz invariant and diffeomorphism invariant. Furthermore $R_{\mu \nu \rho \sigma }\Tilde{R}^{\mu \nu \rho \sigma }$ is topological and is a total derivative; it cannot appear in the Lagrangian just by itself. As a result of the gravitational anomaly \eqref{SJ5}, the axial current is not conserved at a quantum level in spacetimes for which $R_{\mu \nu \rho \sigma }\Tilde{R}^{\mu \nu \rho \sigma } \ne 0$,  for instance, in the presence of chiral gravitational waves (GW)~\cite{Alexander:2004us,Lyth:2005jf,Dorlis:2024yqw}.

In such cases, to ensure the validity of \eqref{dS}, and thus the conservation of the torsion charge, one should add appropriate counterterms to the quantum effective action, order by order in loop perturbation theory (similar to quantum electrodynamics in contorted geometries \cite{Duncan:1992vz}). In a path-integral approach, this can be implemented by inserting the constraint \eqref{dS}
in the form of a delta functional in the relevant path integral of the action \eqref{stotal}, 
\begin{align}\label{dsconstr}
\delta (\partial_\mu S^\mu) \ .
\end{align}
Since the delta-functional is a parity invariant quantity, it 
can be represented in a quantum path integral of the action \eqref{stotal} via a pseudoscalar Lagrange multiplier field $\Phi = (\frac{3}{2\kappa^2})^{1/2}\,b$. Thus, the path-integral over the $S^\mu$ field can be written, in differential-form language (for compactness)~\cite{eguchi},  as~\cite{Duncan:1992vz}:
\begin{align}\label{PIS}
    \mathcal Z &\propto \int \mathcal D S \, \exp\Big(i\, \mathcal S_{\rm total}\Big) \, \delta (\mathbf d\star S) \propto
\int \mathcal D S \, \mathcal D b  \,\exp\Big(i\,\int \Big[\frac{3}{4\,\kappa^2}  \, S \, \wedge \, \star \, S - \frac{3}{4} S \, \wedge \, \star j^5 + (\frac{3}{2\kappa^2})^{1/2}\,b\, \mathbf d\,  \star S\Big]\Big) \nonumber \\ &\propto \int \mathcal D b \, \exp\Big(-i \int \Big[\mathbf d b \wedge \star\, \mathbf d b + \frac{1}{384\, \pi^2\, f_b} \, b \, R_{\mu\nu\rho\sigma}\, \widetilde R^{\mu\nu\rho\sigma}  - \frac{1}{2f_b^2}\, J^5 \wedge \star J^5 \,\Big] \Big) \, , \qquad f_b = \Big(\frac{3\, \kappa^2}{8}\Big)^{-1/2}\,,
\end{align}
where we used \eqref{axial}, \eqref{gravact}.\footnote{The $\propto$ symbol in \eqref{PIS}
is a comprehensive notation for the path-integral (over the graviton $g_{\mu\nu}$  and the remaining torsion component fields, different from $S^\mu$) of the generalised (contorted) scalar-curvature part of the effective action.} In arriving  at the last line, we have 
path-integrated over the non-propagating field $S^\mu$, and integrated by parts the second term in the exponent of the integrand in the second line of \eqref{PIS}, using the anomaly equation \eqref{axialanom}.  Therefore, it follows from \eqref{PIS}, that the effects of the torsion are effectively represented by: (i)  a {\it dynamical} axion-like {\it massless} field $b$ (with canonical kinetic term), which couples to the \emph{gravitationally anomalous term } $R_{\mu\nu\rho\sigma}\, \widetilde R^{\mu\nu\rho\sigma}$, with an axion coupling parameter $f_b$ given in \eqref{PIS}, and (ii) by the presence of (repulsive) four-fermion terms (quadratic in the axial current). Both these effects are characteristic features of all Einstein-Cartan theories~\cite{Cartan:1938ph}.

Generalizations of the ECT exist in the recent literature, which involve coupling of axion-like fields with torsional topological invariants. One such invariant is the Nieh-Yan invariant~\cite{Nieh:1981ww,Nieh:2007zz}, which is a topological term that can be constructed in any spacetime with torsion. Using the first order formalism of GR, the standard Nieh-Yan 4-form 
reads:
\begin{align}\label{NYform}
N=T^{a}\wedge T_{a}-R_{ab}\wedge e^{a}\wedge e^{b}=d(e^{a}\wedge T_{a})\,.
\end{align}
The Latin indices label an orthonormal (tangent-space) frame \cite{Nakahara:206619}. In the absence of torsion, $T^{a}=0$, ${\omega}_{\  b}^{a}=\mathring{\omega}_{\  b}^{a}$. $N$ can only have an effect if it couples (in an effective gravitational action) to a (pseudo) scalar field, which arises if torsion propagates due to coupling with matter. The coupling with matter allows the ECT theory to mimic a Chern-Simons theory~\cite{Duncan:1992vz}. 

In a phenomenological framework it could be possible to add {\it by hand} a linear interaction of the Nieh-Yan invariant with an axion-like field, related to propagating torsion~\cite{Mercuri:2006um,Mercuri:2006wb,Calcagni:2009xz}
which, as shown in~\cite{Mavromatos:2021hai} (see also \cite{iorio}), is equivalent to the string-inspired case~\cite{Duncan:1992vz,Svrcek:2006yi}, where the axion is in the physical spectrum, specifically it is dual to the field strength of the KR field in the massless gravitational multiplet~\cite{str1,str2,pol1,pol2} ({\it cf.} discussion below). Thus, although in the context of pure quantum field theories, promoting the coupling constant of the Nieh-Yan torsional invariant to a (pseudoscalar) field might seem {\it ad hoc}, it acquires an important dynamical meaning in the context of the string-model-independent (or KR) axion of strings~\cite{Duncan:1992vz,Svrcek:2006yi}.\footnote{At this stage it worths mentioning, for completion, that there are also generalisations of the Nieh-Yan  4-form~\cite{Nieh:2018rlg}, associated with the rest (other than the totally antisymmetric) components of the torsion, 
which have been considered in other contexts~\cite{PhysRevD.104.084020}. It is not yet known whether these more general constructions can play a r\^ole in our situation.}

\subsection{String-theory Axions}

The emergence of axion-like fields in 
string-theory~\cite{Duncan:1992vz,Svrcek:2006yi}, parallels that for the Einstein-Cartan theory given above. One considers the bosonic effective action to ${\mathcal O}(\alpha^\prime)$ (where $\alpha^\prime = M_s^{-2}$ is the Regge slope of the string, with $M_s$ the string mass scale, that generally differs from $M_{\rm Pl}$)~\cite{Gross:1987mw,Metsaev:1987zx}:
\begin{align}\label{sea4}
    \mathcal S_B =  \int d^4 x \sqrt{-g} \Big(\frac{1}{2\kappa^2}\, R  - \frac{1}{6} H_{\mu\nu\rho}\, H^{\mu\nu\rho} + \dots \Big) \,,
\end{align}
where the $\dots $ denote other terms (gauge, fermionic matter etc.),  which do not contribute to torsion.\footnote{\color{black} The form of these terms depend highly on the field content of the underlying ten-dimensional (super)string theory~\cite{str1,str2,pol1,pol2}. For instance, the 10-dimensional low-energy effective field theory of type IIA superstring theory reads:
\begin{align}\label{IIA}
S_{IIA}=\int d^{10}x\sqrt{-g} \left( e^{-2\phi}\left[ R+4 ( \nabla_M \phi ) ( \nabla^M \phi ) -\frac{1}{12} H_{M N P}H^{M N P} \right] -\frac{1}{4} F_{M N}F^{M N}-\frac{1}{48} F_{M N P\Sigma}F^{M N P \Sigma}+\ldots \right)
\end{align}
where capital Greek letters $M,N,P,\Sigma$ in \eqref{IIA} denote  ten-dimensional spacetime indices, $F_{MN}$ and $F_{MNP\Sigma}$ are the field strengths associated with the Ramond-Ramond forms of type II superstring theories, $H_{MNP}$ is the field strength of the antisymmetric 2-form field $B_{MN}$: $H_{M N P}=\partial_{M} B_{N P}+\partial_{N} B_{P M}+\partial_{P} B_{M N}$, and $\phi$ is the dilaton, which is massless in simple settings such as Type IIA and IIB supergravities (although it can acquire a mass depending on compactification or supersymmetry breaking mechanisms). The dilaton controls the strength of the string coupling $g_s=\exp(\phi)$, which in turn determines the gauge couplings of the theory, after compactification, and thus its value is important for phenomenology. There are additional fields associated with the supersymmetry partners of the bosonic fields appearing in \eqref{IIA}: gravitino, dilatino, Majorana spinor and Ramond-Ramond, which we ignore, assuming an appropriate supersymmetry breaking scenario at sufficiently high scales, compared to the energy scales we are interested in here. Hence, the supersymmetric partners become massive and decouple from the low-energy spectrum.\color{black}}
We ignore dilaton fields (which are the spin-$0$ part of the massless bosonic gravitational string multiplet~\cite{str1,str2}), assuming that they are stabilised to a constant value, by means of minimisation of an appropriate string-loop-induced potential. 
The quantity 
$H_{\mu\nu\rho}$ is totally antisymmetric in its indices and denotes the field strength of the antisymmetric Kalb-Ramond (KR), spin-one, field $B_{\mu\nu}=-B_{\nu\mu}$ of the massless gravitational multiplet of the closed sector of string theory~\cite{str1,str2}. 
The cancellation of 
gauge and gravitational anomalies in the extra dimensional spacetime of string theories requires the addition of appropriate Green-Schwarz counterterms~\cite{GS} in the effective action. This amounts to the modification of the definition of the three form $H_{\mu\nu\rho}$ from the simple curl of $B$-field, $\mathbf{d} \, B$, as:
\begin{align}\label{csterms}
H &= \mathbf d\,B + \frac{\alpha^\prime}{8\, \kappa} \, \Big(\Omega_{\rm 3L} - \Omega_{\rm 3Y}\Big),  \nonumber \\
\Omega_{\rm 3L} &= \mathring{\omega}^a_{\,\,c} \wedge d\,\mathring{\omega}^c_{\,\,a}
+ \frac{2}{3}  \mathring{\omega}^a_{\,\,c} \wedge  \mathring{\omega}^c_{\,\,d} \wedge \mathring{\omega}^d_{\,\,a},
\quad \Omega_{\rm 3Y} = \mathbf A \wedge  \mathbf d\,\mathbf A + \mathbf A \wedge \mathbf A \wedge \mathbf A,
\end{align}
where, as before, $\mathring{\omega}$ denotes the standard torsion-free spin connection, and $  \mathbf A$ are the non-Abelian gauge fields that generically characterise strings. The $\Omega_{\rm 3L} \, (\Omega_{\rm 3Y})$ are the 
Lorentz (Yang-Mills) CS three-forms~\cite{eguchi}.

The modified $H_{\mu \nu \rho}$ leads to terms in the string effective action  which cancel the anomalies. Formally
 $$\delta L_{anomaly}+\delta L_{Green-Schwarz}=0$$
where $\delta L_{anomaly}$ is the contribution to the anomaly from the gauge fields \cite{GS} and fermions, such as $\rm Tr(\mathbf F\wedge \mathbf F)$ (in differential form notation~\cite{eguchi} for the familiar  chiral gauge anomaly term~\cite{Adler:1969gk}  ${\rm Tr} {\mathbf F}^{\mu\nu}\, \widetilde{\mathbf F}_{\mu\nu} $ (in component form),
with the Tr being taken over (non-Abelian) gauge group indices,  and 
$\widetilde{\mathbf F}_{\mu\nu} = \frac{1}{2} \varepsilon_{\mu\nu\alpha\beta}\, \mathbf F^{\alpha\beta}$ denoting the dual of the (non-Abelian) gauge-field strength. The quantity $\delta L_{Green-Schwarz}$ denotes the contribution from the variation of the modified KR field. The GS mechanism gives evidence that string theory can be a consistent quantum theory of gravity and gauge fields (at least in $10$ dimensions).

The parts of the action \eqref{sea4} which are quadratic in the $H$-field can be absorbed into a  \emph{generalised connection with torsion}: 
\begin{align}\label{connecttors}
{\overline \Gamma}_{\mu\nu}^{\rho} = \mathring{\Gamma}_{\mu\nu}^\rho + \frac{\kappa}{\sqrt{3}}\, {\mathcal H}_{\mu\nu}^\rho  \ne {\overline \Gamma}_{\nu\mu}^{\rho}\,,
\end{align}
where $\mathring{\Gamma}_{\mu\nu}^\rho = \mathring{\Gamma}_{\nu\mu}^\rho$ is the torsion-free Christoffel symbol. This torsion interpretation holds up to and including terms of $\mathcal O({\alpha^\prime}^2)$  in the effective low-energy string-inspired gravitational actions, which are fourth-order in spacetime derivatives ~\cite{Metsaev:1987zx,Duncan:1992vz} ($\alpha^\prime =M_s^{-2}$ is the Regge slope, with $M_s$ the string scale~\cite{str1,str2}). Therefore up to this order, the string case represents a situation of bosonic fields (in contrast to the previous case of Einstein-Cartan theories) in effectively contorted geometries.

There is an analogous constraint to \eqref{SJ5}, that leads to dynamical axion-like fields, once implemented in the path integral.  
By acting with the exterior derivative on \eqref{csterms} we obtain a Bianchi identity :
\begin{align}\label{bianchi}
   \mathbf d H - \frac{\alpha^\prime}{8\, \kappa} \, {\rm Tr}\Big( R \wedge R - \rm \mathbf F \wedge \mathbf F \Big) =0\,.
   \end{align}
 The second term of the left-hand-side of \eqref{bianchi} is the mixed (gravitational and gauge) anomaly term~\cite{Alvarez-Gaume:1983ihn} .The term $R \wedge R$ is proportional to the gravitational anomaly term appearing in \eqref{axialanom}. 
 
 We use a component form of the Bianchi identity~\cite{decesare,anomalies} 
 \begin{align}\label{modbianchi2}
& \varepsilon_{abc}^{\;\;\;\;\;\;\mu}\, {H}^{abc}_{\;\;\;\;\;\; ;\mu} 
 =  \frac{\alpha^\prime}{32\, \kappa} \, \sqrt{-g}\, \Big(R_{\mu\nu\rho\sigma}\, \widetilde R^{\mu\nu\rho\sigma} -
\mathbf F_{\mu\nu}\, \widetilde{\mathbf F}^{\mu\nu}\Big) \,,
\end{align}
where the gravitationally covariant totally antisymmetric Levi-Civita tensor $\varepsilon_{\mu\nu\rho\sigma}$ is defined as~\cite{eguchi}:
\begin{equation}\label{leviC}
\varepsilon_{\mu\nu\rho\sigma} = \sqrt{-g}\,  \epsilon_{\mu\nu\rho\sigma}, \quad \varepsilon^{\mu\nu\rho\sigma} =\frac{{\rm sgn}(g)}{\sqrt{-g}}\,  \epsilon^{\mu\nu\rho\sigma},
\end{equation}
in terms of the (Minkowski) Levi-Civita constant symbol $\epsilon_{\mu\nu\rho\sigma}$.
 The Bianchi identity \eqref{modbianchi2} is analogous to the conservation of the $S$-torsion constraint \eqref{dS}.
Therefore it can be implemented in the $H$-path-integral of the action \eqref{sea4} in a similar way as in the case of Dirac-fermions in contorted spacetimes, examined above, {\it i.e.} via a 
delta-functional constraint:
\begin{align}\label{Bianchilegenfdre}
\delta\Big(\varepsilon_{abc}^{\;\;\;\;\;\;\mu}\, {H}^{abc}_{\;\;\;\;\;\; ;\mu} -  \frac{\alpha^\prime}{32\, \kappa} \, \sqrt{-g}\, \Big(R_{\mu\nu\rho\sigma}\, \widetilde R^{\mu\nu\rho\sigma} -
\mathbf F_{\mu\nu}\, \widetilde{\mathbf F}^{\mu\nu}\Big)\Big)\,.
\end{align}
 Following the procedure used for implementing a constraint in the discussion of Einstein-Cartan theories,  we have the following effective action, with a (canonically-normalised) dynamical axion field $b$ (called string-model-independent or KR axion~\cite{Svrcek:2006yi}), coupled to the $R_{CS}$~\cite{Duncan:1992vz} (in component form):
\begin{align}
	S=\int d^4x \,\sqrt{-g}\,  \left[\frac{R}{2\kappa^2}-\frac{1}{2}(\partial_\mu b)(\partial^\mu b) - A\, b\,\Big(R_{CS} + \mathbf F_{\mu\nu} \widetilde{\mathbf F}^{\mu\nu} \Big) + \dots \right]\,, \quad   A=\sqrt{\frac{2}{3}}\frac{\alpha^\prime}{48\kappa} = 
  \sqrt{\frac{2}{3}}\frac{M_{\rm Pl}}{48\, M_s^2}   \,,
\label{eq:Action1}  
\end{align}
with $R_{CS}$ being the CS gravitational anomaly term appearing on the right-hand-side of \eqref{axialanom}, with the convention defined in \eqref{RCS} in section \ref{sec:CST} below, which we shall use from now on. 
The model \eqref{eq:Action1} constitutes a CS gravity~\cite{Jackiw:2003pm,Alexander:2009tp},

Thus, from the above discussion, it follows that \emph{both} the Einstein-Cartan theory of fermions and the bosonic gravitational theory of closed strings behave in an equivalent way, leading to a dynamical axion $b$ field representing the effective (string case) or fundamental (Einstein-Cartan case) totally-antisymmetric component of the torsion.  For superradiance, we need the axion $b$ to be massive. This is achieved by turning on appropriate potentials for this field, arising from either   world-sheet instantons~\cite{silver} or non-Abelian gauge-group target-space instantons,  during the post-inflationary epoch of cosmological evolution (as in the case of the running-vacuum, string-inspired cosmological model of \cite{anomalies}).  For us, and in \cite{Dorlis:2025zzz} (for which the current article is a non-trivial extension),  the presence of a non-zero axionic mass is important, but not its microscopic origin.

 We also mention that axions need not be related to quantum anomalies. Massive axions can arise from  specific models of string compactification~\cite{Svrcek:2006yi} (compactification axions); these can couple to gCS terms, with coupling constants $f_a$, which are model dependent and differ, in general, from the KR coupling $A$ in \eqref{eq:Action1}. 
 They are also important in cosmology, playing the role of dark matter~\cite{Marsh:2015xka}, which could also be ultralight.

\section{Review of BH Superradiance formalism}\label{app:BHSuperradiance}

\begin{figure}
    \centering
    \includegraphics[width=0.9\linewidth]{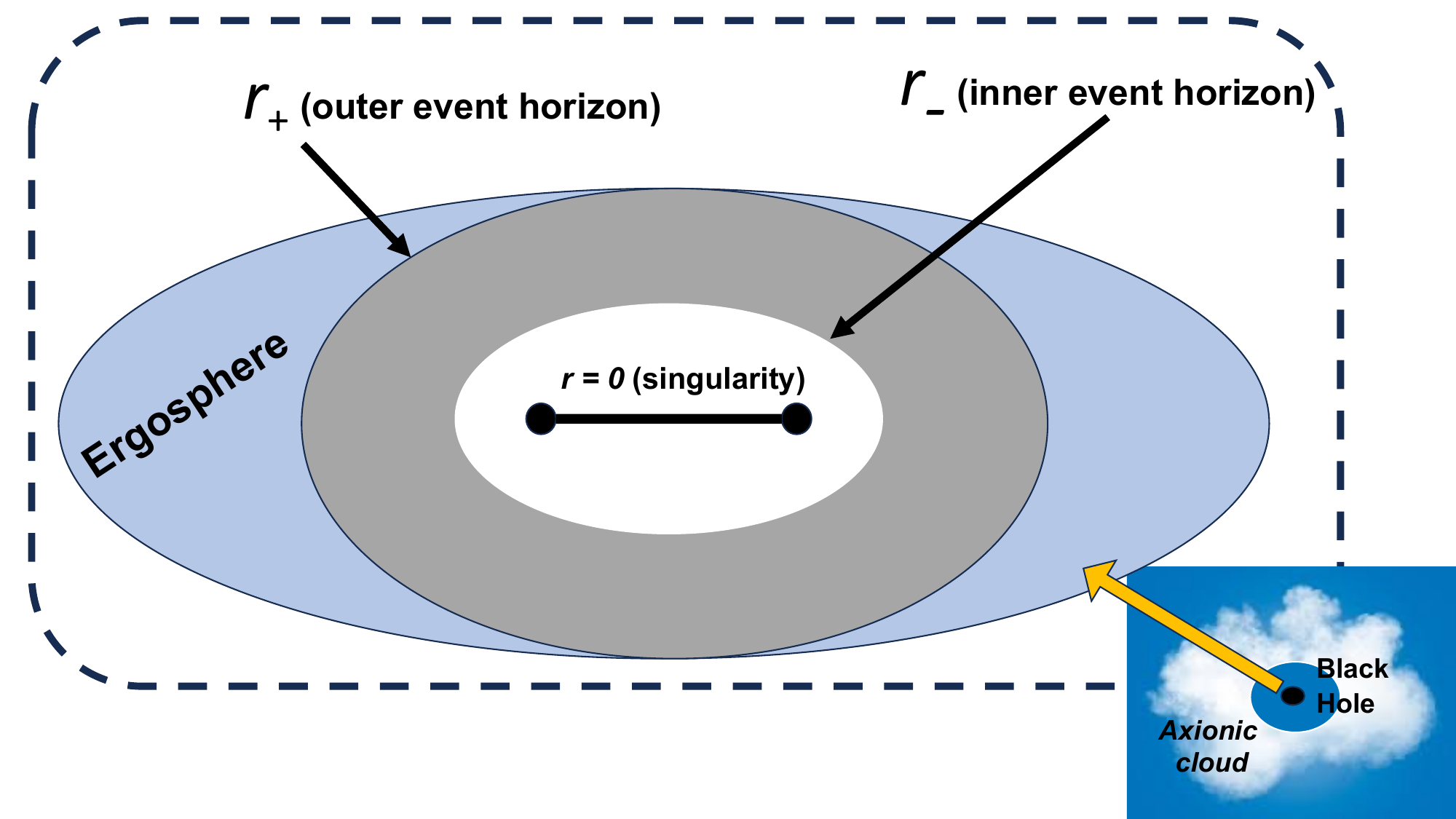}
    \caption{Lower right panel: The Kerr (rotating) BH (black dot) and its axionic cloud. Upper left panel inside the dotted frame: Zoomed-in image of the detailed anatomy of the Kerr BH.}
    \label{axion_cloud}
\end{figure}

In this Appendix, we review in some detail the phenomenon of BH superradiance and its properties, which we made use in this work, and in \cite{Dorlis:2025zzz}.  The most important property, which essentially determines the definition itself of a BH, is the existence of the event horizon, the boundary in spacetime which is responsible for the emergence of causally disconnected regions. Since the BH horizon behaves as a one way membrane in spacetime, it means that vacuum itself is characterized with an intrinsic dissipative mechanism. The other most
important property of rotating BH is the existence of ergoregions, the region between the static limit and the event horizon.  \color{black} In our context, Superradiance will be considered  as a phenomenon in which a (bosonic) classical field interacts with the BH and acquires  an {\it instability}, extracting in this way energy from the BH. In the so called Penrose process,\footnote{\color{black} We stress that, as we have already mentioned in section \ref{sec:BHSuperradiance}, in general, Superradiance is a scattering phenomenon in which the reflected wave
has a larger amplitude in comparison to the incoming wave, and, in this more general sense, there may or may not be instabilities, see, for instance, cases in photonics.\color{black}} rotational energy is extracted from the BH, when classical particles are moving inside the ergoregion/ergosphere \cite{Penrose:1971uk} ({\it cf.}  Figure \ref{axion_cloud}). In the Penrose process, the ergosphere of the Kerr solution~\cite{Kerr:1963ud} is crucial for the energy extraction process, but this is not exactly the case when fields, even classical, are considered.  The distinction of the static limit from the event horizon is a phenomenon of at least of order $\mathcal{O}(\alpha^2)$ ({\it cf.} \eqref{Kerr}), where $\alpha$ is the BH angular momentum over its mass. As such, if the existence of ergoregion was a necessary and sufficient condition for superradiance to occur, then one would expect that the realization happens at least at $\mathcal{O}(\alpha^2)$, too. However, this is not the case as we will see later, since the superradiance condition is at least of order $\mathcal{O}(\alpha)$ (quasi-adiabatic approximation), stemming, although, that rotation is indeed necessary. The conditions for superradiance to occur are studied in \cite{Richartz:2009mi}, where the authors investigate the importance of the event horizon, the dissipation that it induces, while also the significance of the corresponding boundary conditions. \par 
The Kerr line element, in the Boyler-Lindquist coordinates, is given by~\cite{Kerr:1963ud}:
 \begin{equation}
 \begin{aligned}
     ds^2 = -\left(1-\frac{2G \mathcal{M}r}{\Sigma}\right)dt^2 + \frac{\Sigma}{\Delta}dr^2 + \Sigma d\theta^2 + \left(r^2 + \alpha^2 + \frac{2G \mathcal{M}\alpha^2r}{\Sigma}\sin^{2}\theta\right)\, \sin^{2}\theta \, d\varphi^2 - \frac{4 G \mathcal{M}\alpha r\, \sin^{2}\theta }{\Sigma} dtd\varphi \ ,
     \label{Kerr}
     \end{aligned}
 \end{equation}
 where  
 \begin{equation}
 \label{DeltaSigmaAlpha}
      \alpha = \frac{\mathcal{J}_H}{\mathcal{M}} \ , \ \Sigma = r^2 + \alpha^2 \cos^{2}\theta \ \ \text{and} \ \ \Delta = r^2 -2G \mathcal{M}r + \alpha^2 \equiv \left(r - r_+\right)\left(r-r_-\right) \ ,
 \end{equation}
with $\mathcal{M}$ denoting the mass of the BH mass and $\mathcal{J}_H$ the BH's angular momentum, while,
\begin{equation}
    r_\pm= G \mathcal{M} \pm \sqrt{(G \mathcal{M})^2 - \alpha^2}
    \label{r_plus_minus}
\end{equation}
denote the event horizon, $r_+$, and the inner horizon, $r_-$, of the BH ({\it cf.} Figure \ref{axion_cloud}). The radii of \color{black} the ergosurfaces (or radii of the stationary limit surfaces) of the BH, $r^\pm_{erg}$, are defined by the change of $sgn(g_{tt})$, i.e. by $\Sigma-2G \mathcal{M}r=0$, and as such,
\begin{equation}
    r^\pm_{erg}=G \mathcal{M}\pm\sqrt{(G \mathcal{M})^2-\alpha^2\cos^2\theta}\ .
\end{equation}
The study of the superradiant instability when the coupling of the axionic field with the gCS term \eqref{RCS} is considered, is a highly non - trivial task. In such a case, the Einstein's field equations \eqref{grav},\eqref{Axion} in the study of superradiance require a special treatment. In what follows, we begin by considering the coupling as "turned off" to cover analytically the analysis of the BH superradiance in the context of general relativity (GR), and then proceed to address the problems and the solutions when the interaction of the axion with the gCS term  affects the system under consideration.

\subsection{Formation of the Axionic Cloud in the Absence of the CS-Interaction }
In the Kerr background \eqref{Kerr}, 
we first assume the ansatz for the axionic field to take the following form:
\begin{equation}
   b\left(t,r,\theta,\varphi \right)=e^{-i\omega t}e^{im\varphi} S(\theta)R(r)\ .
    \label{axionansatz}
\end{equation}
We note here that, in the study of superradiance, the axion field should be considered as a small perturbation. This means that, since the energy momentum tensor of the axion \eqref{stressb} is a second order quantity in the field $b$, the backreaction of the field to the Kerr geometry would also be a second order effect. Hence, in the context of GR, we can consider the field to propagate in an unperturbed Kerr background. The energy extraction from the BH, will also be of second order (which can be easily obtained as a result of BH mechanics \cite{Bekenstein:1973mi}, where the energy flux from the horizon of the BH will be proportional to the conserved energy momentum tensor of the theory, {\it i.e.} GR with minimally coupled (pseudo-)scalar field, the stress tensor \eqref{stressb}).
The master equation, in the absence of the interaction with the gCS term (i.e. $A=0$) for the Klein-Gordon equation, stemming from \eqref{Axion}, is given by \cite{Brill:1972xj}:
\begin{equation}
    \label{master_equation}
    \begin{aligned}
        &\frac{\partial}{\partial r}\left(\Delta \frac{\partial b}{\partial r}\right) - \frac{\alpha^2}{\Delta}\frac{\partial^2 b}{\partial \varphi^2} - \frac{4 G \mathcal{M} r \alpha}{\Delta}\frac{\partial^2 b}{\partial \varphi \partial t} - \frac{\left(r^2 + \alpha^2\right)^2}{\Delta}\frac{\partial^{2}b}{\partial t^2} - \mu^{2}_b \  r^2  \ b + \frac{1}{\sin\theta}\frac{\partial}{\partial\theta}\left(\sin\theta \frac{\partial b}{\partial \theta}\right)  \\ 
        &+ \frac{1}{\sin^{2} \theta }\frac{\partial^2 b}{\partial \varphi^2} + \alpha^2 \sin^{2}\theta \frac{\partial^2 b}{\partial t^2} - \mu^{2}_b \  \alpha^2  \cos^2 \theta \ b = 0 \ .
    \end{aligned}
\end{equation}

Substituting the ansatz of \eqref{axionansatz} inside \eqref{master_equation}, the {\it homogeneous Klein Gordon equation} \eqref{Axion} can be separated. The equations that govern the evolution of $R_{lm}(r)$ and $S_{lm}(\theta)$ are given by \cite{Brill:1972xj}:
\begin{equation}
    \frac{1}{\sin\theta}\frac{d}{ d\theta}\left[\sin\theta \frac{d}{d\theta}S(\theta)\right] + \left[\alpha^2 \left(\omega^2 - \mu_b^{2}\right)\cos^{2}\theta -\frac{m^2}{\sin^{2}\theta} + \lambda\right]S(\theta)=0 \ ,
    \label{angular_part}
\end{equation}
for the angular part, while the radial part is given by,
\begin{equation}
    \Delta \frac{d}{dr}\left[\Delta \frac{d}{dr}R(r)\right] + \left[\omega^2 \left(r^2 + \alpha^2\right)^{2} - 4\,\alpha\, \omega \ G\mathcal{M}\,  m\,  \ r+ \alpha^2 m^2 -\Delta\left(\mu_b^{2} \ r^2 + \alpha^2 \omega^2 + \lambda\right)\right]R(r)=0 \ .
    \label{radial_full_diffeq}
\end{equation}
The separation constant is denoted by $\lambda$ and corresponds to the eigenvalues of the angular operator,
\begin{equation}
    \hat{\mathbb{L}} = \frac{1}{\sin\theta}\frac{d}{ d\theta}\left[\sin\theta \frac{d}{d\theta}\right] + \left[\alpha^2 \left(\omega^2 - \mu_b^{2}\right)\cos^{2}\theta -\frac{m^2}{\sin^{2}\theta}\right]\ ,
    \label{angularoperator}
\end{equation}
through the following equation,
\begin{equation}
    \hat{\mathbb{L}}\mathcal{P}_{lm}(\theta)=-\lambda \mathcal{P}_{lm}(\theta)\ .
\end{equation}
Massive bosonic fields can form bound states around the BH, which can grow exponentially
 from a seed perturbation through superradiance. This occurs for those modes, for which the frequency $\omega$ in \eqref{axionansatz} becomes complex, 
\begin{equation}
    \omega = \omega_R + i\ \omega_I
    \label{ComplexFrequency}
\end{equation}
with a positive definite imaginary part, 
\begin{equation}
    \omega_I>0\ . 
    \label{superradianceDefinition}
\end{equation}
As we will show later on, superradiance occurs when the condition 
\begin{equation}
\label{superradiance_condition}
    \omega < m \Omega_H
\end{equation}
is satisfied, where $m$ denotes the azimuthal number and $\Omega_H$ is the angular momentum of the outer horizon of the BH,
\begin{equation}
    \Omega_H \equiv \frac{\alpha}{\alpha^2 + r^{2}_+} \ .
\end{equation}
Superradiant instability is amplified when the Compton wavelength of the axion field is comparable with the dimensions of the BH. This relation is expressed through $\mu_b G \mathcal{M}\sim 1$ and this is when the growth of the instability is faster and more efficient \cite{BritoCardoso,arvanitaki}. \par 
However, we will make use of the so-called non-relativistic limit \cite{Detweiller}, given by the following conditions:
\begin{equation}
\text{non-relativistic limit}:\quad  \omega G \mathcal{M}\ll 1 \quad  \text{and}\, \quad \mu_b G \mathcal{M} \ll 1 \label{non-relativistic} \ .
\end{equation}
This approximation, makes use of the separation of scales, i.e. the Compton wavelength of the axion field is much bigger than the size of the BH. Although this approximation produces slower instability rates, superradiance can still be triggered by satisfying the condition $\omega<m\Omega_H$. It is actually proven to be really useful to work in the limits of \eqref{non-relativistic} for a lot of reasons. First of all, it can be used to trace realistic astrophysical environments, with ultra-light bosonic fields (ultra-light fields mean large Compton wavelength, since $\lambda_C \sim 1/\mu_b$), which are the main dark matter candidates. Furthermore, in such a limit, we can deal with the problem analytically, as we shall show later on, providing really useful insights regarding the physical phenomenon. In this limit, the scalar field is more likely to form long-lived bound states around the BH (similar to hydrogen-like energy levels), extracting energy from the BH into these modes.

Assuming from now on the non-relativistic limit, the eigenfunctions of the angular operator \eqref{angularoperator} reduce to the Legendre polynomials,\footnote{In general, notice that we make use of the fact that $\alpha \leq G\mathcal{M}$ and consequently $\alpha \mu_b,\alpha \omega \ll 1$ in the non-relativistic limit.} $P_{lm}(\theta,\varphi)$ with eigenvalues given by $\lambda=l(l+1)$. Notice at this point that this limit mathematically coincides with the slow rotation limit in which one ignores the centrifugal deformations of the spacetime, keeping the leading order $\mathcal{O}(\alpha)$. However, these are physically different schemes, since the slow rotation limit refers only to the spacetime geometric structure, in which the ergosphere is ignored.  At the other scheme, the energy scale of the matter field is constrained to be negligible with respect to the energy scale of the BH. These schemes are different in a highly nontrivial way, from a physical point of view, even if, mathematically, in this case they appear to coincide. Furthermore, since the energy scales of the matter field are far smaller than the energy scale of the BH, issues concerning the back-reaction can be consistently ignored. \par

In what follows, we assume only the non-relativistic limit \eqref{non-relativistic}, as is common in the literature \cite{Detweiller}, implying that the angular part is given by the spherical harmonics, 
\begin{equation}
   e^{i m\varphi} S(\theta) = Y_{lm}(\theta,\varphi) \ ,
\end{equation}
with,
\begin{equation}
    Y^{m}_{l}(\theta,\varphi)=(-1)^{m}\sqrt{\frac{(2l +1)}{4 \pi}\frac{(l-m)!}{(l+m)!}}P^{m}_{l}(\cos\theta)e^{im\varphi} \ , \ \text{for} \ 
\begin{cases}
    l\in \mathbb{N} \\ 
    -l\leq m\leq l
\end{cases}
\end{equation}
satisfying the following orthonormality condition:
\begin{equation}\label{orthonormality_spherical_harmonics}
    \int_{-1}^{1}d(\cos\theta) \int_{0}^{2\pi}d\varphi \ Y^{m}_{l}(\theta,\varphi)\left[Y^{m'}_{l'}(\theta,\varphi)\right]^*  = \delta_{ll'}\delta_{mm'}
\end{equation}
where $\delta_{ij}$ is the Kronecker delta and $d \Omega=\sin\theta \ d\theta d\varphi=-d(\cos\theta)d\varphi$.
The Legendre polynomials are denoted by $P^{m}_{l}(\cos\theta)$, while the radial equation is the same as in \eqref{radial_full_diffeq} with specified $\lambda=l(l+1)$. For the radial part, an analytic solution lacks in the literature and actually seems unlikely to be found. However,  in the non-relativistic limit \eqref{non-relativistic}, an analytic \textit{treatment} is achievable \cite{Detweiller}, through the method of matching asymptotic expansions. The two asymptotic regions that we have to match are the {\it far-zone} from the BH, and the {\it near-horizon-zone}. Through this treatment, we can obtain an estimate of the imaginary part of the frequency analytically, by carefully matching the relative solutions in the {\it intermediate-zone}, which lies between the near and far zones.

\subsubsection{\textbf{The Far Horizon Region}}
The far horizon region formally can be obtained by assuming $r/G\mathcal{M}\gg1$ (and as a consequence, since $\alpha \leq G\mathcal{M}$, this relation implies $r/\alpha\gg 1$). In such a region, the effects of the curvature have to take the form of an effective potential with the radial differential operator in \eqref{radial_full_diffeq} resembling that of the Laplacian operator.  By performing this, while collecting the relevant terms all together, we obtain, 
\begin{equation}
    \frac{1}{r^2}\frac{d}{dr}\left[r^2\frac{d}{dr}R\right] +\omega^2\left\{\left(1+\frac{\alpha^2}{r^2}\right)^2   -4\ m\ \frac{1}{\omega r}\frac{\alpha}{r}\frac{G \mathcal{M}}{r}+m^2\frac{\alpha^2}{r^2}-\left( 1-2\frac{G \mathcal{M}}{r} + \frac{\alpha^2}{r^2}    \right)    \left( \frac{\mu_b^2}{\omega^2} +   \frac{\alpha^2}{r^2} +\frac{l(l+1)}{\omega^2 r^2}     \right)  \right\} R=0\ .
    \label{full_expanded_radial_difeq}
\end{equation}
 Thus, upon assuming r large compare to $G \mathcal{M}$ and thus keep terms  up to $\mathcal{O}(G \mathcal{M}/r)$, we finally obtain,
\begin{equation}
    \frac{d^2}{dr^2}(r R) + \left[\omega^2 - \mu^{2}_b +\frac{2 G \mathcal{M} \mu_{b}^2}{r}-\frac{l(l+1)}{r^2}  \right]r R =0
\end{equation}
in which the term containing the $l(l+1)/\omega^2 r^2$ survives, since the condition $G \mathcal{M}/r\ll1$ does not necessarily imply $1/\omega r\ll 1$.  
Moreover, via the following redefinitions,\footnote{Note here that without loss of generality, we take $k=+i\sqrt{\omega^2-\mu^2}$, implying that  $-\pi\leq \rm{arg}\, x \leq \pi$.}
\begin{align}
    k^2&\equiv \mu_{b}^2 - \omega^2\label{kredefinition}\\
    x&\equiv 2kr\label{xredefinition}\\
    n&\equiv G \mathcal{M}\mu^{2}_b/k\label{nuredefinition}\ ,
\end{align}
the above equation \eqref{full_expanded_radial_difeq} becomes \cite{Detweiller}:
\begin{equation}
    \frac{d^2}{dx^2} \chi_{l}+\left[-\frac{1}{4}+\frac{n}{x} - \frac{l(l+1)}{x^2}\right] \chi_l=0\ .
\label{hydrogen_like_atom_radial_equation}
\end{equation}
in which we defined, 
\begin{equation}
    \chi_l \equiv x R\ .
\end{equation}
We can examine the behavior of \eqref{hydrogen_like_atom_radial_equation} in the respective regions $\vert x\vert \to0$ and $\mathcal{R}e \,(x)\to +\infty$. For the latter, i.e. $\mathcal Re \,(x)  \to +\infty$, we observe that the dominant contribution of the solution of equation \eqref{hydrogen_like_atom_radial_equation} is exponentially suppressed as $\chi_l(x) \sim  e^{\pm \frac{x}{2}}$. As such, one expects that the solution has the form,
\begin{equation}
    \chi_l (x)= f(x)e^{-x/2}\, ,
\end{equation}
with $f(x)$ an arbitrary analytic function. Since it is analytic its contribution to the limit of $x\to +\infty$ has to be some power of $x$, i.e.
\begin{equation}
    \chi_l\to x^c e^{\pm x/2}\ .
\end{equation}
Indeed, inserting the above into \eqref{hydrogen_like_atom_radial_equation}, one obtains that for large $x$ the power has to be $c=\mp n$. Thus, the asymptotic behavior reads, $\chi_l\approx x^{\mp n}e^{\pm x/2}$, which due to boundary conditions only the damped solution is valid,
\begin{equation}
    \chi_l\approx x^{n}e^{-x/2}\ . 
\end{equation}
Taking now the limit $\vert x\vert \ll 1$, we we can verify that it admits solutions of the form $\sim x^\gamma$, with $\gamma=-l$ or $\gamma=l+1$. 
Then, without loss of generality, we assume solutions of the form;
\begin{equation}
    \label{form_of_chi_l}
    \chi_l (x) = e^{- \frac{x}{2}}x^\gamma w_l(x) \ ,
\end{equation}
with $w_l$ a function of $x$ to be determined, satisfying the equation,
\begin{equation}
    x\frac{d^2}{dx^2}w_l(x) + (2\gamma - x)\frac{d}{dx}w_l(x) - \left(\gamma - n\right)w_l(x) =0\, 
    \label{Kummers}
\end{equation}
with the specified boundary conditions, 
\begin{align}
    &w_l(x)\sim x^{-\gamma+n}\ , \ \text{for}\ x\to +\infty\ ,\label{AsymptKum}\\
    &w_l(x)\sim C, \ \text{for}\ x\to 0\ ,
\end{align}
where $C$ is a finite constant.  Equation \eqref{Kummers} is the Kummer's equation, with two independent solutions. The one that satisfies the boundary condition \eqref{AsymptKum} is the Tricomi hyper-geometric function \cite{Mathews}, $ U(\gamma-n, 2\gamma,x)$.
Thus, the general solution reads,
\begin{equation}
    R(x) = e^{-x/2} \left[ \mathcal{C}_1  x^l U(l+1-n, 2l+2,x) +\mathcal{C}_2  x^{-l-1}  U(-l-n, -2l,x) \right] \ ,
\end{equation}
with $\mathcal{C}_1,\mathcal{C}_2$  arbitrary integration constants. However, upon using the Kummer's transformation \cite{Abramowitz,Mathews}, 
\begin{equation}
    U(a,b,z)=z^{1-b} U(1+a-b,2-b,z)\ ,
\end{equation}
we find that, 
\begin{equation}
    x^{-l-1}  U(-l-n, -2l,x) =  x^l U(l+1-n, 2l+2,x) \ ,
\end{equation}
and the solution reduces to, 
\begin{equation}
    R(x)  = \mathcal{C}\ x^l\ e^{-x/2} \ U(l+1-n,2l+2,x)= \frac{\mathcal{C}}{x}\mathcal{W}_{n,l+1/2}(x) \ ,
\end{equation}
where $\mathcal{W}_{n,\ l+1/2}(x)$ is the Whittaker's function \cite{Abramowitz},
\begin{equation}
    \mathcal{W}_{\kappa,\mu}(z)=e^{-z/2}z^{\frac{1}{2}+\mu}U\left(\frac{1}{2}+\mu-\kappa,1+2\mu,z\right),\ \ -\pi<{\rm arg}\ z \leq \pi\ .
\end{equation}
For the hydrogen atom, $l+1-n$ corresponds to the energy levels and has to be a non-positive integer, $l+1-n=-n_r\ ,n_r\in\mathbb{N}$,  in order to have a well defined (non-divergent) limit as $x\to 0$. Then, the integer $n$ denotes the energy levels of the hydrogen atom (i.e. $n$ is the principal quantum number, $n\geq  l+1$). So, since we expect imaginary parts on the frequency \eqref{ComplexFrequency}, this has to be reflected on the imaginary part to the energy levels. Thus, we may introduce an imaginary $\delta n$, through  the relation $n_r+\delta n=n-l-1$,
\begin{equation}
    R(x)  = \mathcal{C}\ x^l\ e^{-x/2} U(-n_r-\delta n, 2l+2,x)\ ,
\end{equation}
with $n_r$ a non-negative integer, $n_r \in\mathbb{N}$. 
The asymptotic behavior for small $\vert x\vert$, reads, 
\begin{equation}
    R(x)\approx \mathcal{C}\left(  x^{-l-1} \frac{\Gamma(2l+1)}{\Gamma(-n_r-\delta n)}  +  x^l \frac{\Gamma(-1-2l)}{\Gamma(-1-2l-n_r-\delta n)}         \right)\ . 
\end{equation}
To the above, since both $\delta n\ll1 $ and $x\ll 1 $, we have to keep up to $\mathcal{O}(\delta n)$ to the first term, while up to $\mathcal{O}(\delta n^0)$ to the second one. Then, through an order of magnitude correspondence $\delta n\sim x^{2l+1}\ll 1$ both terms are of the same order of magnitude. Using the reflection formula \cite{Abramowitz}, 
\begin{equation}
\Gamma(x)\Gamma(-x)=-\frac{\pi}{x\sin(\pi x)} \ ,
\label{reflection_formula}
\end{equation}
we can find ,
\begin{align}
    \frac{1}{\Gamma(-n_r-\delta n)}&= (-1)^{n_r+1} n_r!\, \delta n + \mathcal{O}(\delta n^2)\ ,\\
    \frac{\Gamma(-1-2l)}{\Gamma(-1-2l-n_r)}&=   \frac{(1+2l+n_r)!}{(1+2l)!}(-1)^{n_r} + \mathcal{O}(\delta n)\ . 
\end{align}
Finally, we obtain (remember that $x\equiv 2kr$),
\begin{equation}
    R(x)\approx \mathcal{C}\left[  (-1)^{n_r+1}(2l)!\ n_r! \ \delta n\    (2kr)^{-l-1}        +   (-1)^{n_r} \frac{(1+2l+n_r)!}{(1+2l)!} (2kr)^l\right]\ .
\label{FromFartoNear}
\end{equation}

\subsubsection{\textbf{The Near - Horizon Region}}
Turning back to equation \eqref{radial_full_diffeq}, we concentrate on scales much larger than that of the matter field's energy scales, i.e. $r\ll max\{l/\vert\omega\vert,l/\mu_b\}$, which in view of the non - relativistic limit \eqref{non-relativistic}, is the opposite assumption\footnote{ We would like to leave a comment here, regarding the \textit{near - horizon regime}, expressed through the condition $r\ll max\{l/\vert\omega\vert,l/\mu_b\}$ and its compatibility with the \textit{far - horizon region}, i.e. $r\gg G\mathcal{M}$. Under the assumption that the order of magnitude of $\mu$ and $\omega$ are in the same range, we deduce from \eqref{kredefinition} that the order of magnitude of $\vert k \vert$ is smaller than $\mu_b,\vert \omega \vert$, i.e. the near - horizon region can be written as $k r \ll max\{l \vert k \vert /  \vert\omega\vert,l \vert k \vert /  \mu_b\} < l$. Considering that $l \geq 1$ in the near - horizon zone, we can express the near - horizon treatment through the condition $\vert k \vert r \ll 1$. Hence, from the non - relativistic conditions \eqref{non-relativistic}, we can deduce that $G\mathcal{M} \ll 1/\mu_b \leq l/ \mu_b$ and also $G\mathcal{M} \ll 1/\vert \omega \vert  \leq l/ \vert \omega \vert$, or, combined, $G\mathcal{M} \ll max\left(l/\mu_b,l/\vert \omega \vert \right)$. We point out here that, under the above estimates, the \textit{intermediate - zone} where the matching/overlapping of the solutions will take place is expressed as $G\mathcal{M} \ll r \ll max\left(l/\mu_b,l/\vert \omega \vert \right)$. } with respect to $r\gg G\mathcal{M}$.
Thus, since the latter denotes the far horizon zone, the former should be relevant in a near horizon regime. On account of this, we also assume that $r-r_+\sim\epsilon\ll 1$. We will keep terms only up to first order with respect to the aforementioned small quantities . Hence, for the first term of \eqref{radial_full_diffeq}, we obtain, 
\begin{equation}
\begin{aligned}
    \omega^2(r^2+\alpha^2)^2&= \omega^2((r-r_+)(r+r_+)+r_+^2+\alpha^2)^2 \approx \omega^2(r_+^2+\alpha^2)^2\\
    &\approx \omega^2\left( (G\mathcal{M})^2 +(G\mathcal{M})^2-\alpha^2+2(G\mathcal{M})^2\sqrt{1-\alpha^2/(G\mathcal{M})^2}  +  \alpha^2 \right)^2\\
    &\approx \omega^2 \left(2(G\mathcal{M})^2+2(G\mathcal{M})^2   \sqrt{1-\alpha^2/(G\mathcal{M})^2}  \right)^2\\
    &\approx \left(2\omega (G\mathcal{M}) r_+\right)^2\ , 
    \end{aligned}
    \label{first_term_remaining_part}
\end{equation}
for which, the non-relativistic limit \eqref{non-relativistic} does not necessarily implies that is small, i.e. of $\mathcal{O}(\epsilon^2)$. From the first to the second line, we omitted the terms $(r-r_+)(r+r_+)$, since every term containing this term is of order $\mathcal{O}(\epsilon^2)$, in view of the relativistic limit together with the fact that $r_+,\alpha\sim G\mathcal{M}$ and as such every term containing this term has orders\footnote{In here we have that $r+r_+\sim G\mathcal{M}$.} $\mathcal{O}(\epsilon^4)$ and a term of $\mathcal{O}(\epsilon^3) (r+r_+)\sim \mathcal{O}(\epsilon^3)G\mathcal{M}$, which is subdominant with respect to the final term in \eqref{first_term_remaining_part}, which is $\mathcal{O}(\epsilon^2)(G\mathcal{M})^2$.  For the second term of \eqref{radial_full_diffeq}, we obtain,
\begin{equation}
    -4\ \alpha\ \omega \ G\mathcal{M} \ m \ r = -4\ \ m\  \alpha\ \left( \omega \ G\mathcal{M}  \ (r- r_+)+ \omega \ G\mathcal{M}  \ r_+ \right)\approx-2( m \alpha)(2G\mathcal{M} \ \omega\ r_+)\ . 
\end{equation}
For the final term, we have,
\begin{equation}
-\Delta(\mu_b^2r^2+\alpha^2\omega^2+l(l+1))\approx  -\Delta \omega^2 \left(\frac{\mu^{2}_b}{\omega^2} \ r^2 + \alpha^2 + \frac{l(l+1)}{\omega^2} \right) \approx-\Delta l(l+1)\ .
\end{equation}
 where, we used the assumption that the order of magnitude of $\mu_b$ and $\vert \omega \vert $ is the same, while also the fact that $\alpha\leq G\mathcal{M} \ll max(l/\mu_b,l/\vert \omega \vert)$. 
Moreover, we define the new variable, $z$, by the following relation,
\begin{equation}
    z=\frac{r-r_+}{r_+-r_-}
    \label{newvariable}\ ,
\end{equation}
for which $\Delta= (r_+-r_-)^2z(z+1)$, while,
\begin{equation}
    \Delta\frac{d}{dr}\left[\Delta \frac{d}{dr}R\right]= (r_+-r_-)^2z(z+1)\frac{d}{dz}\left[z(z+1)\frac{d}{dz}R\right]\ .
\end{equation}
Thus, combining all of the above, equation \eqref{radial_full_diffeq} reads \cite{Detweiller}, 
\begin{equation}
    z(z+1)\frac{d}{dz}\left[z(z+1)\frac{d}{dz}R(z)\right] +\left[ P^2 -z(z+1)l(l+1)   \right]R(z)=0
    \label{radia_Near_Horizon}
\end{equation}
where, 
\begin{equation}
    P=\frac{\alpha m - 2G\mathcal{M}\omega r_+}{r_+-r_-}\ . 
    \label{P_Definition}
\end{equation}
Taking the following transformation, 
\begin{equation}
    R(z)=\left(\frac{z}{z+1}\right)^{i P}w(z)\ ,
    \label{Rtransf}
\end{equation}
the above equation reduces to the following equation, 
\begin{equation}
   z(z+1)\frac{d^2}{dz^2}w(z)+(1+2 i P+2z)\frac{d}{dz}w(z)-l(l+1)w(z)=0
    \label{near_to_hypergeometric_diff_Equation}\ .
\end{equation}
Making the change of variable, 
\begin{equation}
z\to y=z+1\ ,
\end{equation}
we obtain the hyper-geometric differential equation (eq. 15.5.1 in \cite{Abramowitz}), 
\begin{equation}
    y(1-y)\frac{d^2}{dy^2}w(y)+(1-2 i P-2y)\frac{d}{dy}w(y)+l(l+1)w(y)=0
    \label{hypergeometric_diff_equation}\ .
\end{equation}
The corresponding hyper-geometric series function is, 
\begin{equation}
F(a,b,c;y)=F(-l,l+1,1-2iP;z+1)  
\label{abc}
\end{equation}
where $a=-l,\ b=l+1\,\text{and}\ c=1-2iP$. We know that, around $y=1 \ (z=0)$, the solution has to be of the form of an ingoing wave, while the solution to the neighborhood of $y=z=\infty$ corresponds to the appropriate one, for which we are about to make the matching with  \eqref{FromFartoNear}. The general solution is therefore given by:
\begin{equation}
    R(z)=\left(\frac{z}{z+1}\right)^{i P} F\left(-l,l+1,1-2iP;z+1\right) \ ,
    \label{general_solution_near}
\end{equation}
where $F(-l,l+1,1-2iP;z+1)$ is any solution of the hyper-geometric equation \eqref{hypergeometric_diff_equation}.
The regions arround the two (regular) singular points $y=1,+\infty$ of the above hyper-geometric equation correspond to the two neighborhoods that we are interested in for determining the solution. In the neighborhood of $y=1$, the solution has to be of the form of an ingoing wave, while the solution to the neighborhood of $y=+\infty$ is the
appropriate one with which we are about to make the matching with \eqref{FromFartoNear}. Two forms of the general solution \eqref{hypergeometric_diff_equation},  appropriate for the expansion around $y=1,+\infty$, are respectively given by \cite{Bateman:100233},
\begin{align}
w(y)=&\ c_1u_2(y)+c_2u_6(y)\label{general_solution_hyper1}\\
w(y)=&\ \tilde{c}_1 u_3(y)+\tilde{c}_2 u_4(y)
\label{general_solution_hyper2} \ .  
\end{align}
in which we follow the convention of \cite{Bateman:100233},
\begin{align}
    &u_2(y)= F(a,b,a+b+1-c;1-y)\\
    &u_6(y)=(1-y)^{c-a-b}F(c-a,c-b,c+1-a-b,1-y)\\
    &u_3(y)=(-y)^{-a}F(a,a+1-c,a+1-b,y^{-1})\\
    &u_4(y)=(-y)^{-b}F(b+1-c,b,b+1-a,y^{-1})
\end{align}
are independent solution of the hyper-geometric equation \eqref{hypergeometric_diff_equation}. The general solution \eqref{general_solution_hyper2} is the appropriate one for the matching with the far horizon regime and  what we actually need to determine, are the the constants $\tilde{c}_{1,2}$, since,
\begin{equation}
    w(z)\approx \tilde{c}_1(-z)^{l}+\tilde{c}_2(-z)^{-l-1}\ ,\ \ z\gg 1\  ,
\end{equation}
which has to be matched with \eqref{FromFartoNear}. The hyper-geometric series appeared in \eqref{general_solution_hyper1} and \eqref{general_solution_hyper2}, have a common domain of existence and as such they are connected by the following linear relations with constant coefficients \cite{Bateman:100233},\footnote{Attention here; in \cite{Bateman:100233} there is a typo in the relation between $u_2(y)$ and $\left(u_3,u_4\right)$. Specifically, the $\Gamma(b)$ in the above denominator is written as $\Gamma(c)$.}
\begin{align}
    &u_2(y)= \frac{\Gamma(a+b+1-c)\Gamma(b-a)}{\Gamma(b+1-c)\Gamma(b)}e^{-i\pi a}u_3(y)
    +\frac{\Gamma(a+b+1-c)\Gamma(a-b)}{\Gamma(a+1-c)\Gamma(a)}e^{-i\pi b}u_4(y)\\
  &u_6(y)= \frac{\Gamma(c+1-a-b)\Gamma(b-a)}{\Gamma(1-a)\Gamma(c-a)} e^{-i\pi(c-b)} u_3(y)+\frac{\Gamma(c+1-a-b)\Gamma(a-b)}{\Gamma(1-b)\Gamma(c-b)}e^{-i\pi(c-a)}u_4(y) \ .
  \end{align}
Using these relations, one can easily obtain, 
\begin{align}
    \tilde{c}_1=&c_1\frac{\Gamma(a+b+1-c)\Gamma(b-a)}{\Gamma(b+1-c)\Gamma(b)}e^{-i\pi a}+c_2\frac{\Gamma(c+1-a-b)\Gamma(b-a)}{\Gamma(1-a)\Gamma(c-a)} e^{-i\pi(c-b)}\\
    \tilde{c}_2=&c_1 \frac{\Gamma(a+b+1-c)\Gamma(a-b)}{\Gamma(a+1-c)\Gamma(a)}e^{-i\pi b}+c_2\frac{\Gamma(c+1-a-b)\Gamma(a-b)}{\Gamma(1-b)\Gamma(c-b)}e^{-i\pi(c-a)}\ .
\end{align}
From the equations \eqref{general_solution_hyper1} and \eqref{Rtransf}, it is clear that near the horizon (remember that $F(a,b,c;0)=1,\forall a,b,c$),
\begin{equation}
    R(z)\approx c_1\, z^{iP} + c_2 \,z^{-iP}\ .
\end{equation}
As such, taking only the ingoing solution for the near horizon behavior $\sim z^{iP}$, requires, on account of the relation $z^{\pm iP}=e^{\pm iP\log z}$, $c_2=0$, and then,  
\begin{align}
\tilde{c}_1=&c_1 (-1)^l\frac{\Gamma(2l+1)\Gamma(1+2iP)}{\Gamma(l+1+2iP)\Gamma(l+1)}    \\
\tilde{c}_2=&c_1(-1)^{l+1}\frac{\Gamma(-2l-1)\Gamma(1+2iP)}{\Gamma(-l+2iP)\Gamma(-l)}\ ,
\end{align}
in which we used \eqref{abc} for the coefficients of the particular hyper-geometric differential equation \eqref{hypergeometric_diff_equation}.
Finally, we find for $ r_+\ll r\ll max\{l/\omega,l/\mu_b\}$,
\begin{equation}
    R(r)\approx c_1\left[\frac{\Gamma(2l+1)\Gamma(1+2iP)}{\Gamma(l+1+2iP)\Gamma(l+1)(r_+-r_-)^l}  r^l  + \frac{\Gamma(-2l-1)\Gamma(1+2iP)(r_+-r_-)^{l+1}}{\Gamma(-l+2iP)\Gamma(-l)} r^{-l-1}\right]\ 
\end{equation}
Thus, the matching with \eqref{FromFartoNear}, yields, 
\begin{align}
    \mathcal{C}\ (-1)^{n_r+1}(2l)!\ n_r! (2k)^{-l-1} \ \delta n &= c_1\  \frac{\Gamma(-2l-1)\Gamma(1+2iP)(r_+-r_-)^{l+1}}{\Gamma(-l+2iP)\Gamma(-l)}\\
   \mathcal{C}\  (-1)^{n_r} \frac{(1+2l+n_r)!}{(1+2l)!} (2k)^l&=c_1\ \frac{\Gamma(2l+1)\Gamma(1+2iP)}{\Gamma(l+1+2iP)\Gamma(l+1)(r_+-r_-)^l} \ .
\end{align}
Thus, by dividing the above equations, we obtain, 
\begin{equation}
    \delta n=- \frac{(2l+1+n_r)!}{n_r!}\frac{\Gamma(-2l-1)\Gamma(l+1)}{\Gamma(-l)(2l)!(2l+1)!\Gamma(2l+1)} \frac{\Gamma(l+1+2iP)}{\Gamma(-l+2iP)}\left[2k(r_+-r_-)\right]^{2l+1}\ .
\end{equation}
The reflection formula of the Gamma function \eqref{reflection_formula}, yields, 
\begin{equation}
    \frac{\Gamma(-2l-1)}{\Gamma(-l)}= (-1)^{l+1}\frac{l!}{(2l+1)!}\ . 
\end{equation}
Moreover, using the standard  relation $\Gamma(1+x)=x\Gamma(x)$, we find retrospectively that, 
\begin{equation}
\Gamma(1+l+2iP)=\Gamma(2iP)\prod_{j=0}^l(j+2iP)\ ,\forall l\in\mathbb{N}\ .
\end{equation}
From the reflection formula we also find, 
\begin{equation}
\begin{aligned}
    \Gamma(-n_r+z)=&-\frac{\pi}{\sin\left[\pi(n_r-z)\right](n_r-z)\Gamma(n_r-z)}=(-1)^{n_r}\frac{\pi}{\sin(\pi z)\Gamma(-z)}\prod_{j=0}^{n_r}(j-z)^{-1}\\
    =&(-1)^{n_r+1}\ z\ \Gamma(z)\prod_{j=0}^{n_r}(j-z)^{-1}=(-1)^{n_r}\Gamma(z)\prod_{j=1}^{n_r}(j-z)^{-1}
   \end{aligned} 
\end{equation} 
in which we substituted,
\begin{equation}
    (n_r-z)\Gamma(n_r-z)=\Gamma(-z)\prod_{j=0}^{n_r}(n_r-z)\ ,
\end{equation}
which can be easily verified retrospectively, as previously.  Thus,
\begin{equation}
    \Gamma(-l+2iP)=(-1)^l\Gamma(2iP)\prod_{j=1}^l(j-2iP)^{-1}
\end{equation}
and gives us, 
\begin{equation}
\frac{\Gamma(l+1+2iP)}{\Gamma(-l+2iP)}= (-1)^l 
 \ 2iP \prod^{l}_{j=1}(j^2 + 4P^2)\ ,
\end{equation}
which finally, yields,
\begin{equation}
\label{delta_nu_expression}
    \delta n =2iP \frac{(2l+1+n_r)!}{n_r!}\left(\frac{l!}{(2l)!(2l+1)!}\right)^2 \left[2k(r_+-r_-)\right]^{2l+1}\prod^{l}_{j=1}(j^2 + 4P^2)\ .
\end{equation}
in agreement with \cite{Detweiller}.

\subsection{The Imaginary Part of the Frequency and Instabilities}

In the next step, we will derive analytically the imaginary part of the frequency \cite{Yang_2023}, and its relation with $\delta n$ of \eqref{delta_nu_expression}. From the redefinitions given in \eqref{kredefinition}, \eqref{xredefinition}, \eqref{nuredefinition}, one can deduce that. 
\begin{equation}\label{help0}
    k^2 = \mu^2_b - \omega^2 = \frac{(G \mathcal{M})^2 \mu^{4}_b}{n^2} = \frac{(G \mathcal{M})^2 \mu^{4}_b}{\left(n_r+l+1+\delta n\right)^2} \ .
\end{equation}
Substituting for the complex quantity $\delta n = \delta n_R + i \delta n_I$, we obtain:
\begin{equation}\label{help1}
    \mu^2_b - \omega^2 = \frac{(G \mathcal{M})^2 \mu^{4}_b}{n^2} = \frac{(G \mathcal{M})^2 \mu^{4}_b}{\left(n_r+l+1+\delta n_R + i \delta n_I\right)^2} \ .
\end{equation}
The modulus of $\delta n$ is a small quantity, which is translated to $\vert \delta n_R \vert , \vert \delta n_I \vert \ll 1$. Expanding \eqref{help1} to the first order of these small quantities, and also writing the complex frequency in the form $\omega = \omega_R + i \ \omega_I$, one obtains the following relation for the real part of the frequency:
\begin{equation}
\begin{aligned}\label{real_part_frequency}
    \omega_R\approx  \ \mu_b \sqrt{1-\frac{(G \mathcal{M})^2 \mu^{2}_b}{\left(n_r+l+1\right)^2}} \overset{\mu_b G \mathcal{M} \ll 1}{\approx} \mu_b \left(1-\frac{1}{2}\frac{(G \mathcal{M})^2 \mu^{2}_b}{\left(n_r+l+1\right)^2}\right) \ ,
\end{aligned}
\end{equation}
where, from the last line, one can see that $\omega_R < \mu_b$. This means that the particle is at a \textit{quasibound state}, i.e. an imaginary part to the energy appears, yielding the strength of the instability. 

For the imaginary part of the frequency we get:
\begin{equation}\label{imaginary_part_frequency_help0}
\begin{aligned}
    \omega_I   \approx  \frac
    {\delta n_I}{G \mathcal{M}}\left(\frac{G \mathcal{M} \mu_b}{n_r+l+1}\right)^3 \left[1 + \frac{(G \mathcal{M})^2 \mu^{2}_b}{2\left(n_r+l+1\right)^2}\right]\approx \frac{\delta n_I}{G \mathcal{M}}\left(\frac
    {\mu_b G \mathcal{M}}{n_r+l+1}\right)^3\ll \omega_R \ ,
\end{aligned}
\end{equation}
where, in the last step, we kept only terms up to the dominant contribution with respect to the small quantity $\mu_b G \mathcal{M}$. Now, in order to find $\omega_I$, we need to calculate the imaginary part of \eqref{delta_nu_expression}, denoted by $\delta n_I$. By looking at the expression for $\delta n$, i.e. \eqref{delta_nu_expression}, we can see that this quantity depends on the product of two other complex (by definition) quantities, $P=P_R + i P_I$ and $k=k_R + ik_I$. In the final expression, relative products between the real and imaginary parts of these two quantities will appear, and what is important is to determine which one is dominant regarding its contribution to $\delta n_I$. 

One can easily see that $\vert k_R \vert \gg \vert k_I \vert$ as well as $\vert P_R \vert \gg \vert P_I \vert$. Hence, we can conclude that the dominant contribution to the imaginary part $\delta n_I$ comes from  products between $k_R$ and $P_R$, which are given by: 
\begin{align}
    \label{real_part_P}P_R \approx & \ \frac{\alpha m - 2 \mu_b \ G \mathcal{M} r_+}{r_+ - r_-} \ ,  \ m \in \mathbb{Z}\\
    \label{real_part_k}k_R \approx & \ \frac{G \mathcal{M} \mu_b}{n_r+l+1}\mu_b \ ,
\end{align}
where we used \eqref{real_part_frequency} and substituted $\omega \approx \mu_b$, in accordance with the non-relativistic limit $\mu_b G \mathcal{M} \ll 1$. The positivity of $P_R$, using eq. \eqref{real_part_P}, leads to the following condition:
\begin{equation}
    \left(\frac{G \mathcal{M}}{a}\right)<\frac{m}{2 \mu_b  r_+} \ , \  m \in \mathbb{Z} .
\end{equation}
Hence, for the imaginary part of the frequency, we obtain the following expression \cite{Detweiller} from \eqref{imaginary_part_frequency_help0}:
\begin{equation}
    \label{imaginary_part_frequency_help}
    \omega_I \approx \frac{(G \mathcal{M})^2 \mu^{3}_b}{\left(n_r+l+1\right)^3}\delta n_I \ ,
\end{equation}
where, as we mentioned above, $\delta n_I$ comes from \eqref{delta_nu_expression} after substituting $P,k$ with \eqref{real_part_P} and \eqref{real_part_k} respectively. The derivation of the result is presented below:
\begin{equation}
\begin{aligned}
   \label{imaginary_part_frequency_help2} \omega_I \approx & \  \frac{(G \mathcal{M})^2 \mu^{3}_b}{\left(n_r+l+1\right)^3}2 P_R \  2^{2l+1} \ k^{2l+1}_R\left(r_+ - r_-\right)^{2l+1}\frac{\left(2l+n_r+1\right)!}{n_r!}\left[\frac{l!}{(2l)!(2l+1)!}\right]^2\prod^{l}_{j=1}\left(j^2 + 4P^{2}_R\right) \\
    =& \  2^{2l+2}\frac{(G \mathcal{M})^2 \mu^{3}_b}{\left(n_r+l+1\right)^3}\left(\frac{\alpha m - 2 \mu_b \ G \mathcal{M} r_+}{r_+ - r_-}\right)\left(\frac{(G \mathcal{M})^{2l+1}\mu^{4l+2}_b}{\left(n_r+l+1\right)^{2l+1}}\right)\left(r_+ - r_-\right)^{2l+1} \frac{\left(2l+n_r+1\right)!}{n_r!}\times \\
    \times & \left[\frac{l!}{(2l)!(2l+1)!}\right]^2\prod^{l}_{j=1}\left(j^2 + 4\left(\frac{\alpha m - 2 \mu_b \ G \mathcal{M} r_+}{r_+ - r_-}\right)^{2}\right) \\ 
    \nonumber \overset{\eqref{r_plus_minus}}{=}& \ 2^{2l+2}\frac{G \mathcal{M}^{2l+3} \mu^{4l+5}_b}{\left(n_r+l+1\right)^{2l+4}} G \mathcal{M}\left(\frac{am}{G \mathcal{M}}-2\mu_b r_+\right) \frac{\left(2l+n_r+1\right)!}{n_r!}\left[\frac{l!}{(2l)!(2l+1)!}\right]^2\times \\
    \times & \prod^{l}_{j=1}\left(j^2 \ 4(G \mathcal{M})^2 \left(1-\frac{\alpha^2}{(G \mathcal{M})^2}\right) + 4(G \mathcal{M})^2 \left(\frac{\alpha m}{G \mathcal{M}} - 2\mu_b r_+\right)^{2}\right) \\ 
    \nonumber = & \ 2^{2l+2}\frac{(G \mathcal{M})^{2l+4} \mu^{4l+5}_b}{\left(n_r+l+1\right)^{2l+4}}\left(\frac{am}{G \mathcal{M}}-2\mu_b r_+\right)  \frac{\left(2l+n_r+1\right)!}{n_r!}\left[\frac{l!}{(2l)!(2l+1)!}\right]^2\times \\
    \times & 2^{2l} (G \mathcal{M})^{2l}\prod^{l}_{j=1}\left(j^2 \ \left(1-\frac{\alpha^2}{(G \mathcal{M})^2}\right) +  \left(\frac{\alpha m}{G \mathcal{M}} - 2\mu_b r_+\right)^{2}\right)
\end{aligned}
\end{equation}
and finally, we arrive at the desired expression:
\begin{equation}
\begin{aligned}
\label{imaginary_part_frequency}
    \omega_I \approx \mu_b (\mu_b \ G \mathcal{M} )^{4l+4} \left(\frac{am}{G \mathcal{M}}-2\mu_b r_+\right)\frac{2^{4l+2}\left(2l+n_r+1\right)!}{\left(n_r+l+1\right)^{2l+4} \ n_r!}\left[\frac{l!}{(2l)!(2l+1)!}\right]^2\times \\
     \times \prod^{l}_{j=1}\left(j^2 \ \left(1-\frac{\alpha^2}{(G \mathcal{M})^2}\right) +  \left(\frac{\alpha m}{G \mathcal{M}} - 2\mu_b r_+\right)^{2}\right) \ ,
\end{aligned}
\end{equation}
in agreement with the result of \cite{Detweiller}.
\begin{figure}
    \centering
    \includegraphics[width=0.9\linewidth]{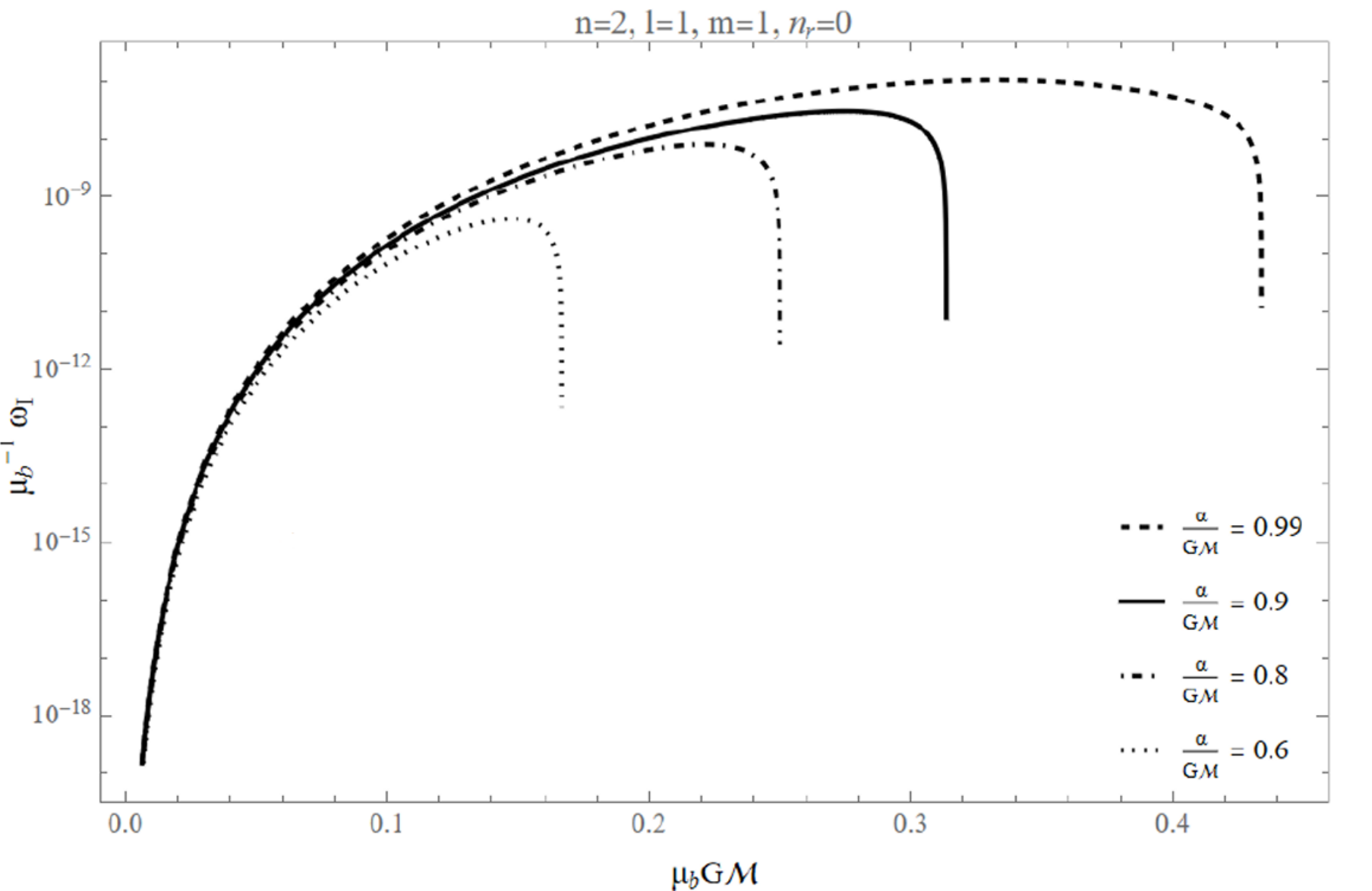}
    \caption{The growth rate stemming from eq.\eqref{imaginary_part_frequency} for the dominant ''$2p$-state", for different values of the BH spin ratio $\alpha/(G \mathcal{M})$.}
    \label{growth_rate}
\end{figure}
One can see from the expression of $\omega_I$, that this quantity is directly proportional to $P_R$, given by equation \eqref{real_part_P}, multiplied by a positive quantity, since each term apart from $P_R$ of \eqref{imaginary_part_frequency_help2} is positive-definite. So, the sign of the imaginary part of the frequency $\omega_I$ is completely determined by the sign of $P_R$. By a simple check at \eqref{real_part_P}, as we mentioned earlier, the first term in the numerator is much larger than the second one, due to the non-relativistic limit $\mu_b G \mathcal{M} \ll 1$, and this means that the sign of $\omega_I$, is completely determined by the sign $P_R$. Moreover, in the case of the Schwarzschild BH ($\alpha=0$), $P_R=-2G \mathcal{M}\mu_b<0$, i.e. it is strictly negative and independent of the azimuthal state, imposing that only damping modes exist, signifying the close relationship between rotation and superradiance in the case of BH. In this respect, $P_R>0$ denotes the superradiance condition, which reads,
\begin{equation}
    \omega_R <\mu_b < m \Omega_H 
    \label{superadiance_condition}
\end{equation}
where
\begin{equation}
\Omega_H \equiv \frac{\alpha}{\alpha^2 + r^{2}_+}
\label{Omega_horizon}
\end{equation}
is the angular momentum of the outer horizon of the BH. We mention here that this condition has been also derived by Bekenstein \cite{Bekenstein:1973mi} from first principles and more simple considerations as the Hawking's area theorem and the assumption for the validity of the null energy condition for the scalar field at the event horizon. The dominant mode is the $2p$-axion state, corresponding to 
\begin{equation}
    l=m=1, \  n_r=0\quad \text{and} \quad n=2 \ ,
\end{equation}
for which the superradiance rate is given by:
\begin{equation}
    \begin{aligned}
\omega_I(2p)=\mu_b\left(\frac{\alpha}{G\mathcal{M}}\right)\frac{\left(\mu_b G\mathcal{M}\right)^8}{24}
    \end{aligned}
    \label{2p_imaginary_omega}
\end{equation}
in agreement with \cite{Detweiller}, under the assumption of the non - relativistic approximation and also considering highly rotating BHs.

\color{black}
\section{Takagi mode decomposition, Covariance Matrix and System Reduction} 
\label{appC:takagi}

In this Appendix  we discuss the so-called Takagi decomposition~\cite{Takagi1933} (or, more correctly Autonne-Takagi, since earlier Autonne  also presented a similar decomposition~\cite{autonne}), which, together with the 
Bloch-Messiah/Euler decomposition, 
is one of the several matrix decompositions applied to symmetric or symplectic matrices encountered in quantum optics~\cite{Houde:2024mkj}, but also in our entangled-graviton situation, discussed in section \ref{sec:Multimode}.
\color{black} As explained in section \ref{sec:Takagi}, this method also proves essential in connecting our theoretical findings regarding correlations of entangled squeezed graviton states to practical spectral modes of interferometric detectors. \color{black} 

The  kernel in the exponent of the squeezing operator \eqref{Scattering_Multimode_Squeeze_GR} appears in the form of a quadratic Hamiltonian $\mathcal H$, which is responsible for  generating multimode squeezing. Such a structure is common in multimode squeezing of photons in quantum optics~\cite{multimode} and produces, for example, the observed squeezing in parametric down conversion~\cite{PhysRevA.31.2409}. Since all such $\mathcal H$ lead to symplectic transformations of phase space, the squeezing kernel is  connected to the symplectic group. 
\color{black}
It will be useful to place our discussion in terms of a finite set $\{a_i^{\dagger},a_i\}_{i=1,\cdots, n}$ of quantum oscillator creation and annihilation  and write the squeezing operator of the form \eqref{Scattering_Multimode_Squeeze_GR}, as follows,
\begin{equation}
\label{discretekernel}
\mathcal{S}=\exp\left[\dfrac{1}{2}\hat{\alpha}^+ \mathcal{G} \hat{\alpha}^\dagger-h.c.\right] =\exp\left[\dfrac{1}{2}\sum _{I,J}\left( \mathcal{G}_{IJ}\widehat{a_{I}}^{\dagger}\widehat{a}_{J}^{\dagger}-\mathcal{G}_{IJ}^{\ast }\widehat{a}_{I}\widehat{a}_{J}\right)\right],\ \ I,J=1,\dots,N\ ,
\end{equation}
where $\hat a = (\hat a_1, \dots , \hat a_N)$ denote an $N-$ dimensional row of annihilation operators and 
$\hat a^+ = (a^\dagger_1, \dots a_N^\dagger)$ denotes the corresponding row of creation operators, to be distinguished from the $N$-dimensional column vector of creation operators, $\alpha^\dagger=(\alpha^+)^T$, where $T$ denotes matrix transposition. In other words, the $+$ operation corresponds to just taking the hermitian conjugate of each element, but we use such notation to incorporate the fact that the elements are operators rather than c-numbers. Moreover, $\mathcal{G}$ is the  Fock space squeezing kernel, which is a complex and symmetric matrix. It is important to mention that the diagonal elements of $\mathcal
G$ are connected with single mode squeezing for each mode, while the off-diagonal ones induce entanglement between the modes,  thus providing its  multi-mode squeezing features.

\subsection{Autonne-Takagi Decomposition}\label{subsec:ATdecomp}

A useful linear algebra technique to study multimode squeezing is due to Autonne~\cite{autonne} and Takagi~\cite{Takagi1933}, which states that, for any complex and  symmetric matrix $M\in \mathbb{C} ^{l\times l}$, there exist a unitary matrix, $W$, and a \emph{non-negative} diagonal matrix, $\Sigma$,
\be\label{diagLambda}
\Sigma =\begin{pmatrix}
\sigma_{1} &  &  &  \\
 & \sigma_{2} &  &  \\
 &  & \ddots  & \sigma_{l}
\end{pmatrix}\,,\quad \sigma_i \ge 0\,, \, \quad i=1, \dots, l\,,
\ee
such that,
\be\label{takagidec}
M=W\, \Sigma \, W^T\,.
\ee
We stress at this point that it is the $W^T$ and not $W^\dagger$ that appears in the Takagi decomposition mapping \eqref{diagLambda}. 

In the following, we shall apply this transformation to multimode squeezing in the Fock basis, in order to relate the physical modes corresponding to $\hat{\alpha}_i$, with the modes, $\hat{b}_i$, defined due to the above transformation. Applying the decomposition \eqref{takagidec} to $\mathcal{G}$ in \eqref{discretekernel}, we obtain, 
\begin{equation}\label{stag}
    \mathcal{S}=\exp\left[\frac{1}{2}(\alpha^+W)\,\Sigma\,(W^T\alpha^\dagger)-h.c.\right]\ , 
\end{equation}
from which we define the modes by the creation operators,
\begin{equation}\label{bmodes}
    \hat{b}^+=\alpha^+W,
\end{equation}
and the corresponding annihilation operators, 
\begin{equation}\label{bmodesannil}
    \hat{b}=\hat{\alpha}W^\star,
\end{equation}
since $(W^\dagger)^T=W^\star$ and the elements of $W$ are c-numbers. 
Thus, we obtain the following diagonalized form of \eqref{discretekernel},
\begin{equation}
    \mathcal{S}(\Sigma)=\exp\left[\dfrac{1}{2}\sum ^{N}_{i=1}\sigma_{i}\left( \widehat{b}_{i}^{\dagger}\right) ^{2}-h.c.\right]
\end{equation}
which leads to $N$ independent single mode squeezed states, that in component form are given by $\hat{b}^\dagger_i=W_{ij}\hat{\alpha}_j^\dagger$. In bi-partite systems, the modes associated with $b_i$  are known as Schmidt modes~\cite{Schmidt1907, Fabre:2019nza, Houde:2024mkj}. In our context, though, which goes beyond simple bipartite systems, we prefer to call them  Takagi-Autone-Schmidt (TAS) modes. Since the relation between the physical modes and the transformed TAS modes is linear, they share the same vacuum state, $|0\rangle$, since,

\begin{equation}
    \hat{b}_i\vert0\rangle=W_{ij}^\star\hat{\alpha}_i\vert0\rangle=0, \ i=1,\dots,N\ .
\end{equation}
For the excited states, we acquire, 
\begin{equation}
    \hat{b}^\dagger_i\vert 0\rangle =\sum_j W_{ij}\hat{\alpha}^\dagger_j \vert 0\rangle ,
\end{equation}
implying that a TAS excitation corresponds to a superposition of the field's physical excitations.
Moreover, since $W$ is unitary, we have that $\hat{\alpha}^\dagger_i\hat{\alpha}_i\rightarrow \hat{b}_i^\dagger \hat{b}_i$, implying that the free Hamiltonian reads $\hbar \omega_i\hat{b}_i^\dagger \hat{b}_i$. Thus, the TAS modes are \textit{dynamically} decoupled harmonic oscillators, each being in a single mode squeezed state. Specifically, in the $i$-th mode, the squeezing operator reads,
\begin{align}\label{ithschmidt}
S_i(\sigma_i) = \exp\left( \frac{1}{2} \sigma_i \hat{b}_i^{\dagger 2} - \frac{1}{2} \sigma_i \hat{b}_i^2   \right) \,,
\end{align}
and acts on the vacuum to produce the single-mode squeezed vacuum,
\begin{align}\label{sqvacithschmidt}
\ket{r_ie^{i\theta}} = S_i(\sigma_i) \ket{0}, 
\end{align}
with $\sigma_i=r_ie^{i\theta}$, the complex squeezing parameter. On defining  quadrature operators,
\be\label{quadrxpb}
\widehat{\widetilde{x}}_i = \frac{1}{\sqrt{2}} (\hat{b}_i + \hat{b}_i^\dagger), \qquad
\widehat{\widetilde{p}}_i = \frac{1}{\sqrt{2}i} (\hat{b}_i - \hat{b}_i^\dagger)\,,
\ee
 in terms of which the squeezing operator \eqref{ithschmidt} acquires a symplectic structure in ``b-mode-phase-space'' $(\widetilde x,\widetilde p)$  of the form:
 \begin{equation}
 S(\sigma_i) = \exp\Big(-i\sigma_i \,\hat{\widetilde{x}}_i \, \hat{\widetilde{p}}_i \Big)\,,
 \end{equation}
 we obtain in the squeezed vacuum:
\be\label{sqvaCXPB}
\langle \hat{\widetilde{x}}_i^2 \rangle = \frac{1}{2} \left[ \cosh(2r)-\cos\theta~\sinh(2r) \right], \qquad
\langle \hat{\widetilde{p}}_i^2 \rangle = \frac{1}{2} \left[ \cosh(2r)+\cos\theta~\sinh(2r) \right]\,.
\ee
The squeezed vacuum state in the Fock basis is:
\be\label{sqvacfock}
\ket{r_i} = \frac{1}{\sqrt{\cosh r_i}} \sum_{n=0}^\infty \frac{\sqrt{(2n)!}}{n!} \left(-\frac{1}{2} \tanh r_i\right)^n \ket{2n}\,.
\ee
Only even particle numbers appear. The mean photon number is given by:
\begin{align}\label{meanphot}
\langle \hat{n}_i \rangle = \sinh^2 r_i\,.
\end{align}
As we have discussed in the text, section \ref{sec:number_of_gravitons}, a similar expression characterises approximately the mean squeezed-graviton modes squeezing parameter. 

The total multi-mode state is analyzed as a simple product of independent single mode squeezed states, 
\begin{equation}
    \vert \Psi\rangle =\bigotimes_i S_i(\sigma_i)\vert 0\rangle \ .
    \label{single_mode_squeezed_states}
\end{equation}
With this form, calculations are readily simplified, since they are reduced to computations regarding the TAS modes alone, which are trivial to obtain, due to the structure of \eqref{single_mode_squeezed_states}. Specifically, calculations are reduced to the following standard correlations, 
\begin{align}
   & \langle \hat{b}^\dagger_m\rangle=\langle \hat{b}_m\rangle=0\ ,\label{b_expectation}\\
   & \langle \hat{b}^\dagger_m\hat{b}_n\rangle= \langle \hat{b}_m\hat{b}^\dagger_n\rangle-1=\sinh^2(\sigma_n)\ \delta_{mn}\ ,\label{bbdagger_expectation}\\
   &  \langle \hat{b}_m\hat{b}_n\rangle= \langle \hat{b}^\dagger_m\hat{b}^\dagger_n\rangle=\cosh(\sigma_n)\sinh(\sigma_n)\ \delta_{mn}\ ,\label{bb_expectation}
\end{align}
where, in this case, the expectation values have been taken with respect to the multi-mode squeezed state~\eqref{single_mode_squeezed_states}. Then, since $\hat{b}_m$ and $\hat{b}^\dagger_m$ are zero-mean operators, due to Wick's theorem, all their higher-order correlators   can be calculated from the second-order ones. Specifically, every odd power correlations vanish, {\it e.g.} $\langle \hat{b}_m\hat{b}_n\hat{b}_l\rangle=0$. For higher (even) order correlations, reduction to second order is obtained, for instance, 
\begin{equation}
\begin{aligned}
    \langle\hat{b}_m\hat{b}_n\hat{b}_k\hat{b}_l\rangle&= \langle\hat{b}_m\hat{b}_n\rangle\langle\hat{b}_k\hat{b}_l\rangle+\langle\hat{b}_m\hat{b}_k\rangle\langle\hat{b}_n\hat{b}_l\rangle+\langle\hat{b}_m\hat{b}_l\rangle\langle\hat{b}_n\hat{b}_k\rangle\\
    &=\cosh(\lambda_m)\ \left[\sinh(\lambda_k)\delta_{mn}\delta_{kl}+\sinh(\lambda_n)(\delta_{mk}\delta_{nl}+\delta_{ml}\delta_{nk})\right]
    \,,\end{aligned}
\end{equation}
where we used \eqref{bbdagger_expectation} and
\eqref{bb_expectation}.

\subsection{Covariance Matrix of Gaussian states}\label{subsec:covmatrix}

Actually, the above statement follows the more general one, according to which every Gaussian state can be fully characterized by its first moment (mean), $\langle\hat{\xi}\rangle$, and its covariance matrix, $\sigma_{ij}$, which is defined though the quadratures~\cite{RevModPhys.84.621},
\begin{align}\label{xicorr}
\hat{\xi} = (\hat{x}_1, \hat{p}_1, \ldots, \hat{x}_{N}, \hat{p}_{N})^T , 
\quad \text{with} \quad 
\hat{x}_j = \frac{1}{\sqrt{2}} (\hat{a}_j + \hat{a}_j^\dagger), \quad
\hat{p}_j = \frac{1}{\sqrt{2}i} (\hat{a}_j - \hat{a}_j^\dagger), \quad i,j=1, \dots, N\,,
\end{align}
as,
\begin{equation}
\sigma_{ij} = \frac{1}{2} \langle \{\Delta\hat{\xi}_i, \Delta\hat{\xi}_j\} \rangle \ ,
\label{covariance_matrix_definition}
\end{equation}
where $\{,\}$ denotes anti-commutation, $\Delta\hat{\xi}_i=\hat{\xi}_i-\langle\hat{\xi}_i\rangle$, and the expectation value is taken with respect to the Gaussian state at hand. This treatment constitutes the phase space representation. The quadratures, obey the usual commutation relations of conjugate variables  (in natural units, $\hbar=1$), 
\begin{equation}
    [x_i,p_j]=i\delta_{ij},\  [x_i,p_j]=0\ \text{and}\  [p_i,p_j]=0\ ,
\end{equation}
and can be summarized to the following expression,
\begin{equation}
    [\hat{\xi}_i,\hat{\xi}_j]=i\Omega_{ij},
    \label{xi_commutation_relation}
\end{equation}
where $\Omega$ is the $2N\times2N$ matrix,
\begin{equation}
    \Omega=\begin{pmatrix}
  0 & \mathbb{I}_N \\
  -\mathbb{I}_N &0
    \end{pmatrix}\ ,
    \label{Omega_Matrix}
\end{equation}
satisfying $\Omega^T=\Omega^{-1}=-\Omega$. In the following, we present the inequality that relates $\Omega$ and the covariance metric, which is the seed of the uncertainty principle. Consider $\Delta\hat{\xi}_i\Delta\hat{\xi}_j$, break it into its symmetric (anti-commutator) and anti-symmetric (commutator) parts and take its expectation values with respect to an arbitrary state, $\vert\phi\rangle$, 
\begin{equation}
    \langle\Delta\hat{\xi}_i\Delta\hat{\xi}_j\rangle_\phi =\frac{1}{2}\langle \{\Delta\hat{\xi}_i,\Delta\hat{\xi}_j\}\rangle_\phi+\frac{1}{2}\langle[\Delta\hat{\xi}_i,\Delta\hat{\xi}_j]\rangle_\phi\ . 
\end{equation}
Since, $\vert\tilde{\phi}\rangle =\Delta\hat{\xi}_j\vert\phi\rangle$ is a vector on the Hilbert space has to has positive inner product, $\langle\tilde{\phi}\vert\tilde{\phi}\rangle\geq0$, implying, 
\begin{equation}
   \frac{1}{2}\langle \{\Delta\hat{\xi}_i,\Delta\hat{\xi}_j\}\rangle_\phi+\frac{1}{2}\langle[\Delta\hat{\xi}_i,\Delta\hat{\xi}_j]\rangle_\phi\ \geq 0\, . 
\end{equation}
The first term in the above inequality is the covariance matrix, $\sigma_{ij}$. In view of the commutation relations \eqref{xi_commutation_relation}, we arrive at the following matrix inequality, 
\begin{equation}
    \sigma+\frac{i}{2}\Omega\geq 0.
    \label{lower_limit_uncertainty_principle}
\end{equation}
Combining with Cauchy-Schwarz inequality, $\vert\langle\phi\vert\tilde{\phi}\rangle\vert^2\leq \langle\phi\vert\phi\rangle\langle\tilde{\phi}\vert\tilde{\phi}\rangle$, the above implies the well known uncertainty principle. In other words, \eqref{lower_limit_uncertainty_principle} represents the lower limit for the product between the variances of conjugate variables. The covariance matrix represents the state dependent correlation between the conjugate variables, which introduces a stronger inequality for the uncertainty principle.  Notice that in the limit of no correlation, the usual $1/2$ (in natural units, $\hbar=1$) lower limit is obtained.

\subsection{System Reduction}\label{subsec:reduct}

The phase space representation enforces us with an elegant way of tracing out modes in the case of Gaussian states. Suppose that we divide the system in two parts, say $A$ and $B$. Then, we can always rearrange the quadratures and bring them in the following form, 
\begin{equation}
    \hat{\xi}_{AB}=(\hat{\xi}_A,\hat{\xi}_B)^T.
\end{equation}
Consequently, the covariance matrix reads, 
\begin{equation}
    \sigma=\begin{pmatrix}
        \sigma_A & \sigma_{AB}\\
        \sigma_{AB}^T & \sigma_B
    \end{pmatrix}\ ,
\end{equation}
where $\sigma_A\ (\sigma_B)$ is the sub-block of correlations between the quadratures of system A (B), while $\sigma_{AB}$ contains their mutual correlations,
\begin{equation}
    (\sigma_A)_{ij}=\frac{1}{2}\langle \{\hat{\xi}^{(A)}_i,\hat{\xi}^{(A)}_i\}\rangle, \ (\sigma_B)_{ij}=\frac{1}{2}\langle \{\hat{\xi}^{(B)}_i,\hat{\xi}^{(B)}_i\}\rangle \ \ \text{and}\ \ (\sigma_{AB})_{ij}=\frac{1}{2}\langle \{\hat{\xi}^{(A)}_i,\hat{\xi}^{(B)}_i\}\rangle\ .
\end{equation}
The total density matrix can be written as $\hat{\rho}_{AB}=\hat{\rho}(\langle\xi_{AB}\rangle,\sigma_{AB})$ and if we trace out the system $B$, we formally obtain the reduced density matrix,
\begin{equation}
    \hat{\rho}_A=Tr_B(\hat{\rho}_{AB})\ .
\end{equation}
However, this is proven quite impractical in most situations. However, dealing with Gaussian states, in the phase space representation, simplifies the situation enormously. To this end, we first mention that tracing out modes in a Gaussian state, the remaining state is also Gaussian, and fully determined by its mean and covariance matrix. In phase space representation, it is proven that the reduced density matrix can be obtained by simply keeping only the sub-block of system $A$ and its corresponding mean, $\langle\hat{\xi}_A\rangle$. In other words, 
\begin{equation}
    \hat{\rho}_A=\hat{\rho}(\langle\hat{\xi}_A\rangle,\sigma_A).
\end{equation}
For a derivation of this result, using the Wigner quasi-probability distribution, see~\cite{bishop2006pattern}. Suppose now that we are interested in (or we have access to) only one mode, say $i=0$. Then, the mean and the covariance matrix are given as follows,
\begin{equation}
  \langle\hat{\xi}_0\rangle=\begin{pmatrix}
      \langle\hat{x}_0\rangle\\
      \langle\hat{p}_0\rangle
  \end{pmatrix} ,\ \ \   \sigma_0= \begin{pmatrix}
        \langle \hat{x}_0^2  \rangle &  \frac{1}{2}\langle\{ \hat{x}_0,\hat{p}_0   \}\rangle\\
         \frac{1}{2}\langle\{ \hat{p}_0,\hat{x}_0   \}\rangle &  \langle \hat{p}_0^2   \rangle
    \end{pmatrix}\ ,
    \label{single_mode_reduced}
\end{equation}
with the expectation values taken with respect to the Gaussian state. Since, by definition, the quadratures are linear combinations of creation/annihilation operators, the above correlations are reduced to the calculation of correlations between $\hat{\alpha}_0$ and $\hat{\alpha}_0^\dagger$. Thus, specifying the Autonne-Takagi decomposition~\eqref{takagidec}, $(W,\{\sigma_n\})$, we can fully specify the reduced state of any subsystem. 
In order to demonstrate this result, we consider the case of the reduced system \eqref{single_mode_reduced}, as part of the multi-mode (pure) state, $\vert\Psi\rangle$, given by \eqref{single_mode_squeezed_states}. Firstly, we trivially obtain that the reduced system is zero-mean,
\begin{align}
    \langle\hat{\xi}_0\rangle_\Psi=0\ ,
\end{align}
since $\langle\hat{\alpha}_0\rangle_\Psi=\langle\hat{\alpha}^\dagger\rangle_\Psi=0$, in view of~\eqref{b_expectation}. For the elements of the covariance matrix in~\eqref{single_mode_reduced} we obtain,
\begin{align}
   &  \langle \hat{x}_0^2\rangle_\Psi=\frac{1}{2}+\langle \hat{N}_0\rangle_\Psi+Re\{\langle\hat{\alpha}^2\rangle_\Psi\}\\
   & \langle\hat{p}_0^2\rangle_\Psi=\frac{1}{2}+\langle \hat{N}_0\rangle_\Psi-Re\{\langle\hat{\alpha}^2\rangle_\Psi\}\\
   & \frac{1}{2}\langle\{x_0,p_0\}\rangle_\Psi=Im\{\langle\hat{\alpha}^2\rangle_\Psi\}
\end{align}
 Using~\eqref{bbdagger_expectation} and \eqref{bb_expectation}, we express every relevant term with respect to the TAS decomposition $(W,\sigma_n)$, as follows,
\begin{align}
    & \langle \hat{N}_0\rangle_\Psi=\sum_i\vert W_{i0}\vert^2\sinh^2\sigma_i\\
    &Re\{\langle \hat{\alpha}_0^2\rangle_\Psi\}=\sum_iRe\{W_{i0}^2\}\sinh\sigma_i\cosh\sigma_i\\
     &Im\{\langle \hat{\alpha}_0^2\rangle_\Psi\}=\sum_iIm\{W_{i0}^2\}\sinh\sigma_i\cosh\sigma_i \,.
\end{align}

\subsection{Symplectic Transformations and von Neumann Entropy}\label{sec:appD}

As already mentioned, every Gaussian state is characterized by its first moments and its covariance matrix, since every higher order correlation is expressed with respect to them. Thus, every Gaussian state is described by a density matrix, $\rho$, which is fully determined by $\langle \hat{\xi}\rangle$ and the state's  covariance matrix $\sigma$,
\begin{equation}
    \hat{\rho}=\hat{\rho}(\langle\xi\rangle,\sigma)\  .
\end{equation}
The inequality \eqref{lower_limit_uncertainty_principle} guarantees the positivity of the density matrix (non-negative eigenvalues). For the vacuum state:
\begin{equation}
    \langle\hat{\xi}\rangle_{vac}=0\ \ \text{and}\ \ \sigma_{vac}=\frac{1}{2}I,
    \end{equation}
while for a thermal state, with mean particle number  $n_{th}$, 
\begin{equation}
    \langle\hat{\xi}\rangle_{th}=0\ \ \text{and}\ \ \sigma_{th}=\left(n_{th}+\frac{1}{2}\right)I.
    \end{equation}
Thus, the density matrix of the vacuum is denoted as $\hat{\rho}_{vac}=\rho(0,I/2)$, while that of a thermal state, with mean particle number, $n_{th}=n$, as,
\begin{equation}
    \hat{\rho}_{th}(n)=\hat{\rho}(0,\sigma_{th}) \,.
\end{equation}
The covariance matrix for a single mode squeezed state, with complex squeezing parameter, as in \eqref{ithschmidt}, reads, 
\begin{equation}
    \sigma_{sq}=\frac{1}{2}\begin{pmatrix}
        \cosh(2r)-\cos\theta\sinh(2r) & -\sin\theta\sinh(2r)\\
        -\sin\theta\sinh(2r) & \cosh(2r)+\cos\theta\sinh(2r) 
    \end{pmatrix}\ ,
\end{equation}
in which we see that a non-trivial phase leads to correlation between the conjugate variables. Such correlations lead to a tilted ellipse in the phase space. For $\theta=0$, we obtain the well known result of a horizontal ellipse, with $\langle x^2\rangle=e^{-2r}/2$ and $\langle p^2\rangle=e^{2r}/2 $.

By definition, the covariance matrix $\sigma$ is a $2N\times2N$ matrix, which is real, positive definite and symmetric. This is a crucial property, implying that $\sigma$ is subjected to \textit{Williamson's theorem}~\cite{Williamson1936OnTA,RevModPhys.84.621}, which states that for any real, positive definite, even dimensional  $2N\times2N$ matrix, there exists a symplectic matrix $S\in Sp(2N,\mathbb{R})$, such that, 
\begin{equation}
    \sigma=S^T\sigma_{sym}\ S
\end{equation}
where $\nu_i\geq1/2, \ \forall i$, are the symplectic eigenvalues of $\sigma$, while $\sigma_{sym}$, acquires the diagonal form, 
\begin{equation}
    \sigma_{sym}=diag(\nu_1,\nu_1,\nu_2,\nu_2\dots,\nu_N,\nu_N)\ .
\end{equation}
Symplectic transformations is the group of quadrature transformations,
\begin{equation}
    \hat{\xi}^\prime=S\hat{\xi},
    \label{symm_transformation}
\end{equation}
that preserve the CCRs, i.e.
\begin{equation}
    S\Omega S^T=\Omega
\end{equation}
and corresponds to a Gaussian unitary, that is, a unitary transformation, $U_S$,  
\begin{equation}
    \hat{\rho}\rightarrow\hat{\rho}^\prime=U_S\ \hat{\rho} \ U^\dagger_S,
\end{equation}
that preserves the Gaussianity of the state. What is important here is that the set $\{\nu_i\}$ fully determines the Gaussian state and can be found by taking the absolute value of the spectrum of $ i\sigma\Omega$. For a comprehensive introduction and more details for the topic discussed, we refer the reader to~\cite{Demarie:2012hxi}.\par 
An arbitrary density matrix, $\hat{\rho}(\langle\xi\rangle,\sigma)$, can be written in terms of a zero mean, $\rho(0,\sigma)$, via the action of a displacement operator, $D(\langle\xi\rangle)$, 
\begin{equation}
    \hat{\rho}(\langle\xi\rangle,\sigma)=D(\langle\xi\rangle)\hat{\rho}(0,\sigma)D^\dagger(\langle\xi\rangle)\ . 
\end{equation}
Moreover, in view of the symplectic transformation \eqref{symm_transformation}, we observe that if $\langle\hat{\xi}\rangle=0\rightarrow\langle\hat{\xi}^\prime\rangle=0$, {\it i.e.} a zero-mean state remains a state of zero mean under a symplectic transformation. Hence, according to Williamson's theorem, a general Gaussian state can be decomposed as, 
\begin{equation}
    \hat{\rho}(\langle\xi\rangle,\sigma)=D(\langle\xi\rangle)U_S\hat{\rho}(0,\sigma_{sym})U_S^\dagger D^\dagger(\langle\xi\rangle)
\end{equation}
Remarkably, the zero-mean state associated with $\sigma_{sym}$ corresponds to a thermal bath of mean particle numbers $n_i=\nu_i-1/2$, 
\begin{equation}
    \hat{\rho}(0,\sigma_{sym})=\bigotimes_i\hat{\rho}_{th}\left(\nu_i-\frac{1}{2}\right)\  .
\end{equation}
Consequently, the von Neuman entropy, $S=-{\rm Tr}[\hat{\rho} \log(\hat{\rho})]$, is given by, 
\begin{equation}
    S(\rho_{sym})=\sum_i\left[\left( \nu_i + \frac{1}{2} \right) \log \left( \nu_i + \frac{1}{2} \right)
- \left( \nu_i - \frac{1}{2} \right) \log \left( \nu_i - \frac{1}{2} \right)\right]\ .
\label{entropy}
\end{equation}
The above entropy monotonically increases and vanishes only in the limit of $\nu_i\rightarrow1/2$, for all $i=1,2,\dots N$. In this sense, the symplectic spectrum of the covariance matrix 
indicates when a state is pure or thermal and \eqref{entropy} quantifies the pertinent information loss.\par

\color{black}

\begin{center}\underline{\emph{A relevant Example}: Two-mode Squeezed States} \end{center}

For us ({\it cf.} \eqref{Scattering_Multimode_Squeeze_GR},  \eqref{GLeqGR} and \eqref{EPRGRsymm}) \color{black} the most interesting example is for $n=2$, corresponding to two annihilation $\widehat a_i\,, i=1,2$ (and the corresponding creation) operators, with the canonical two-mode squeezed state generated by the squeezing operator:
\begin{align}\label{examplesq2}
S_2(r) = \exp\left[ r \left( \hat{a}_1^\dagger \hat{a}_2^\dagger - \hat{a}_1 \hat{a}_2 \right) \right]=\exp\left[\sum_{ij}M_{ij}\hat{a}_i^\dagger\hat{a}_j^\dagger-h.c\right], \quad r > 0\,,
\end{align}
acting on the vacuum:
\begin{align}\label{examples2qpsi}
\ket{\Psi} = S_2(r) \ket{0}_1 \otimes \ket{0}_2.
\end{align}
\color{black} In this case, we have ({\it cf.} \eqref{takagidec}):
\begin{align}\label{matrixMW}
  M = r \begin{pmatrix} 0 & \,\, 1 \\1 & \,\, 0 \end{pmatrix} \, ,\quad W = \frac{1}{\sqrt{2}} \begin{pmatrix} 1 & \,\, i  \\1 & \,\, -i \end{pmatrix} \, , \quad 
  \Sigma = r \begin{pmatrix} 1 & \,\, 0 \\0 & \,\, 1 \end{pmatrix} \, ,
\end{align}
and from \eqref{bmodes}
we obtain the following Takagi-Schmidt modes:,
\begin{equation}
\label{bmodesexample}
\begin{aligned}
    &\hat{b}_1=\frac{1}{\sqrt{2}}\left( \hat{\alpha}_1+\hat{\alpha}_2    \right)\\
    &\hat{b}_2=-\frac{i}{\sqrt{2}}\left( \hat{\alpha}_1-\hat{\alpha}_2 \right)\ .
\end{aligned}
\end{equation}
Thus, the two mode squeezed vacuum can be written as two independent single mode squeezed states of the same squeezing parameters, $\sigma_1=\sigma_2=r$,
\begin{equation}
    \vert \Psi\rangle=S_2(r)\vert 0\rangle =S_{b_1}(r)S_{b_2}(r)\vert 0\rangle,
\end{equation}
where $\vert 0\rangle = \ket{0}_1 \otimes \ket{0}_2$ and 
\begin{equation}
    \begin{aligned}
        &S_{b_1}(r)=\exp\left[ \frac{r}{2}(\hat{b}^\dagger_1)^2-\frac{r}{2}\hat{b}_1^2  \right]\\
        &S_{b_2}(r)=\exp\left[ \frac{r}{2}(\hat{b}^\dagger_2)^2-\frac{r}{2}\hat{b}_2^2 \right]\ .
    \end{aligned}
\end{equation}
\color{black}
Then the covariance matrix of the two-mode squeezed state is:
\begin{align}\label{covmatrexample2}
\sigma =
\frac{1}{2}
\begin{pmatrix}
\cosh(2r) & 0 & \sinh(2r) & 0 \\
0 & \cosh(2r) & 0 & -\sinh(2r) \\
\sinh(2r) & 0 & \cosh(2r) & 0 \\
0 & -\sinh(2r) & 0 & \cosh(2r)
\end{pmatrix}.
\end{align}
The full covariance matrix $\sigma$ can be measured using homodyne detection in all modes and quadratures.

To compute the entanglement entropy of mode 1, we trace out mode 2. The reduced covariance matrix is the upper-left $2 \times 2$ block of the matrix in \eqref{covmatrexample2}:
\begin{align}\label{subsystem1}
\sigma_{(1)} =
\frac{1}{2}
\begin{pmatrix}
\cosh(2r) & 0 \\
0 & \cosh(2r)
\end{pmatrix}
= \frac{\cosh(2r)}{2} I_2.
\end{align}
For a single-mode covariance matrix \( \sigma_{(1)} \in \mathbb{R}^{2 \times 2} \), the \textbf{symplectic eigenvalue} \( \nu \) is given by:
\[
\nu = \sqrt{\det \sigma_{(1)}}.
\]
Since:
\[
\det \sigma_{(1)} = \left( \frac{\cosh(2r)}{2} \right)^2,
\]
we obtain:
\begin{align}\label{nudef}
\nu = \frac{1}{2} \cosh(2r).
\end{align}
The von Neumann entropy of the single-mode 1 Gaussian state \cite{RevModPhys.84.621,Serafini:2017rrn,
adesso2014continuous} with symplectic eigenvalue \( \nu \ge \tfrac{1}{2} \) is:\footnote{\color{black} In information theory it is customary to define the von Neumann entropy of a Gaussian state \eqref{vNentr} via the binary logarithm $\rm log_2(x)$, defined as $\rm log_{2}(x) = y $, such that $x=2^y$, which is related to the natural logarithm 
via $\rm log_2(x) = \rm ln(x)/ln2 = ln(x)/0.693$.
\color{black}}
\begin{align}\label{vNentr}
S(\nu) =f(\nu)\equiv \left( \nu + \frac{1}{2} \right) \log \left( \nu + \frac{1}{2} \right)
- \left( \nu - \frac{1}{2} \right) \log \left( \nu - \frac{1}{2} \right).
\end{align}

In our case:
\begin{align}\label{entrexample2}
S(r) = f\left( \nu = \frac{1}{2} \cosh(2r) \right).
\end{align}
This entropy is zero when \( r = 0 \), and increases with the squeezing \( r \), reflecting increased entanglement between modes.

The understanding of the squeezing kernel for the axion cloud source allows, in principle, multimode squeezed light to be injected into the gravitational detectors which mirror the Schmidt modes from our gravitational source. This could lead to greater sensitivity in measuring noise in the gravitational system of quantum origin. The decomposition into Takagi modes~\cite{Takagi1933} is a first step in using quantum information protocols to study entanglement.

\subsection{Special multimode squeezed states related to two-mode GW squeezed states}\label{subsec:2mode}

From our detailed calculations in this article we have found some correlations in the multimode squeezing kernel for axion decay into two gravitons in the CS theory and the axion coupled to GR through its kinetic term, leading to a two axion and two graviton vertex. These processes are the gravitational analogues of parametric down conversion and four wave mixing. Our calculations (in the axion context)  show qualitatively similar behaviours  to those in the optical context. \\
\par

\underline{{\it Parametric down-conversion type squeezing kernel} } \\

\par
For a  squeezing kernel
\be
M\left( k, k'\right) \propto \delta ^{\left( 3\right) }\left( \vec{k}+\vec{k'}\right) 
\ee
there is a powerful simplification since 
\be
\widehat{H}=\dfrac{1}{2}\int d^{3}k\left( \zeta \left( \vec{k}\right) \widehat{a}_{k}^{+}\widehat{a}_{-k}^{+}+h.c.\right), 
\ee
reflecting strong pairwise correlation. This is effectively many two mode correlations,i.e. the system decomposes into independent two-mode squeezing operations for each pair $(\vec k, -\vec k)$. We can take over the analysis from our previous discussion of Takagi decomposition, since it is now block diagonal with known analytic blocks.\\

\underline{\emph{Four-wave mixing Kernel}} \\

This case has also a relatively simple and physically intuitive kernel structure. In the momentum or frequency domain, the kernel in FWM reflects the momentum and energy conservation conditions that arise from the nonlinear interaction of four bosonic fields.
\be
\widehat{H}_{FWin}\propto \int d^{3}k_{1}d^{3}k_{2}d^{3}k_{3}d^{3}k_{4}\delta^{(3)}\left( k_{1}+k_{2}-k_{3}-{k}_{4}\right) \mathcal{G}\left( k_{1},k_{2},k_{3},{k}_{4}\right) \widehat{a}^{\dagger}_{k_{1}}\widehat{a}^{\dagger}_{k_{2}}\widehat{a}^{\dagger}_{k_{3}}\widehat{a}^{\dagger}_{k_{4}}) 
\ee
Consequently a pair of input modes,  with momenta $\vec k_{3},\vec {k}_{4}$, can be  converted into into a pair of output photons with momenta $\vec k_{1}$ and $\vec {k}_{2}$. If (i) $\mathcal{G}$ is sharply peaked around a central mode $\vec {k}_{p}$ and (ii) the modes $\vec {k}_{3,4}$ are classical with complex amplitudes $\alpha_{\vec k_{3}}$ and $\alpha_{\vec k_{4}}$ then 
\[
\alpha_{\vec k_{3,4}} \propto \delta^{(3)}(\vec k_{3,4}-\vec k_{p}).
\]
and 
\[
\mathcal{G}\left( k_{1},k_{2},k_{3},{k}_{4}\right) \to \hat{\mathcal{G}}\left( k_{1},k_{2}\right)
\propto \delta^{(3)}(\vec k_{1}+\vec k_{2}-2\vec k_{p}). \]
On writing  $\vec{k_1}=\vec k_{p}+\vec {q}$ and $\vec{k_2}=\vec k_{p}+\vec {q'}$ we see that we are dealing with symmetric deviations from $\vec k_{p}$ and the momentum conserving delta function is $\delta^{(3)}(\vec q+\vec q')$.  Hence it is akin to two-mode squeezing. This construction also applies to parametric down conversion where the formulae for $\vec k_{1,2}$ change to $\vec{k_1}=\vec k_{p}/2+\vec {q}$ and $\vec{k_2}=\vec k_{p}/2+\vec {q'}$.

\color{black}



\bibliographystyle{apsrev4-2}
\bibliography{squeezing_rev.bib}  

\end{document}